\documentclass[aps,prd,a4paper,showpacs,12pt]{article}
\usepackage{multicol}
\usepackage{subfigure}
\usepackage {graphicx}
\usepackage{color}
\usepackage{xcolor}
\usepackage{threeparttable}
\usepackage{amsfonts}
\usepackage{times}
\usepackage{setspace}
\usepackage{balance}
\usepackage{lastpage}
\usepackage{amsthm}
\usepackage[tbtags]{amsmath}

\usepackage{slashed}
\usepackage{nomencl}
\usepackage{pstricks}
\usepackage{pstricks-add}
\usepackage{pst-plot,pstricks-add}

\usepackage{amsmath,amsfonts,amssymb,simplewick}
\usepackage{float}
\usepackage{amsmath,amssymb,amsfonts,epsfig,graphicx,euscript}%
\usepackage{hyperref}
\hypersetup{bookmarks=true, 
bookmarksnumbered=true,
linktoc=page, 
pdfstartview={FitH}, 
colorlinks=true, 
citecolor=red, 
filecolor=blue, 
linkcolor=blue, 
urlcolor=blue}
\usepackage{amsmath}
\usepackage{amsfonts}
\usepackage{graphicx}
\usepackage{caption}
\usepackage{float}
\usepackage[all]{hypcap}
\numberwithin{equation}{section}
\usepackage{cite}
\usepackage{calc}
\usepackage{mathtools}
\newlength{\spacer}
\newsavebox{\mybox}

\DeclareMathOperator{\Tr}{Tr}

\DeclareMathOperator{\csch}{csch}

\DeclareMathAlphabet{\mathpzc}{OT1}{pzc}{m}{it}
\usepackage{ragged2e}
\renewcommand{\thefootnote}{\fnsymbol{footnote}}

\begin{document}
	\begin{center}
		{\large{\textbf{Casimir free energy for massive scalars: a comparative study of various approaches}}}
		\vspace*{1.5cm}
		\begin{center}
		{\bf M. Sasanpour\footnote{m\_sasanpour@sbu.ac.ir} and S. S. Gousheh\footnote{ss-gousheh@sbu.ac.ir}}\\
		\vspace*{0.5cm}
		{\it{Department of Physics, Shahid Beheshti University, Evin, Tehran, 1983969411, Iran }}\\
		\vspace*{1cm}
	\end{center}
\end{center}
	\begin{center}
		\today
	\end{center}

\renewcommand*{\thefootnote}{\arabic{footnote}}
\setcounter{footnote}{0}
\date{\today}
\textbf{Abstract:}
We compute the Casimir thermodynamic quantities for a massive real scalar field between two parallel plates with the Dirichlet boundary conditions, using three different general approaches and present explicit solutions for each. The Casimir thermodynamic quantities include the Casimir Helmholtz free energy, pressure, energy, and entropy. The three general approaches that we use are based on the fundamental definition of Casimir thermodynamic quantities, the analytic continuation method, and the zero temperature subtraction method. Within the analytic continuation approach, we use two distinct methods which are based on the utilization  of the zeta function  and the Schl\"{o}milch summation formula. We include the renormalized versions of the latter two approaches as well, whereas the first approach does not require one. Within each general approach, we obtain the same results in a few different ways to ascertain the selected cancellations of infinities have been done correctly. We show that, as expected, the results based on the zeta function and the Schl\"{o}milch summation formula are equivalent. We then do a comparative study of the three different general approaches and their results and show that they are in principle not equivalent to each other, and they yield equivalent results only in the massless case. In particular, we show that the Casimir energy calculated only by the first approach has all three properties of going to zero as the temperature, mass of the field or the distance between the plates increases.
Moreover, we show that in this approach the Casimir entropy reaches a positive constant in the high temperature limit, which can explain the linear term in the Casimir free energy.

\vspace*{25pt}


\section {Introduction}
\indent

The Casimir effect, predicted by Hendrik Casimir in 1948~\cite{r1Cas.}, is a direct consequence of the zero-point energy of the quantum fields and has played an important role in various branches of physics such as particle physics~\cite{r32Particle., r322Particle., r323Particle., r324Particle.}, condensed matter and laser physics~\cite{r33Cond., r332Cond., r333Cond., r334Cond.}, nanotechnology~\cite{r33Nano., r332Nano., r333Nano., r334Nano.}, string theory~\cite{r34String., r342String., r343String.}, and cosmology~\cite{r35Cosmo., r352Cosmo., r353Cosmo., r354Cosmo.}. This effect appears when a system is subject to nontrivial boundary conditions, background fields such as solitons, or nontrivial space-time backgrounds. In the experimental aspect, Sparnaay was the first to attempt to observe the Casimir effect~\cite{r3Spar.}, but Lamoreaux et al.~\cite{r4Lamo.} were the first to measure the Casimir force with acceptable precision. For a comprehensive review, see for example~\cite{r32Particle., r30Kimb., r302Kimb., r31Bord2.}. 

In this paper, we explore the differences between three of the commonly used general approaches for calculating the finite temperature Casimir effects for the bosonic case, as has been done for the fermionic case in~\cite{r26Goushe.}. Our reference general approach is based on the fundamental definition of the Casimir thermodynamic quantities which for the Casimir Helmholtz free energy, for example, is the difference between the infinite vacuum Helmholtz free energies of systems subject to the constraints and the corresponding ones that are free from them, both being at the same temperature. We shall henceforth refer to this as the fundamental approach. The second approach is based on the analytic continuation methods, for which we include the zeta function method and the Schl\"{o}milch summation formula method, as two distinct representatives. The third approach is based on the zero temperature subtraction method. We also include the renormalized versions of the latter three methods, and shall refer to them collectively as the zeta function approach (ZFA), the Schl\"{o}milch formula approach (SFA), and the zero temperature subtraction approach (ZTSA), respectively. As mentioned above, both ZFA and SFA are representatives of the analytic continuation approach. As is well known, the result of analytic continuations is unique, and one of the questions that we want to address here, similar to the fermionic case~\cite{r26Goushe.}, is whether this unique result is the physically acceptable one that we seek for the Casimir thermodynamic quantities of the bosonic case.

In order to be concrete, we concentrate on an illustrative example for the bosonic case. Our choice is a massive real scalar field confined between two parallel plates with the Dirichlet boundary condition. In this paper, we present calculations for the Casimir thermodynamic quantities within each of the general approaches mentioned above, and present their results in explicit forms. Moreover, to ascertain the validity of our results, we present or outline a few different ways of obtaining the same results within each general approach. We then do a comparative study of the three general approaches and their results. Before we start with the computations, we briefly review the historical development of the finite temperature Casimir effects and the use of various approaches.


The fundamental definition of the zero temperature Casimir energy, as stated by Casimir in 1948, is the difference between the zero point energies of the system with and without the constraints. Finite temperature Casimir effect was first introduced by Lifshitz~\cite{r2Lif.} in 1956, who calculated the attractive force between two parallel dielectric plates at finite temperature, by introducing fluctuating electromagnetic field. At high temperatures, the Casimir pressure was found to be proportional to the temperature. This term was subsequently denoted as the classical term\footnote{It was named the classical term, since it did not have any factors of $\hbar$ ~\cite{r31Bord2., r20Lokh., r302Kimb.}. In this paper we present an alternative justification for this name.}. Later on, Mehra~\cite{r5Mehr.} in 1967, used the Helmholtz free energy, which we shall henceforth refer to simply as the free energy, to calculate the thermal correction to the zero temperature Casimir pressure for a conducting cubic cavity. In that paper, the Casimir pressure was calculated as the difference between the pressure inside and outside of the cube, both being at the same temperature. His results also included the classical term at high temperatures.

The next major work on thermal corrections is due to Brown and Maclay~\cite{r6Brown.} in 1969, who calculated the electromagnetic stress-energy tensor between two conducting parallel plates. Using the image-source construction, they obtained the components of the tensor as thermodynamic variables, without any divergent terms. However, for the first time, the final results for both the Casimir pressure and energy density included terms due to the black-body radiation which are proportional to $T^4$. 

In a series of papers from 1976 to 1980, Dowker et al.~\cite{r8Dowk76., r8Dowk78., r8Dowk80.} calculated the vacuum expectation value of the stress-energy tensor at finite temperature using the Green function formalism for a scalar field in curved space-time. They used three different renormalization schemes to obtain finite results. First, they subtracted the $(0,0)$ temperature-spatial mode. Second, they used a `Casimir renormalization' as the difference between free energies before and after constructing the boundary, both being at the same temperature, to compute the heat kernel coefficients. This is analogous to the fundamental approach. Third, they subtracted the contribution of the free Green function at the zero temperature, which they referred to as `the standard flat space renormalization prescription'. This is equivalent to ZTSA. The high temperature limit of $\langle T_{00}\rangle $ in their first and third work had terms proportional to $T^4$ and $T$, while Casimir free energy in their second work had terms proportional to $T^3$, $T$ and $T \ln T$. 

In 1978, Balian and Duplantier~\cite{r9Balian.} defined and used the fundamental definition of the Casimir free energy for the electromagnetic field in a region bounded by thin perfect conductors with arbitrary smooth shapes. The high temperature limit of their results for parallel plates was proportional to $T$, while for the enclosures included an additional term proportional to $T\ln T$.

In 1983, Ambj$\o$rn and Wolfram~\cite{r10Wolf.} computed the Casimir energy and entropy for scalar and electromagnetic fields in a hypercuboidal region, using the generalized zeta function along with the reflection formula as an analytic continuation technique. They showed that the high temperature limit of the Casimir energy for the scalar field in a rectangular cavity in $3+1$ dimensions includes terms proportional to $T^4$, $T^2$ and $T\ln T$. In 1991, Kirsten~\cite{r12Kris.} computed the heat kernel coefficients for the grand thermodynamic potential for a massive bosonic field in hypercuboids in n-dimensions subject to the Dirichlet boundary condition, using the zeta function, and in four dimensions obtained terms proportional to $T^4$, $T^3$, $T^2$, $T$ and $T \ln T$. 

In 1992, Elizalde and Romeo~\cite{r51Elizald92.} calculated the high and low temperatures expansions of the free energy for a massive scalar field in hypercuboids of arbitrary dimensions, using multidimensional Epstein zeta functions. They indicated that, as stated in~\cite{r11Plunien.}, to calculate the Casimir free energy, one has to subtract the free energy of the unconstrained boson field, which would eliminate only the $T^4$ term at high temperatures. In 2008, Geyer et al.~\cite{r17Geyer.} suggested a renormalization procedure to calculate the finite temperature free energy, which would supplement the use of zeta function. They stated that the use of zeta function does not include all necessary subtractions, and the terms proportional to powers of $T$ higher than the classical terms obtained in the high temperature limit from the heat kernel method, have to be subtracted. Subsequently, in 2009, Bordag et al.~\cite{r31Bord2.} presented a general picture of the renormalization for the Casimir free energy within the ZFA. They used the heat kernel coefficients for subtraction of the extra terms at low and high temperature limits. 

Most of the work mentioned above have used ZFA. In 2018, Mo and Jia~\cite{r29Junji.} used SFA to calculate the thermal correction of the Casimir free energy for an electromagnetic field in a conducting rectangular box. To eliminate the high temperature divergences, they defined a renormalized free energy by subtracting the free black-body term along with any possible terms proportional to $T^2$ and $T^3$, with reference to Geyer's work~\cite{r17Geyer.}. They showed that after removing these terms, the high temperature limit of their expression for the Casimir free energy included terms proportional to $T \ln (T)$ and $T$. They further removed the $T \ln (T)$ term, reasoning that it is independent of the geometry of the boundary.



As is apparent from the historical outline presented here for the bosonic cases, and also presented in \cite{r26Goushe.} for the fermionic cases, the zeta function has been used extensively in the calculations of the Casimir effects to evaluate the sums over the regular spatial and Matsubara modes, and often as an analytical continuation technique. In some methods, the zeta function is used explicitly to calculate the Casimir thermodynamic quantities, e.g.\ in~\cite{r51Elizald92.}, or implicitly, e.g.\ in the Schl\"{o}milch's formula, or as a supplementary part. Examples of the latter include the heat kernel method and the Bogoliubov transformation where the zeta function is used to evaluate the final summations. The spatial modes of the massive and massless scalar fields with the usual Dirichlet or Neumann boundary conditions are both regular, and the generalized zeta function can be used for summing over them.

As mentioned before, in this paper we compute the Casimir effects for a massive real scalar field between two parallel plates at finite temperatures by three different  general approaches, {\it i.e.}, the fundamental approach, the ZTSA, and the analytic continuation approach, represented here by the ZFA and the SFA. Within each general approach, we display or outline multiple ways of computing the same physical quantities and use various methods and computational techniques to ascertain the selected and delicate cancellations of divergent sums and integrals have been done correctly, all yielding equivalent results. These methods include the Poison summation formula, the Abel-Plana formula, and the Principle of the Argument theorem. For all three approaches mentioned above, we calculate the results for both the massive and massless cases, and show that the massless limits of the massive cases always coincide with the massless cases.

First we use the fundamental approach to calculate all of the Casimir thermodynamic quantities, and   show that the Casimir free energy and pressure decrease linearly with $T$ at high temperatures, while the Casimir energy goes to zero and the Casimir entropy goes to a positive constant.
This linear temperature term can be looked upon as stemming from the nonzero value of Casimir entropy according to the thermodynamic relation $F_{\mbox{\scriptsize Casimir}}(T,L) =E_{\mbox{\scriptsize Casimir}}(T,L) - TS_{\mbox{\scriptsize Casimir}}(T,L)$, which is a reason for calling it the classical term.
We also show that at any temperature all of the Casimir thermodynamic quantities go to zero in the large mass and plate separation limits, as expected. Indeed, in this approach, the subtraction of the thermodynamic quantities of the constrained and unconstrained systems at the same temperature yields the correct extensive results, including the mentioned limits, without having to label any terms in the resulting expressions for the Casimir quantities as unphysical, and subsequently removing them by hand.

Next we calculate the Casimir thermodynamic quantities using ZFA and SFA, both being in the category of analytic continuum approach, and show that, as expected, their results are equivalent. Finally we use the ZTSA. We then show that the results obtained by the three different general approaches are in principle not equivalent. The results of ZFA, SFA, and ZTSA contain extra nonpolynomial terms in variables $T$ and $m$, as compared to those of the fundamental approach. To calculate the renormalized versions of ZFA, SFA, and ZTSA, we first obtain the high temperature limits of these extra nonpolynomial terms, both directly and by the heat kernel method up to and including the black-body term $T^d$ in $d$ space-time dimensions. The expansion of the results for ZFA and SFA yield terms proportional to $T^4$, $T^3$, $T^2$, $T\ln (T)$, $T$ and $\ln (T)$, and for ZTSA yield $T^4$, $T^2$, $T$ and $\ln (T)$. As we shall show, all even powers of $T$, including  $\ln (T)$, are due the thermal free energy of the free case which has not been subtracted, and the odd powers of $T$, except for the classical term, are due to a  nonextensive term generated by the analytic continuation. We then subtract terms with powers of $T$ greater or equal to two, in accordance with the renormalization programs introduced\cite{r17Geyer.,r29Junji.}.The remaining linear term is the correct classical term, while the remaining $T\ln (T)$ and $\ln (T)$ terms are unphysical. Therefore the differences between the results of ZFA, ZTSA and the fundamental approach for the massive case are not resolved by the renormalization programs of the former two. On the other hand, as we shall show, the extra terms in the massless case are simple polynomials and can be removed by the renormalization programs introduced~\cite{r17Geyer.,r29Junji.}, or cancel out in the piston method for the Casimir pressure~\cite{r17Geyer.,r29Teo., r26Cheng.,r27Khoo.}. Hence, for the massless case, the results of the fundamental approach, the ZFA (supplemented with the piston approach or its renormalized version), and the renormalized version of ZTSA are all equivalent. We like to emphasize that the fundamental approach does not require any supplementary renormalization program.


The outline of the paper is as follows. In Sec.\ \ref{Helmholtz free energy}, we present two forms for the free energy of a real scalar field at finite temperature, which is subject to the Dirichlet boundary conditions at two plates but is otherwise free, starting with the path integral formalism. In Sec.\ \ref{FCasimir}, we calculate the Casimir free energy for a massless scalar field using the fundamental approach and show that the Casimir free energy and pressure decrease linearly at high temperatures and go to zero at the large plate separations. In Sec.\ \ref{zeta}, we calculate the Casimir free energy and pressure of a massless scalar field using the ZFA, the SFA, and the ZTSA, obtaining identical results for the first two. These two have extra $T^4$ and $T^3$ terms, while the ZTSA only includes the extra $T^4$ term.
We then show how the renormalization program subtracts these extra terms, yielding the correct results, based on the fundamental approach. In Sec.\ \ref{massive}, we consider a massive scalar field as the simplest nontrivial example, and calculate the Casimir free energy, pressure, energy, and entropy using the fundamental approach. We show that, as expected, they all go to zero in the large mass or large plate separation limits. More importantly, we show that the Casimir free energy and pressure decrease linearly with $T$ at high temperatures, while the Casimir energy goes to zero and the Casimir entropy goes to a positive constant. In Sec.\ \ref{Epstein}, we calculate the Casimir free energy and pressure for the same massive scalar problem as in Sec.\ \ref{massive}, using the other two general approaches, {\it i.e.}, the analytic continuation approach, represented by the ZFA and the SFA, and the ZTSA, including their renormalized versions. We show that, as expected, the results of the SFA, and the ZFA are equivalent. However, we show that none of the four different sets of results that we obtain in  Sec.\ \ref{Epstein} is equivalent to that based on the fundamental approach.  In particular we show that the contain different nonpolynomial functions of $m$ and $T$, which cannot be removed, even in the high temperature limit, by the renormalization programs thus far devised. As a side note, we present the condition under which the piston method would yield the correct Casimir pressure. Finally, in Sec.\ \ref{summary}, we present our conclusions.

\section {The Helmholtz Free Energy}\label{Helmholtz free energy}
\indent
Historically, the first and most commonly-used approach to thermal field theory is the imaginary-time formalism. This approach has its roots in the work of Felix Bloch in 1932, who noticed the analogy between the inverse temperature and imaginary-time~\cite{r36Bloch.}, which led to the so-called temperature Green functions with purely imaginary-time arguments. In 1955, Matsubara presented the first systematic approach to formulate quantum field theory at finite temperature by the imaginary-time formalism, using the Wick rotation~\cite{r37Mats.}. The discrete frequencies in this formalism are known as the Matsubara frequencies. In 1957, Ezawa et al.\ extended Matsubara's work to the relativistic quantum field theory~\cite{r38Ezawa.}. They discovered the periodicity (anti-periodicity) conditions for the Green function of boson (fermion) fields, the generalization of which became known as the KMS (Kubo~\cite{r38Kubo.} (1957), Martin and Schwinger~\cite{r38MaSch.} (1959)) condition. In the 1960s, Schwinger~\cite{r38Schwin.}, Keldysh~\cite{r38Keld.}, and others~\cite{r38Mill., r382Mill., r383Mill.} developed the real time formalism for the finite temperature field theory. The latest development of this formalism was presented by Takahashi and Umezawa~\cite{r38Umeza., r382Umeza.}, based on an operator formulation of the field theory at finite temperature, which is called thermofield dynamics. Since then, many subjects in finite temperature field theory, e.g., thermal Ward-Takahashi relations, KMS relations, and renormalization procedure, have been studied and are reported in, for example,~\cite{r40Kapusta., r41Bellac., r41Khanna., r42Lands., r42Lain.}. 

In this paper, we use the Matsubara formalism to study the Casimir effect for a free real scalar field confined between two parallel plates at finite temperature. In this formalism, a Euclidean field theory is obtained by a Wick rotation on the time coordinate, $t \to  - i \tau$, such that the Euclidean time $\tau$ is confined to the interval $\tau \in [0 , \beta]$, where $\beta=(kT)^{-1}$ ~\cite{r37Mats., r41Bellac.}. The partition function in the path integral representation becomes:
\begin{equation}\label{s1}
Z = \int\limits_{\begin{array}{l}
\scriptstyle \phi (\beta , \vec{r} ) =   \phi (0 , \vec{r})\hfill
\end{array}}
 D\phi   \exp\left( - \int_{0}^\beta  d\tau  \int {d^3 x} \mathcal{L}_E \right) .
\end{equation}
For a free scalar field, this expression simplifies to
\begin{eqnarray}\label{s2}
Z &=& \det \left[\left( \frac{- \partial _{ E}^2 + m^2}{\mu^2} \right)^{- \frac{1}{2}}\right],
\end{eqnarray}
where $\mu$ is an arbitrary mass scale introduced for dimensional reasons. It should not appear in acceptable final expressions for physically measurable quantities.
Using the partition function given by Eq.~(\ref{s2}), the free energy is obtained as
\begin{eqnarray}\label{s3}
F =  - \frac{\ln(Z)}{\beta } &=&  \frac{T}{2} \ln \left[ \det \left(\frac{ - \partial _{ E}^2 + m^2 }{\mu^2}\right) \right] = 
\frac{T}{2}  \Tr \left[ \ln\left(\frac{ P_E^{2} + m^2 }{\mu^2}\right) \right].
\end{eqnarray}
The trace in Eq.~(\ref{s3}) indicates the summation over eigenvalues of Klein-Gordon operator in the momentum-space representation. Moreover, the modes of zero-component of momentum, i.e., the Matsubara frequencies, are discrete due to the KMS periodicity condition on the finite $\tau$ interval: 
\begin{equation}\label{s3a}
 \omega_{n_0} = \frac{2 n_0 \pi}{\beta}, \;\; \mbox{where}   \; \;  ({n_0} =0, \pm 1, \pm 2, \pm 3, ... )  .
 \end{equation}
 %
%

We impose the Dirichlet boundary condition at the plates, as follows
\begin{equation}\label{s3b} 
  \Phi (x) | _{(z = {z_j})}  = 0.
\end{equation}
We consider the plates to be located at $ z =- \frac{L}{2} $  and $z =\frac{L}{2}$, and obtain the following condition for the discrete spatial modes in $z$ direction:
\begin{equation}\label{s1000}
f(k_{n_1}) := \sin(k_{n_1} L) = 0.
\end{equation}
Note that the modes for both the massive and massless cases are regular, {\it i.e.}, equally spaced, and given by 
\begin{equation} \label{s1b}
k_{n_1} = \frac{{n_1} \pi}{ L}   \;\;\;\;      ({n_1} = 1, 2, 3,....) .
\end{equation}

Using Eqs.\ (\ref{s3}, \ref{s3a}, \ref{s1b}), the expression for the free energy becomes
\begin{eqnarray}\label{s04}
F_{\mbox{\scriptsize bounded}}(T,L) &=& \frac{T A}{2} \int \frac{d^2  K_T}{ {\left(2 \pi\right)}^2}  \sum\limits_{{n_0} =  - \infty }^{\infty}  {\sum\limits_{n_1=1}^{\infty} \ln \left[ \frac{{\left( \frac{2{n_0} \pi }{\beta } \right)}^2  + \omega _{{n_1} , {K_T}}^2}{\mu^2} \right]},
\end{eqnarray}
where $\omega _{{n_1} , {K_T}} = \sqrt{ \left( \frac{{n_1} \pi }{L } \right)^2 + K_T^2 + m^2}$, and $A$ denotes the area of the plates. We shall refer to this expression for free energy as the first form.
A very commonly-used alternate expression for the first form, which has an embedded analytic continuation, is the following\footnote{This is obtained by replacing the logarithm using
$$\ln\left(\frac{A}{\mu^2}\right)=-\lim\limits_{s \to 0}  \frac{\partial }{\partial s}  \left[\frac{1}{\mu^{-2s}\Gamma(s)}\int_{0}^{\infty} dt e^{-t A} t^{s-1} \right],$$ 
where we have assumed that $A$ has mass dimension two, as in Eq.~(\ref{s04}). We like to emphasize that the above replacement includes an analytic continuation. To trace it, we note that the expression for the logarithm is obtained by first using the identity $\ln\left(\frac{A}{\mu^2}\right)=-\lim\limits_{s \to 0}  \frac{\partial }{\partial s}  \left(\frac{A}{\mu^2}\right)^{-s}$. Next, $A^{-s}$ has been replaced using the Euler integral representation of the Gamma function, {\it i.e.}, $\Gamma(s)=\int_0^{\infty} dt e^{-t} t^{s-1}$. This integral is finite for $s>0$, while it admits an analytic continuation for $s\leq0$.}
\begin{equation}\label{s4}
F _{\mbox{\scriptsize bounded}}(T,L)= - \frac{T A}{2} \int \frac{d^2 K_T}{\left(2 \pi \right)^2} \sum\limits_{n_0 =  - \infty }^{\infty} \sum\limits_{n_1=1}^{\infty} \lim\limits_{s \to 0}   \frac{\partial }{\partial s}  \int_0^\infty  
		\frac{e^{ - t \left[ {\left( \frac{2{n_0} \pi}{\beta} \right)}^2 + 
				\omega _{{n_1} , {K_T}}^2 \right]}}{\mu^{- 2 s} \Gamma (s)   t^{1 - s}}   dt .
\end{equation}
The integral over $t$, accompanied by the operation $\lim\limits_{s \to 0}   \frac{\partial }{\partial s}$, embodies the analytic continuation. This is important since in this paper we intend to keep track of all analytic continuations. The result of this expression depends, in principle, on the order in which the sums and integrals are performed. However, if we do the integral over $t$ last we obtain the analytic continuation of this expression, which is certainly finite and unique, regardless of the order of other sums and integral. If we do the integral over $t$ at any step other than last and we are interested in the analytic continuation of this expression, we are going to need a supplementary analytic continuation at the end. 

Integrating over the transverse momenta, we obtain
\begin{equation}\label{s004}
F _{\mbox{\scriptsize bounded}}(T,L)=-\frac{ T A}{8 \pi}  \sum\limits_{n_0 =  - \infty }^{\infty} \sum\limits_{n_1=1}^{\infty} \lim\limits_{s \to 0}   \frac{\partial }{\partial s}  \int_0^\infty  
		\frac{e^{ - t \left[ {\left( \frac{2{n_0} \pi}{\beta} \right)}^2 + 
				\omega _{{n_1}}^2 \right]}}{{\mu^{- 2 s}} \Gamma (s)   t^{2 - s}}   dt  , 
\end{equation}				
where $\omega_{n_1}=\sqrt{\left( \frac{{n_1} \pi}{L} \right)^2 + m^2}$. If we now integrate over $t$, we obtain the form which is almost invariably used in the ZFA:
	\begin{equation}\label{SZeta}
		F _{\mbox{\scriptsize bounded}}(T,L)= -\frac{TA}{8\pi} \sum\limits_{n_0 =  - \infty }^{\infty} \sum\limits_{n_1 = 1}^{\infty} \lim\limits_{s \to 0}   \frac{\partial }{\partial s}  \frac{{\mu^{2 s}}}{s-1}  \left[ \left( \frac{2{n_0} \pi}{\beta} \right)^2 + 
		\omega _{{n_1}}^2 \right]^{1 - s} .
\end{equation}

On the other hand, one can perform the sum over the Matsubara frequencies in the original expression for the free energy given in Eq.~(\ref{s04}), using the Principle of the Argument theorem~\cite{r49Ahlf.}, to obtain the usual form in statistical mechanics~\cite{r31Bord2.}: 
\begin{equation}\label{s5}
F_{\mbox{\scriptsize bounded}} =\frac{ A}{2} \int  \frac{d^2  K_T}{{\left( 2 \pi \right)}^2}   \sum\limits_{n_1=1}^{\infty}  \left[  \omega _{{n_1} , {K_T}} + 
2 T \ln\left( 1 - e^{ - \beta \omega _{{n_1} , {K_T}}} \right) \right] ,
\end{equation}
which we shall refer to as the second form of the free energy. This expression and the original first form given by Eq.~(\ref{s04}), do not contain any embedded analytic continuation. One advantage of this form is that the contribution of the zero temperature part is separated from the thermal correction part.


\section {The Casimir Free Energy for a Massless Scalar Field}\label{FCasimir}

\indent 
In this section, we calculate the Casimir free energy, using its fundamental definition, for a free real scalar field between two parallel plates, separated by a distance $L$, with the Dirichlet boundary conditions. 
In Sec.~\ref{massive}, we generalize to the massive case, and verify that, as expected, its massless limit coincide with the results of this section. As mentioned above, the fundamental definition of $F_{\mbox{\scriptsize Casimir}}$ is the difference between the free energy of the system in the presence of nonperturbative conditions or constraints and the one with no constraints, both being at the {\em same} temperature $T$ and having the same volume. The nonperturbative conditions or constraints include boundary conditions, background fields such as solitons, and nontrivial space-time backgrounds. For cases where the constraints are in the form of non-trivial boundary conditions, the free cases can be defined as the cases in which the boundaries have been placed at spatial infinities. For the latter cases, the fundamental definition can be written as
\begin{equation}\label{s6}
F_{\mbox{\scriptsize Casimir}}(T,L) = F_{\mbox{\scriptsize bounded}}(T,L) - F_{\mbox{\scriptsize free}}(T,L) ,
\end{equation}
where the dependence of $F_{\mbox{\scriptsize free}}$ on $L$ simply denotes the restriction of the volume of space considered. We expect this dependence to be linear for simple extensive thermodynamic quantities, such as $F_{\mbox{\scriptsize free}}(T,L)$.
%
%
	
To calculate the free energy for a massless scalar, we use the first form presented in Eq.~(\ref{s004}), and obtain:
\begin{equation}\label{s7}
F _{\mbox{\scriptsize bounded}}(T,L)=-\frac{TA}{8 \pi}  \sum\limits_{n_0 =  - \infty }^{\infty} \sum\limits_{n_1 = 1}^{\infty} \lim\limits_{s \to 0}   \frac{\partial }{\partial s}  \int_0^\infty  
\frac{e^{ - t \left[ {\left( \frac{2{n_0} \pi}{\beta} \right)}^2 + 
\left( \frac{{n_1} \pi}{L}\right)^2  \right]}}{ {\mu^{- 2 s}}\Gamma (s)  t^{2-s}}   dt  .
\end{equation}
First, we use the Poisson summation formula\footnote{The Poisson summation formula (see, for example,~\cite{r43Stein., r432Stein., r43Gasq., r432Gasq.}) for a continuous and bounded function $f$ on $\mathbb{R}$ can be expressed as
\begin{equation}
\sum\limits_{{n} =  - \infty }^{\infty}  f(n)  =\sum\limits_{{m} =  - \infty }^{\infty} \int_{- \infty}^{\infty}  dx f(x) e^{- i 2 \pi m x}   \nonumber .
\end{equation}}
for the sum over Matsubara frequencies, as follows
\begin{equation}\label{s7bbPoisson}
 \sum\limits_{{n_0} =  -\infty }^{\infty} e^{ - t  \left(\frac{2{n_0} \pi }{\beta} \right)^2}  =
\frac{\beta}{2\sqrt{ \pi  t}} + \frac{\beta}{\sqrt{\pi  t}} \sum\limits_{{n_0} = 1}^\infty  e^{ - \frac{{n_0}^2   \beta^2}{4 t} } ,
\end{equation}
and we evaluate the integral over $t$, to obtain
\begin{eqnarray}\label{s7bb}
\hspace{-3mm}F_{\mbox{\scriptsize bounded}}(T,L) &=&-\frac{A}{16 \sqrt{\pi^3}}  \lim\limits_{s \to 0}   \frac{\partial }{\partial s} \frac{{\mu^{ 2 s}}}{\Gamma (s)} \sum_{n_1=1}^{\infty} \left\lbrace \Gamma \left(s - \frac{3}{2}\right) \left(\frac{\pi n_1}{L}\right)^{3 - 2s} + \right. \nonumber \\
&& \left. 4 \sum_{n_0=1}^{\infty}  \left(\frac{2 \pi n_1}{n_0 L \beta}\right)^{\frac{3}{2} - s} K_{\frac{3}{2}-s }\left(\frac{n_0 n_1 \pi}{TL}\right)\right \rbrace.
\end{eqnarray}
Next, we evaluate $\lim \limits_ {s \to 0} \partial / \partial s $ only for the second term of Eq.~(\ref{s7bb}) which is finite, and express the result in a form in which the zero temperature part is separated from the thermal correction part, as follows 
\begin{eqnarray}\label{s8}
\hspace{-8mm}F_{\mbox{\scriptsize bounded}}(T,L) &=& F_{\mbox{\scriptsize bounded}}(0,L)+\Delta  F_{\mbox{\scriptsize bounded}}(T,L),  \mbox{ where} \nonumber\\
\hspace{-8mm}F_{\mbox{\scriptsize bounded}}(0,L) &=& -\frac{A}{16 \sqrt{\pi^3}}  \lim\limits_{s \to 0}   \frac{\partial }{\partial s} \frac{\Gamma \left(s - \frac{3}{2}\right)}{{\mu^{- 2 s}}\Gamma (s)} \sum\limits_{n_1=1}^{\infty} \left(\frac{n_1 \pi}{L}\right)^{3 - 2s},  \mbox{ and }\nonumber\\
\hspace{-8mm}\Delta F_{\mbox{\scriptsize bounded}}(T,L) &=&- A \sqrt{\frac{T^3 }{2L^3}} \sum\limits_{n_0=1}^{\infty} \sum\limits_{n_1=1}^\infty \left(\frac{n_1}{n_0}\right)^{\frac{3}{2}} K_{\frac{3}{2}}\left(\frac{n_0 n_1 \pi}{TL}\right).
\end{eqnarray}
%
%

On the other hand, the free energy of the unconstrained case, which is considered at the same temperature $T$ and same volume $V=AL$, can be computed using any of the forms presented in Sec.~\ref{Helmholtz free energy}. However, to ultimately use the fundamental definition, it is important to use the same form and same order of summations and integrations as used to calculate the free energy of the bounded configuration, which in this case is Eq.~(\ref{s004}). Hence, we use the following expression for the free case,
\begin{eqnarray}\label{s9}
&&\hspace{-8mm}F_{\mbox{\scriptsize free}}(T,L)  = -\frac{ T A}{8\pi}   \sum\limits_{{n_0} =  - \infty }^{\infty} \int_0^\infty {\frac{L dk}{ \pi}} \lim\limits_{s \to 0} \frac{\partial }{\partial s}  \int_0^\infty 
\frac{e^{ - t  \left[\left(\frac{2{n_0} \pi }{\beta } \right)^2 +  k^2 + m^2 \right]} }{ {\mu^{- 2 s}}\Gamma (s) t^{2 - s}} dt ,
\end{eqnarray}
with $m=0$ in this case. Performing the same procedure as for the bounded case, we obtain the free energy of the free case. Below, we again express the result in a form in which the zero temperature part is separated from the thermal correction part, as follows
\begin{eqnarray}\label{s10aa}
F_{\mbox{\scriptsize free}}(T,L) &=& F_{\mbox{\scriptsize free}}(0,L) + \Delta F_{\mbox{\scriptsize free}}(T,L) , \mbox{ where} \nonumber \\
F_{\mbox{\scriptsize free}}(0,L)&=& -\frac{A}{16 \sqrt{\pi^3}}  \lim\limits_{s \to 0}   \frac{\partial }{\partial s} \frac{\Gamma \left(s - \frac{3}{2}\right)}{{\mu^{- 2 s}}\Gamma (s)} \int_{0}^{\infty} {\left(\frac{k' \pi}{L}\right)^{3 - 2s}} dk' , \mbox{ and}\nonumber \\
\Delta F_{\mbox{\scriptsize free}}(T,L) &=&  - A \sqrt{\frac{T^3 }{2L^3}}  \sum\limits_{n_0=1}^\infty \int_{0}^{\infty} dk' \left(\frac{k'}{n_0}\right)^{\frac{3}{2}} K_{\frac{3}{2}}\left(\frac{n_0 k' \pi}{TL}\right),\nonumber\\ 
&=& -\frac{AL \pi^2}{90} T^4 ,
\end{eqnarray}
where $k' =\frac{Lk}{\pi}$. As can be seen from Eq.~(\ref{s10aa}), $F_{\mbox{\scriptsize free}}(0,L)$, being the continuum version of $F_{\mbox{\scriptsize bounded}}(0,L)$ given in Eq.~(\ref{s8}), is also divergent, while the thermal correction term of the free case is finite. The last equality, {\it i.e.}, $\Delta F_{\mbox{\scriptsize free}}(T,L)=-ALT^4 \pi^2/90$, which is easily obtained by first evaluating the integral and then evaluating the resulting finite sum over the Matsubara modes by $\zeta(4)$, shows that $\Delta F_{\mbox{\scriptsize free}}(T,L)$ for the massless case is simply the black-body radiation term.
Substituting our results for $F_{\mbox{\scriptsize bounded}}(T,L)$, given in Eq.~(\ref{s8}), and $F_{\mbox{\scriptsize free}}(T,L)$, given in Eq.~(\ref{s10aa}), into Eq.~(\ref{s6}), we obtain the following expression for $F_{\mbox{\scriptsize Casimir}}(T,L)$ based on its fundamental definition,
\begin{eqnarray}\label{s10aaa}
&&\hspace{-8mm}F_{\mbox{\scriptsize Casimir}}(T,L) = -\frac{A}{16 }  \lim\limits_{s \to 0}   \frac{\partial }{\partial s} \frac{\Gamma \left(s - \frac{3}{2}\right) \pi^{\frac{3}{2}-2s}}{{\mu^{- 2 s}} \Gamma (s) L^{3-2s}} \left[\sum\limits_{n_1=1}^\infty \left(n_1\right)^{3 - 2s}-\int_{0}^{\infty} \left(k'\right)^{3 - 2s} dk' \right]-\nonumber \\
&&\hspace{-8mm} A \sqrt{\frac{T^3 }{2L^3}}  \sum\limits_{n_0=1}^\infty \frac{1}{\sqrt{n_0^3}}\left[ \sum\limits_{n_1=1}^\infty\sqrt{n_1^3}  K_{\frac{3}{2}}\left(\frac{n_0 n_1 \pi}{TL}\right) - \int_{0}^{\infty} dk' \sqrt{\left(k'\right)^3} K_{\frac{3}{2}}\left(\frac{n_0 k' \pi}{TL}\right)\right] .
\end{eqnarray}
Finally, using the Abel-Plana formula (see, for example,~\cite{r43Sahra.})\footnote{The simplest form that is needed here is the following
\begin{equation}
	\sum\limits_{n = 1}^\infty  f\left(n\right)  - \int_a^\infty f(x)  dx =\frac{1}{2 i} \left[\int_{a}^{a - i{\infty}} f(z) \left(\cot (\pi z)-i\right) dz - \int_{a}^{a + i{\infty}} f(z) \left(\cot (\pi z)+i\right)dz\right], \nonumber 
	\end{equation}
where $0<a<1$. This indicates that we need to use a regularization here: We consider the free case to be a limiting form of the bounded case in which the distance between the plates, denoted by $L_1$, goes to infinity. Then $a=  L/(\pi L_1)\rightarrow 0$, and the above expression simplifies to, 
\begin{equation}\label{abelplana}
\lim\limits_{a \to 0}\left[\sum\limits_{n = 1}^\infty  f\left(n\right)  - \int_a^\infty f(x)  dx\right] =\lim\limits_{a \to 0} i \int_{0}^{\infty} \frac{f \left(a + it\right) - f \left(a - it\right)}{e^{2 \pi t} - 1} dt . 
\end{equation}}
%
%
for the sum over the spatial modes and evaluating $\lim \limits_ {s \to 0} \partial / \partial s $\footnote{We have used $\lim\limits_{s \to 0} \frac{\partial }{\partial s} \frac{f(s)}{\Gamma (s)} = f(0)$, when $ f(s) $ is an analytic function for $s<1$.}, the divergent terms completely cancel and we obtain the free energy between the two plates. Below, we express the results in a form in which the zero temperature part is separated from the thermal correction part, as follows\footnote{If we start with Eq.~(\ref{SZeta}) for the massless case, {\it i.e.}, $\omega_{n_1}=k_{n_1}= n_1 \pi /L$, and use the Abel-Plana summation formula, the bounded and free cases both include  $F_{\mbox{\scriptsize free}}(0,L)=\frac{A L}{16 \pi^2} \lim\limits_{s \to 0} \left[\frac{\partial}{\partial s}\frac{{\mu^{2s}}}{s-1} \int_0^\infty dk k^{3-2s}\right]$, which is now explicitly divergent. However, this makes no difference in the fundamental approach since, upon subtraction, they again cancel each other yielding the same result as given by Eq.~(\ref{s11}).}
\begin{eqnarray}\label{s11}
F_{\mbox{\scriptsize Casimir}}(T,L) = - \frac{ A \pi^2}{1440 L^3}  - \frac{2 A L T^4}{ \pi^2}  \sum\limits_{{n_0} = 1 }^{\infty}
\sum\limits_{{n_1} = 1}^{\infty} \frac{1}{\left[n_0^2 +\left( 2 n_1 T L\right)^2 \right]^2} .
\end{eqnarray}
We can now compute the sum over ${n_0}$ to obtain,
\begin{eqnarray}\label{s12}
 F_{\mbox{\scriptsize Casimir}}(T,L) = - \frac{A T}{16 \pi L^2}  \sum\limits_{{n_1} = 1}^\infty  \frac{1}{n_1^3} \left[ \coth\left( 2 \pi {n_1}T L \right) +\left( 2 \pi {n_1}T L \right)  \csch^2 \left( 2 \pi {n_1}T L \right) \right].
\end{eqnarray}
The zero temperature limits of Eqs.~(\ref{s11}, \ref{s12}) yield the following well known result $F_{\mbox{\scriptsize Casimir}}(0,L) =E_{\mbox{\scriptsize Casimir}}(0,L) =-\pi^2 A/(1440L^3)$. The high temperature limit of  $F_{\mbox{\scriptsize Casimir}}(T,L)$ is
\begin{equation}\label{Fm0HT}
		F_{\mbox{\scriptsize Casimir}}(T,L) \xrightarrow{TL\gg 1} - \frac{A \zeta(3)}{16 \pi L^2}T,
\end{equation}
which shows the classical behavior. One can also start with the second form of the free energy given by Eq.~(\ref{s5}) and use the Abel-Plana summation formula to obtain exactly the same expression for the Casimir free energy as given by Eq.~(\ref{s12}) (see Appendix \ref{appendixC:FCasimir}). 

Another powerful method which can also be used to calculate the free energies, besides the Abel-Plana formula, is based on the Principle of the Argument theorem. In Appendix \ref{appendixB:The Argument Principle}, we use this method to calculate the Casimir free energy within the fundamental approach in four different ways, starting with Eqs.~(\ref{SZeta}, \ref{s004}, \ref{s5}), the results of which are identical to those displayed in Eqs.~(\ref{s11}, \ref{s12}, \ref{C4}).

%
%

\begin{center}
	\begin{figure}[h!] 
		\includegraphics[width=13.5cm]{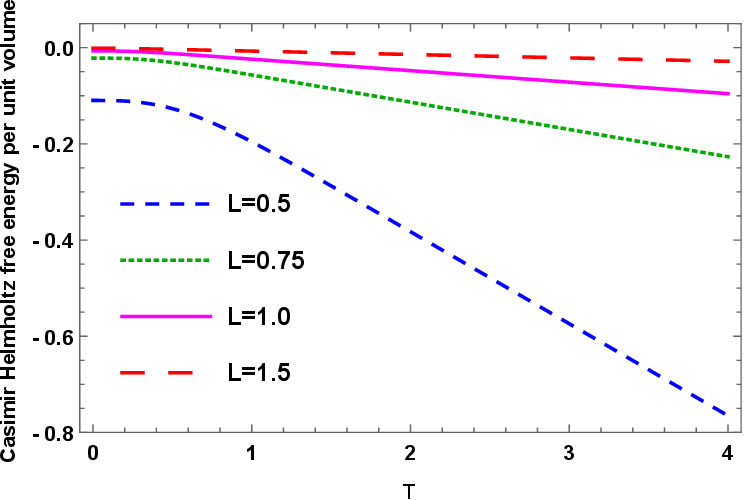}
		\caption{\label{fp5} \small
			The Casimir free energy per unit volume for a massless real scalar field between two parallel plates as a function of temperature for various values of plate separations $L = \{ 0.5,0.75,1.0,1.5\} $.
	As an illustration of the magnitudes, for $L=1 \mu m$, each unit of the horizontal axis is equivalent to $458 K$, and each unit of the vertical axis is about $0.24 mPa$.}
	\end{figure}
\end{center}

In Fig.~(\ref{fp5}), the Casimir free energy is plotted as a function of temperature for various values of $L$. As can be seen from this figure, $F_{\mbox{\scriptsize Casimir}}(T,L)$ is always negative, and goes to zero as $L$ increases. Moreover, this shows that there is a classical term proportional to $T$ for the massless scalar fields between two parallel plates as the temperature increases, which, as we shall show, also holds for the massive scalar fields. Note that the subtraction of the free case at the same temperature amounts to the complete cancellation of zero-temperature infinities and the black-body term, without the need for any extra renormalization program, with the classical term remaining as the leading high temperature term.

\begin{center}
\begin{figure}[h!] 
\includegraphics[width=13.5cm]{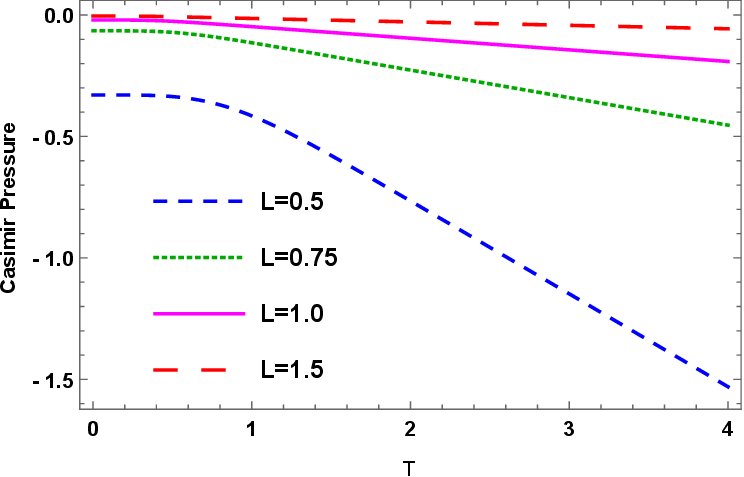}
\caption{\label{fp34} \small
The Casimir pressure for a massless real scalar field between two parallel plates as a function of temperature for various values of plate separations $L = \{ 0.5,0.75,1.0,1.5\} $.
As an illustration of the magnitudes, for $L=1 \mu m$, each unit of the horizontal axis is equivalent to $458 K$, and each unit of the vertical axis is about $0.66 mPa$.}
\end{figure}
\end{center}

Having obtained the Casimir free energy, one can easily calculate all other thermodynamic quantities such as the Casimir pressure, energy, and entropy. For example the Casimir pressure is given by,
\begin{eqnarray}\label{s13}
&&\hspace{-11mm} P_{\mbox{\scriptsize Casimir}} (T,L) =  -\frac{1}{A}\frac{\partial}{\partial L} F_{\mbox{\scriptsize Casimir}}(T,L) =- \frac{T}{8 \pi  L^3}  \sum\limits_{{n_1} = 1}^\infty  
	\frac{1}{n_1^3 }  \left[ \coth\left( 2 \pi {n_1}T L \right) + \right.     \nonumber\\
&&\hspace{-13mm}\left. \left( 2 \pi {n_1}T L \right)  \csch^2 \left( 2 \pi {n_1}T L \right)+  
	\left( 2 \pi {n_1}T L \right)^2 \coth\left( 2 \pi {n_1}T L \right) \csch^2 \left( 2 \pi {n_1}T L \right) \right] . 
\end{eqnarray}
Moreover, one can calculate directly the Casimir pressure based on its fundamental definition, as given by~\cite{r2Lif., r5Mehr.}, which is the difference between the pressure in the regions between the two plates and outside the plates, both regions being at the same temperature. To this end, we consider two inner plates enclosed within two outer plates, as the distance between the latter two goes to infinity, and obtain the same result as given by Eq.~(\ref{s13}). By integrating the expression for the pressure over the distance between two plates, at fixed temperature, the Casimir free energy can be calculated, yielding the same result as given by Eq.~(\ref{s12}), without any extra terms. In Fig.~(\ref{fp34}), the Casimir pressure is plotted as a function of temperature for various values of $L$. As can be seen from this figure, $P_{\mbox{\scriptsize Casimir}}(T,L)$ is always negative, and goes to zero as $L$ increases. Furthermore, as can be shown from Eq.~(\ref{s13}) or inferred from Eq.~(\ref{Fm0HT}), in the high temperature limit the classical term is $- \frac{ \zeta(3)}{8 \pi L^3}T$.

\section {Massless Scalar Fields and the Generalized Zeta Function}\label{zeta}
\indent
In this section, we consider the zeta function approach (ZFA) which is commonly used for computing the Casimir free energy for massless real scalar at finite temperature. To fully explore this approach, we consider four different ways of using the zeta function and show that they yield equivalent results. Moreover, we shall compute the Casimir free energy using the Schl\"{o}milch formulas approach (SFA)~\cite{r29Junji.}, as the second representative of the analytic continuation approach, and show that its results are equivalent to those of the ZFA. Finally we use the zero temperature subtraction approach (ZTSA)~\cite{r22Rav.}, and show that its results are not equivalent to those of the ZFA and SFA. Moreover, as we shall show, none of these results are equivalent to the one obtained in the last section, based on the fundamental definition of the Casimir free energy, since they contain extra terms. The results of the analytic continuation approaches have two extra terms, {\it i.e.},  $T^3$ and $T^4$ terms, while the ones of ZTSA have only an extra $T^4$ term. We then illustrate how the renormalization procedure for these approaches in this trivial example yields the correct results, which we take to be those obtained using the fundamental approach. Moreover, we show that if we calculate the free energies of both the bounded and free cases using the zeta function and subtract them according to the fundamental approach all extra terms cancel and we obtain the correct results.

%
%

For the computation of the Casimir free energy using the ZFA, we use only the first form given by Eq.~(\ref{s4}), as expressed in Eq.~(\ref{SZeta}). For the massless case we have $\omega_{n_1}=k_{n_1}= n_1 \pi /L$. To use the generalized Epstein zeta function to calculate the sums, we express the sum over $n_0$  modes as a sum over positive integers and a zero mode:
\begin{eqnarray}\label{s21}
F (T,L)&=& -\frac{T A}{8 \pi }  \lim\limits_{s \to 0} \frac{\partial }{\partial s}  \frac{{\mu^{2 s}}}{s-1} \left[ \sum\limits_{n_{1} =  1}^{\infty }  \left(  \frac{ n_1^2 \pi^2 }{L^2 }\right)^{1 - s}  +   \right.   \nonumber\\
&& \left. 2\sum\limits_{n_{0} =  1 }^\infty \sum\limits_{n_{1} =  1 }^\infty\left( \frac{4 n_0^2 \pi^2 }{\beta^2} + \frac{n_1^2 \pi^2 }{ L^2 }\right)^{1 - s}   \right] .
\end{eqnarray}
In first three methods, we use this form of the Helmholtz free energy. The first term can be easily calculated using the Riemann zeta function. Below we use the generalized zeta function in three different ways to compute the double summation in the second term.

For our first method, we use the homogeneous generalized Epstein zeta function to do simultaneously the double summations for the second term in Eq.~(\ref{s21}). In this case, the analytic continuation is rendered by the reflection formula (see Appendix \ref{appendixA:The zeta function}). Here, we present the final result as follows (see Appendix \ref{appendixA:The zeta function})
\begin{eqnarray}\label{s23}
F_{\mbox{\scriptsize Zeta}}(T,L) =  F_{\mbox{\scriptsize  Casimir}}(T,L)+ \Delta F_{\mbox{\scriptsize free}}(T,L) +\frac{A \zeta (3)}{4 \pi} T^3 .
\end{eqnarray}
We have denoted the Casimir free energy obtained by this method as $F_{\mbox{\scriptsize Zeta}}$ to distinguish it from the one obtained using the fundamental definition, which we have simply denoted by $F_{\mbox{\scriptsize Casimir}}$ and is given by Eq.~(\ref{s12}). As shown in this equation, $F_{\mbox{\scriptsize Zeta}}$ has two extra terms. The first extra term is $\Delta F_{\mbox{\scriptsize free}}(T,L)$ which is the thermal correction term of the massless free case, i.e., the black-body term proportional to $T^4$, and given by Eq.~(\ref{s10aa}). The last term is an extra L-independent $T^3$ term and does not contribute to the pressure. As shown in Appendix~\ref{appendixA:The zeta function}, this extra nonextensive term is precisely minus one half of the contribution of the zero spatial mode, which is disallowed by the Dirichlet boundary conditions. However, the zero temperature limit of the above expression gives the correct result $F_{\mbox{\scriptsize  Casimir}}(0,L)=E_{\mbox{\scriptsize Casimir}}(0,L) = - \pi^2 A/(1440L^3)$.


For our second method, we use the inhomogeneous form of the Epstein zeta function to first sum over the Matsubara modes for the second term in Eq.~(\ref{s21}), obtaining (see Appendix \ref{appendixA:The zeta function})
\begin{eqnarray}\label{s21b}
F_{\mbox{\scriptsize Zeta}}(T,L) &=& - \frac{A}{16 \sqrt{\pi^3}}  \lim\limits_{s \to 0} \frac{\partial }{\partial s} \frac{\Gamma\left(s - \frac{3}{2} \right)}{ {\mu^{- 2 s}}\Gamma (s)} \sum\limits_{n_1 = 1}^\infty \left(\frac{n_1 \pi}{L}\right)^{ 3 - 2s}+ \frac{A \zeta (3)}{4 \pi} T^3 - \nonumber \\
&&\frac{A T^3}{4\pi } \sum\limits_{n_0 = 1}^\infty   \frac{ \coth\left( \frac{\pi n_0}{2 T L} \right)  + \frac{ \pi n_0}{2 T L} \csch^2 \left( \frac{\pi n_0}{2 T L} \right)}{n_0^3 }.
\end{eqnarray}
The first term on the right hand side has a sum over the spatial modes which is divergent. We can express this sum in terms of the zeta function $\zeta (2s -3)$ and use its analytic continuation\footnote{We have used: $\zeta(-3)=1/120$.} to obtain,
\begin{eqnarray}\label{s21bgd}
\hspace{-2mm}F_{\mbox{\scriptsize Zeta}}(T,L) = - \frac{A \pi^2}{1440 L^3} + \frac{A \zeta (3)}{4 \pi} T^3 - \frac{A T^3}{4\pi } \sum\limits_{n_0 = 1}^\infty   \frac{ \coth\left( \frac{\pi n_0}{2 T L} \right)  + \frac{ \pi n_0}{2 T L} \csch^2 \left( \frac{\pi n_0}{2 T L} \right)}{n_0^3 }.
\end{eqnarray}
This expression for $F_{\mbox{\scriptsize Zeta}} (T,L)$ is equivalent to Eq.~(\ref{s23}), once we use the expression for $F_{\mbox{\scriptsize Casimir}} (T,L)$ given by Eq.~(\ref{C4}). In this case, the black-body $T^4$ term does not appear explicitly in the final result, but it is embedded in the high temperature limit of the last term of Eq.~(\ref{s21bgd}).

For our third method, we use the inhomogeneous form of the Epstein zeta function to first sum over the spatial modes of the free energy given by Eq.~(\ref{SZeta}), and obtain (see Appendix \ref{appendixA:The zeta function})
\begin{eqnarray}\label{s21c}
\hspace{-2mm}F_{\mbox{\scriptsize Zeta}} (T,L)&=& \frac{ AT}{8 \pi}  \lim\limits_{s \to 0} \frac{\partial }{\partial s} \frac{{\mu^{2 s}}}{\Gamma(s)} \sum\limits_{n_{0} = 1}^{\infty } \left[\Gamma(s-1) \left(  \frac{ 2 n_0 \pi }{\beta}\right)^{2 - 2 s} +\right.    \nonumber\\ 
&&\left. \frac{L\Gamma\left(s - \frac{3}{2} \right)}{\sqrt{\pi}}\left(  \frac{ 2 n_0 \pi }{\beta}\right)^{3 - 2 s}\right] + F_{\mbox{\scriptsize Casimir}} (T,L),
\end{eqnarray}
where $F_{\mbox{\scriptsize  Casimir}}(T,L)$ is given by Eq.~(\ref{s12}). The first and second terms on the right-hand side have sums over temperature modes which are divergent and can be expressed in terms of zeta function $\zeta(2s - 2)$ and $\zeta(2s - 3)$, respectively. Applying the analytic continuation of the zeta function embedded in its reflection formula\footnote{We have used:   $\zeta(2s-2) \Gamma(s-1) = \Gamma\left(\frac{3}{2}-s\right) \zeta(3 - 2s) \pi^{2s - \frac{5}{2}}$

\hspace{2.3cm}  $\zeta(2s-3) \Gamma(s-\frac{3}{2}) = \Gamma\left(2-s\right) \zeta(4 - 2s) \pi^{2s - \frac{7}{2}}$
}, we obtain a finite result which is identical to Eq.~(\ref{s23}).

For our fourth method, we use the homogeneous generalized zeta function to do simultaneously the summations. To do this case, we express the sum over the spatial modes of Eq.~(\ref{SZeta}) as one half of the difference between the sum over all integers and the zero mode to obtain the following form for the free energy:
\begin{eqnarray}\label{s21dd}
F (T,L)&=& \frac{T A}{16 \pi }  \lim\limits_{s \to 0} \frac{\partial }{\partial s}  \frac{{\mu^{2 s}}}{s-1}  \sum\limits_{n_{0} =  -\infty}^{\infty } \left[ \left(  \frac{4 n_0^2 \pi^2 }{\beta^2 }\right)^{1 - s}  -   \right.   \nonumber\\
&& \left. \sum\limits_{n_{1} =  -\infty }^\infty \left( \frac{4 n_0^2 \pi^2 }{\beta^2} + \frac{n_1^2 \pi^2 }{ L^2 }\right)^{1 - s}   \right] .
\end{eqnarray} 
We use the generalized homogeneous zeta functions to do both the single-sum and simultaneously the double-sum terms in Eq.~(\ref{s21dd}). For both cases, the analytic continuation is rendered by the reflection formula (see Appendix~\ref{appendixA:The zeta function}). After computing the sums and simplifying, the expression that we obtain for $F_{\mbox{\scriptsize Zeta}}(T,L)$ is identical to Eq.~(\ref{s23}), where the expression for $F_{\mbox{\scriptsize Casimir}}$ is given by Eq.~(\ref{s12}). Hence, we have shown that all four different ways of using the zeta function yield equivalent results.


As stated above, we also obtain the Casimir free energy using the Schl\"{o}milch formula approach (SFA)~\cite{r29Junji.}, as a second representative of the analytic continuation approach. In fact, this approach is used for obtaining only the thermal corrections of the Casimir effect, and the zero temperature part should be calculated separately using other methods, {\it e.g.}, the ZFA. For the computation of the Casimir free energy in this approach, we use the second form of the free energy given by Eq.~(\ref{s5}). In the massless case, we have $\omega_{n_1, K_T}=\sqrt{K_T^2 +\left(\frac{n_1 \pi}{L}\right)^2}$ and use the generalized Schl\"{o}milch formulas to sum over the spatial modes for the thermal corrections part of the free energy to obtain (see Appendix~\ref{appendixB:The Schlomilch formula})
\begin{eqnarray}\label{210c}
F_{\mbox{\scriptsize SFA}}(T,L)   &=& F_{\mbox{\scriptsize Zeta}}(T,L), 
\end{eqnarray}
where the expression for $F_{\mbox{\scriptsize Zeta}}(T,L)$ is given by Eq.~(\ref{s23}), within which $F_{\mbox{\scriptsize Casimir}}$ is given by Eq.~(\ref{s12}). This is the expected results since both ZFA and SFA are members of the analytic continuation approach.

Next, we calculate the Casimir free energy using the zero temperature subtraction approach (ZTSA)~\cite{r22Rav.}. This approach is defined by
\begin{equation}\label{s16}
F_{\mbox{\scriptsize ZTSA}}(T,L) = F_{\mbox{\scriptsize bounded}} (T,L) - F_{\mbox{\scriptsize free}} (0,L)  .
\end{equation}
Adding an subtracting $\Delta F_{\mbox{\scriptsize free}}(T,L)$ and using definitions given by Eqs.~(\ref{s6},\ref{s10aa}), we can write the following alternative expression for $F_{\mbox{\scriptsize ZTSA}}(T,L)$
\begin{eqnarray}\label{s16000}
		F_{\mbox{\scriptsize ZTSA}}(T,L) =  F_{\mbox{\scriptsize Casimir}}(T,L) + \Delta F_{\mbox{\scriptsize free}}(T,L).
\end{eqnarray}
This equality holds for both the massless and massive cases. Using Eqs.~(\ref{s23},\ref{210c}) we have for the massless case
\begin{eqnarray}\label{s16001}
 F_{\mbox{\scriptsize ZTSA}}(T,L)=F_{\mbox{\scriptsize Zeta}}(T,L) - \frac{A \zeta(3)}{4 \pi} T^3=F_{\mbox{\scriptsize SFA}}(T,L) - \frac{A \zeta(3)}{4 \pi} T^3
\end{eqnarray}

\begin{center}
	\begin{figure}[h!]
		\includegraphics[width=13.5cm]{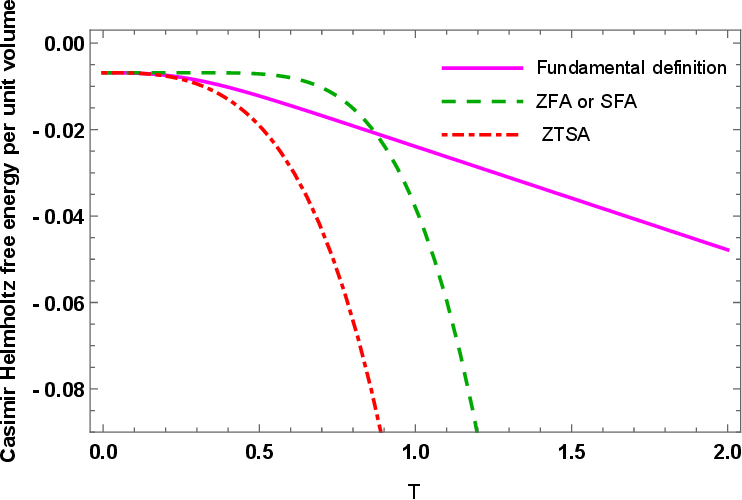}
		\caption{\label{fp6} \small
	The Casimir free energy per unit volume for a massless real scalar field between two parallel plates as a function of temperature with fixed plate separation $L = 1.0$. The solid line is for the one obtained via the fundamental definition, the dashed line is for the ones obtained by the zeta function approach ZFA, or the Schl\"{o}milch formulas approach SFA, and the dotdashed line is for the one obtained by the zero temperature subtraction approach ZTSA.}
	\end{figure}
\end{center}

We display the results of the three general approaches in Fig.~(\ref{fp6}). As can be seen from this figure, the free energy obtained via the ZFA, SFA, or ZTSA decreases as $T^4$  at high temperatures, while the one obtained via the fundamental definition only decreases linearly. The temperature dependence of ZFA and SFA differs from that of ZTSA due to the $T^3$ term. In fact, this extra term makes $F_{\mbox{\scriptsize Zeta}}(T,L)$ a nonextensive quantities.

\begin{center}
	\begin{figure}[h!] 
		\includegraphics[width=13.5cm]{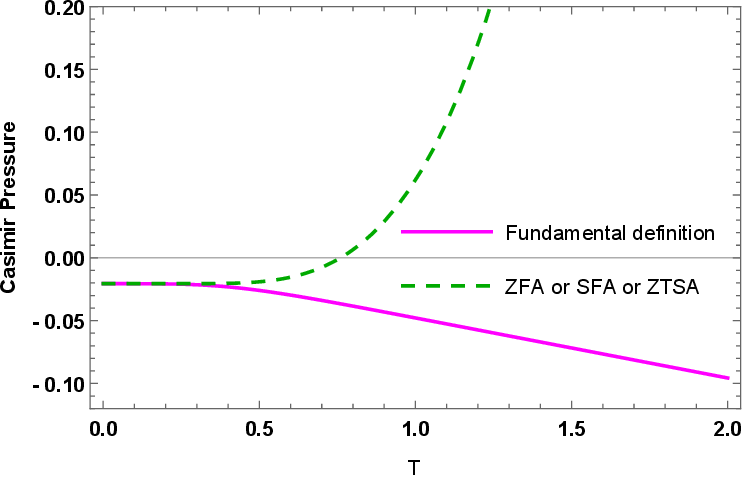}
				\caption{\label{fp7} \small
			The Casimir pressure for a massless scalar field between two parallel plates as a function of temperature with fixed plate separation $L = 1.0$. The solid line is for the pressure obtained via the fundamental definition, and the dashed line is for the pressure obtained by the zeta function approach ZFA, the Schl\"{o}milch formula approach SFA, or the zero temperature subtraction approach ZTSA.}
	\end{figure}
\end{center}

One can now easily calculate all other thermodynamic quantities using the expressions obtained for the free energies by the ZFA, SFA, or ZTSA. For example, calculation of pressure, using first part of Eq.~(\ref{s13}), yields
\begin{eqnarray}\label{s19}
P_{\mbox{\scriptsize Zeta}}(T,L) = P_{\mbox{\scriptsize SFA}}(T,L)=P_{\mbox{\scriptsize ZTSA}}(T,L)=	P_{\mbox{\scriptsize Casimir}}(T,L) +\Delta P_{\mbox{\scriptsize free}}(T),
\end{eqnarray}
where an expression for $P_{\mbox{\scriptsize  Casimir}}(T,L)$ is given by Eq.~(\ref{s13}), and $\Delta P_{\mbox{\scriptsize free}}(T) = ( {\pi}^2/90) T^4$ is the thermal correction to the pressure of the free case. As shown in Eq.~(\ref{s19}), the Casimir pressures obtained using the analytic continuation approach is identical to the one obtained using the zero temperature subtraction approach, since the extra $T^3$ term in the former is $L$ independent. However, none of these results are equivalent to the Casimir pressure obtained by the fundamental approach, since they all contain  $\Delta P_{\mbox{\scriptsize free}}(T)$, which is the black-body term in the massless case. In Fig.~(\ref{fp7}), we compare the pressure obtained using the ZFA, the SFA, or the ZTSA, given by Eq.~(\ref{s19}), with the Casimir pressure obtained based on the fundamental definition given by Eq.~(\ref{s13}). As can be seen, the pressure obtained using the ZFA is negative, corresponding to attractive forces, at low temperatures and becomes positive, corresponding to repulsive forces, at high temperature, while the Casimir pressure is always negative and decreases as linearly at high temperatures.

As mentioned in the Introduction, it has been recognized that the ZFA might yield additional unphysical terms, and renormalization programs have been devised to eliminate them. The most common program is to subtract the polynomials in $T$ appearing in the large temperature limit, with exponents higher that one corresponding to the classical term~\cite{r17Geyer., r29Junji.}. These are usually calculated using the heat kernel coefficients. In the massless case, the mentioned polynomial includes $T^3$ and $T^4$ terms, while for the pressure there is only the $T^4$ term. In this case the renormalization program yields the correct results, based on the fundamental approach. Specifically, the removal of $\Delta F_{\mbox{\scriptsize free}}(T,L) = - ( A L \pi^2/90) T^4$ and $(A\zeta (3)/(4 \pi))T^3$ from the expression for $F_{\mbox{\scriptsize Zeta}}(T,L)=F_{\mbox{\scriptsize SFA}}(T,L)$ in Eq.~(\ref{210c}), the removal of $\Delta F_{\mbox{\scriptsize free}}(T,L)$ from the expression for $F_{\mbox{\scriptsize ZTSA}}(T,L)$ in Eq.~(\ref{s16001}), and the removal of $\Delta P_{\mbox{\scriptsize free}}(T) = ( {\pi}^2/90) T^4$ from the expression for $P_{\mbox{\scriptsize Zeta}}(T,L)=P_{\mbox{\scriptsize SFA}}(T,L)=P_{\mbox{\scriptsize ZTSA}}(T,L)$ in Eq.~(\ref{s19}), yield the correct the results. We like to emphasize that these extra unphysical terms appear in the results of the ZFA, the SFA, and the ZTSA for different reasons. In the two former cases, they are left out by the embedded analytic continuation, and in the latter case, the $\Delta F_{\mbox{\scriptsize free}}(T, L)$ is left out by its definition. As we have shown, in the massless case, these extra terms are simple polynomials in $T$ which can be easily removed by the renormalization programs that have been devised. In the next section, we solve the massive case using the fundamental approach, which subtracts the corresponding thermodynamic quantities of the free case from those of the bounded case. We find that the thermal corrections to the free case are no longer simple polynomials that can be removed by any renormalization programs thus far devised to supplement ZFA, SFA, and ZTSA.

\section {The Casimir Free Energy for a Massive Scalar Field}\label{massive}
\indent
In this section, we calculate the Casimir free energy, using its fundamental definition as given by Eq.~(\ref{s6}), for a massive real scalar field confined between two parallel plates with the Dirichlet boundary conditions at finite temperature. Then, we calculate other Casimir thermodynamic quantities, including pressure, energy, and entropy, and show that all of them are finite and go to zero as the mass, or $L$ increases. Moreover, by increasing temperature, the Casimir free energy and pressure decrease linearly, the Casimir entropy goes to a nonzero constant, and the Casimir energy goes to zero. In the next section, we compute the Casimir free energy using ZFA, SFA and ZTSA, and compare the results. 

We start with the first form of the free energy given by Eq.~(\ref{s004}), use the Poisson summation formula on the Matsubara frequencies given by Eq.~(\ref{s7bbPoisson}), evaluate the integral over $t$, and obtain
\begin{eqnarray}\label{s241} 
F_{\mbox{\scriptsize bounded}}(T,L) = -\frac{A}{ 4\sqrt{\pi^3}} \sum\limits_{{n_1}=1}^\infty  \lim\limits_{s \to 0} \frac{\partial}{\partial s} \frac{{\mu^{2 s}}}{\Gamma (s)} \left[\frac{\Gamma \left(s - \frac{3}{2}\right)}{4} \left(\omega_{n_1}\right)^{3 - 2s}+ \right.\nonumber \\
\left. \sum\limits_{{n_0} = 1}^\infty \left(\frac{2\omega_{n_1}}{n_0 \beta}\right)^{\frac{3}{2}-s}   K_{\frac{3}{2}-s} \left(\beta {n_0} \omega_{n_1}\right) \right], 
\end{eqnarray}
where $\omega_{n_1}=\sqrt{\left(\frac{n_1 \pi}{L}\right)^2 + m^2}$. The spatial modes $k_{n_1}$ are the roots of $f(k_{n_1})$ in Eq.~(\ref{s1000}), which are regular, {\it i.e.}, they are equally spaced due to the Dirichlet boundary conditions, contrary to the massive fermionic case~\cite{r26Goushe.}. To evaluate the sum over the spatial modes, we use the Principle of the Argument theorem and after simplifying (see Appendix \ref{appendixB:The Argument Principle}), we can express the free energy of the bounded region as
\begin{eqnarray}\label{s1003}
&&\hspace{-8mm}F_{\mbox{\scriptsize bounded}}(T,L) = - \frac{A }{4\sqrt{\pi^5}} \lim\limits_{s \to 0} \frac{\partial}{\partial s}\frac{{\mu^{2 s}}}{\Gamma (s)} \left\lbrace  L \frac{\Gamma \left(s - \frac{1}{2}\right) }{ 2} \int_{ 0}^{ \infty}p^{2 - 2s} \omega(p) dp	-\right.\nonumber \\
&&\hspace{-8mm}\left. \sum\limits_{n_0 = 1}^{\infty} \left(\frac{2T}{n_0 }\right)^{\frac{1}{2} - s}L \int_{0}^{\infty}  p^{\frac{3}{2} - s} J_{\frac{1}{2} - s}  \left(n_0 \beta p\right) \omega(p) dp+ \int_{0}^{\infty} \ln\left( 1 - e^{ - 2 L \omega(p)}\right)\times\right.  \nonumber \\
&&\hspace{-6mm}\left. \left[\frac{\Gamma \left(s - \frac{1}{2}\right) }{2}p^{2 - 2s} - \sum\limits_{n_0 = 1}^{\infty}  \left(\frac{2T}{n_0}\right)^{\frac{1}{2} - s}  p^{\frac{3}{2} - s} J_{\frac{1}{2} - s}  \left(n_0 \beta p\right) \right]   dp \right \rbrace ,
\end{eqnarray}
where $\omega(p)=\sqrt{p^2 + m^2}$. Only the first term of the above expression contains a divergent integral. Therefore, for the other terms, which include the logarithm function and the Bessel function, we evaluate $\lim \limits_ {s \to 0} \partial / \partial s $, and after simplifying we obtain
%
%
\begin{eqnarray}\label{s24} 
&&\hspace{-8mm}F_{\mbox{\scriptsize bounded}}(T,L) = -\frac{AL}{ 8\sqrt{\pi^5}} \lim\limits_{s \to 0} \frac{\partial}{\partial s}  \left[\frac{\Gamma \left(s - \frac{1}{2}\right) }{{\mu^{-2 s}} \Gamma (s)} \int_{ 0}^{ \infty}p^{2 - 2s} \omega(p) dp \right]- \frac{AL m^2}{ \pi^2} \Bigg\{\frac{T^2}{2}\times\nonumber \\
&&\hspace{-8mm} \sum\limits_{{n_0} = 1}^\infty \frac{K_2 \left(n_0 \beta m\right)}{  n_0^2 } + \frac{1}{8 L^2}\sum\limits_{{n_1} = 1}^\infty \frac{K_2 \left(2 n_1 m L\right)}{ n_1^2} +
T^2  \sum\limits_{{n_0} = 1}^\infty \sum\limits_{{n_1} = 1}^\infty \frac{K_2 \left(\beta m \omega_{n_0, n_1}\right)}{\left(\omega_{n_0, n_1}\right)^2} \Bigg\},
\end{eqnarray}
where $\omega_{n_0, n_1} =\sqrt{n_0^2+\left(2 n_1 TL\right)^2}$.

Since we are going to use the fundamental definition of the Casimir free energy, we also need to calculate the free energy of the free massive case at finite temperature. To this end, we start with the first form of the free energy as given by Eq.~(\ref{s4}) and follow the same procedure as in the bounded case. That is, we use the Poisson summation on the Matsubara frequencies, evaluate the integral over $t$, and evaluate $\lim \limits_ {s \to 0} \partial / \partial s $ for the finite parts. Then, we can 
express the free energy of the free case as a zero temperature part and a finite temperature correction part as follows 
\begin{eqnarray}\label{s25}
F_{\mbox{\scriptsize free}}(T,L) &=& F_{\mbox{\scriptsize free}}(0,L) +\Delta F_{\mbox{\scriptsize free}}(T,L) \mbox{, where} \nonumber\\
F_{\mbox{\scriptsize free}}(0,L) &=& -\frac{AL}{ 8\sqrt{\pi^5}} \lim\limits_{s \to 0} \frac{\partial}{\partial s} \left[ \frac{\Gamma \left(s - \frac{1}{2}\right)}{{\mu^{-2 s}}\Gamma (s)}\int_0^\infty  k^2 \left[\omega(k)\right]^{1 - 2s} dk\right], \mbox{and} \nonumber \\
\Delta F_{\mbox{\scriptsize free}}(T,L) &=& - \frac{A L T^2 m^2}{2 \pi^2}
\sum\limits_{{n_0} = 1}^\infty \frac{K_2 \left(n_0 \beta m\right)}{  n_0^2 } , 
\end{eqnarray}
where $\omega(k)=\sqrt{k^2 + m^2}$. The first two terms of $F_{\mbox{\scriptsize bounded}}(T,L)$, given by Eq.~(\ref{s24}), are equivalent to the two terms of $F_{\mbox{\scriptsize free}}(T,L)$, given by Eq.~(\ref{s25}). The second terms are actually identical, while the first terms contain equivalent divergent integrals, which we compute using  dimensional regularization, with fixed $s$, obtaining
\begin{eqnarray}\label{F10}
F _{\mbox{\scriptsize free}}(0,L) &=&- \frac{AL}{16\sqrt{\pi^5}} \lim\limits_{s \to 0} \frac{\partial }{\partial s} \frac{{\mu^{2s}}}{\Gamma (s)}  \left[m^{4-2s}  \Gamma \left(\frac{3}{2}-s\right)\Gamma \left(s-2\right) \right]\nonumber \\
&=&- \frac{AL}{16\sqrt{ \pi^5}} \lim\limits_{s \to 0} \frac{\partial }{\partial s} \frac{ m^4}{\Gamma (s)}\left[\frac{\sqrt{\pi}}{4s} -\frac{ \sqrt{\pi}}{8}\left( 4 \ln\left(\frac{m}{\mu}\right) -3\right)\right] \nonumber \\
&=&-\frac{A L m^4}{(128 \pi^2)}  \left[ 3 - 4 \ln \left(\frac{m}{\mu}\right)\right] .
\end{eqnarray}
Notice how the divergent term proportional to $s^{-1}$ which appears in the second line is eliminated. This result is finite due to the analytic continuation embedded in expression that we have used for the first form of the free energy Eq.~(\ref{s4}), as mentioned in Sec.~\ref{Helmholtz free energy}.

Now, using the fundamental definition, as expressed in Eq.~(\ref{s6}), these four terms cancel each other upon subtraction, and we obtain the following expression for the Casimir free energy for a massive scalar field confined between two plates
\begin{eqnarray}\label{s27}
F_{\mbox{\scriptsize Casimir}}(T,L) &=&- \frac{AL m^2}{\pi^2} \sum\limits_{n_1 = 1}^\infty \Bigg\{ \frac{K_{2} \left(2 n_1 m L\right)}{8 L^2 n_1^2} +\nonumber \\
&&T^2\sum\limits_{n_0=1}^{\infty}  \frac{K_{2} \left(m \beta \sqrt{n_0^2+\left(2 n_1 TL\right)^2}\right)}{\left[n_0^2+\left(2 n_1 TL\right)^2\right]} \Bigg\}   .
\end{eqnarray}
The zero temperature and finite temperature correction parts, {\it i.e.}, $F_{\mbox{\scriptsize Casimir}}(0,L)$ and $\Delta F_{\mbox{\scriptsize Casimir}}(T,L)$, are associated with the two terms in Eq.~(\ref{s27}), respectively.
This expression is our main result for Casimir free energy which is obtained by its fundamental definition. One advantage of fundamental approach is that the free parts cancel out, whether they are finite or not.

The high temperature limits of the our results for $F_{\mbox{\scriptsize bounded}}(T,L)$, $F_{\mbox{\scriptsize free}}(T,L)$, and $F_{\mbox{\scriptsize Casimir}}(T,L)$ are important, particularly when we compare them to the analogous results obtained by the other approaches in the next section. Below we display the results including the massless limit of the latter,
\begin{eqnarray}\label{HighTFA}
F_{\mbox{\scriptsize bounded}}(T,L)&\xrightarrow[T \gg m]{TL\gg 1}&  - \left( \frac{ A L \pi^2}{90 } \right) T^4 + \left( \frac{ A L m^2 }{24 } \right) T^2 +\frac{ A L m^4}{32 \pi^2 }  \ln \left(\frac{4 \pi T}{m}\right)- \nonumber\\
&&\left[\frac{A}{4}  \sum\limits_{n_1=1}^\infty \sqrt{\frac{m^3}{L \pi^3 n_1^3}} K_{\frac{3}{2}} \left(2 n_1 m L\right) \right]T, \nonumber\\
F_{\mbox{\scriptsize free}}(T,L) &\xrightarrow[T \gg m]{TL\gg 1}&  - \left( \frac{ A L \pi^2}{90 } \right) T^4 + \left( \frac{ A L m^2 }{24 } \right) T^2 + \frac{ A L m^4}{32 \pi^2 }  \ln \left(\frac{4 \pi T}{m}\right),\nonumber\\
F_{\mbox{\scriptsize Casimir}}(T,L) &\xrightarrow[T \gg m]{TL\gg 1}& - \left[\frac{AL}{4} \sum\limits_{n_1 = 1}^\infty \left(\frac{m}{\pi L n_1}\right)^{\frac{3}{2}} K_{\frac{3}{2}} \left(2 n_1 m L\right) \right] T \nonumber\\
&\xrightarrow{mL\ll 1}&- \frac{A \zeta(3)}{16 \pi L^2}T.
\end{eqnarray}
Notice that in the massless limit, the coefficients of the $T^2$ and $\ln T$ terms go to zero, while only the $T^4$ and $T$ terms remain.

Below we outline a few alternative derivations of our main result, {\it i.e.}, $F_{\mbox{\scriptsize Casimir}}(T,L)$ given by Eq.~(\ref{s27}), with details presented in appendices. As mentioned above, the spatial modes for the massive scalar field are also regular, and hence the sum over $n_1$ of the bounded case given by Eq.~(\ref{s241}) can be also calculated using the generalized Abel-Plana summation formula, as an alternative method. After subtracting the free case, we obtain an expression for $F_{\mbox{\scriptsize Casimir}}(T ,L)$ which is identical to the above expression (see Appendix \ref{appendixC:FCasimir}). Moreover, we can start with the same first form of the free energy given by Eq.~(\ref{SZeta}) as used in the computations by the ZFA, and use the Abel-Plana summation formula for both sums over $n_0$ and ${n_1}$ modes. After simplifying, we obtain exactly the same expression given by Eq.~(\ref{s27}) (see Appendix \ref{appendixC:FCasimir}). 
 
One can easily show that using the second form of the free energy given by Eq.~(\ref{s5}), one obtains exactly the same expression as in Eq.~(\ref{s27}). We show the details of this computation, in which we utilize dimensional regularization, in Appendix \ref{appendixD:The Dimensional Regularization}. 
As we have shown, the free energies of both the bounded and free cases contain $F_{\mbox{\scriptsize free}}(0,L)$ which is in principle divergent. Its value, obtained using the second form in Appendix \ref{appendixD:The Dimensional Regularization}, is proportional to $\Gamma(-2)$ which is divergent. This is due to the fact that the second form given by Eq.~(\ref{s5}), similar to the expression of the first form given by Eq.~(\ref{s04}) and in contrast to the expression of the first form given by Eq.~(\ref{s4}), does not have an embedded analytic continuation. 
One of the advantages the fundamental definition is that $F_{\mbox{\scriptsize free}}(0,L)$ terms cancel upon subtraction of $F_{\mbox{\scriptsize free}}(T,L)$ from $F_{\mbox{\scriptsize bounded}}(T,L)$,  whether they are infinite or have been rendered finite by analytic continuations. The other advantage is that both $F_{\mbox{\scriptsize bounded}}(T,L)$ and $F_{\mbox{\scriptsize free}}(T,L)$ contain $\Delta F_{\mbox{\scriptsize free}}(T,L)$ which also cancel upon subtraction, regardless of whether it is a simple polynomial of $T$ or not.

When using the fundamental approach, we have implicitly assumed that the contributions to the Casimir free energy coming from the regions outside of the bounded region cancel with the corresponding contributions of the free case. We use the Boyer method to ascertain this cancellation in Appendix \ref{appendixE:The Boyer method}.

In Fig.~(\ref{fp8}), we plot the Casimir free energy of a massive scalar field for various values of mass. As can be seen, the Casimir free energy goes to zero rapidly as mass of the scalar field increases and decreases linearly as the temperature increases with a slope that depends on the mass, in accordance with Eq.~(\ref{HighTFA}). As can be seen directly from Eq.~(\ref{s27}), the Casimir free energy goes to zero rapidly as $L$ increases, as well. Moreover, as can be seen from Fig.~(\ref{fp8}), and can be shown easily from Eq.~(\ref{s27}), the massless limit of our result for the massive case coincides exactly with the massless case given by Eq.~(\ref{s12}). The zero temperature limit of $F_{\mbox{\scriptsize Casimir}}(T,L)$, given by Eq.~(\ref{s27}), yields the following well known result, 
\begin{eqnarray}\label{s28}
\hspace{-9mm}F_{\mbox{\scriptsize Casimir}}(0,L) =E_{\mbox{\scriptsize Casimir}}(0,L) = - \frac{A m^2}{8 \pi^2 L} \sum\limits_{n_1=1}^{\infty}    \frac{K_{2} \left(2 n_1 m L\right)}{n_1^2} .
\end{eqnarray} 
\begin{center}
\begin{figure}[h!] 
\includegraphics[width=13.5cm]{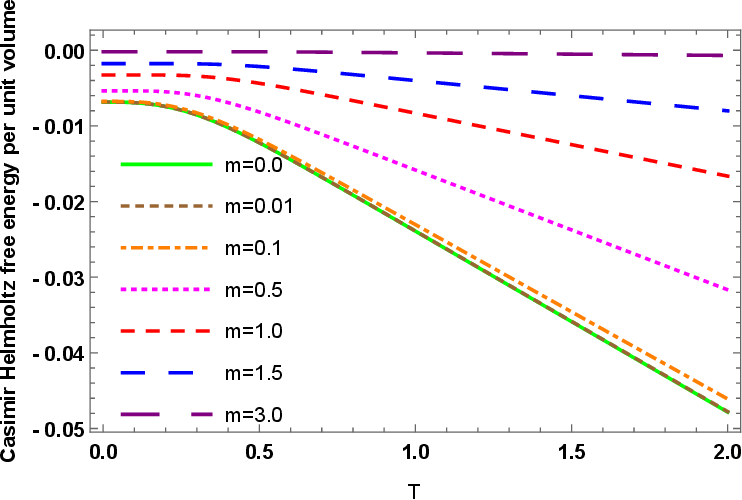}
\caption{\label{fp8} \small
The Casimir free energy per unit volume for a massive real scalar field between two parallel plates as a function of temperature with fixed plate separation $L = 1.0$, for various values of mass $m = \{ 0.0,0.01,0.1,0.5,1.0,1.5,3.0 \} $. Note that the Casimir free energy goes to zero as the mass increases and decreases linearly as the temperature increases.}
\end{figure}
\end{center}

\begin{center}
\begin{figure}[h!] 
\includegraphics[width=13.5cm]{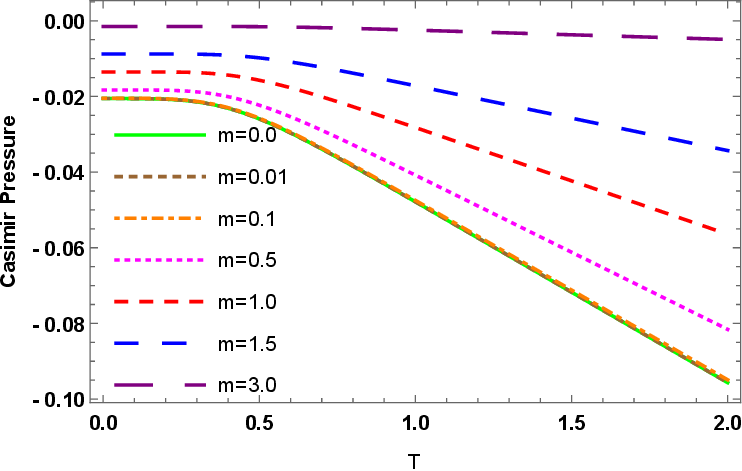}
\caption{\label{fp9} \small
The Casimir pressure for a massive real scalar field between two parallel plates as a function of temperature with fixed plate separation $L = 1.0$, for various values of mass $m = \{ 0.0,0.01,0.1,0.5,1.0,1.5,3.0 \} $. Note that the Casimir pressure goes to zero as the mass increases and decreases linearly as the temperature increases.}
\end{figure}
\end{center}

Now, one can obtain other thermodynamic quantities including, the Casimir pressure, Casimir energy, and Casimir entropy from the expression we have obtained for the Casimir free energy in Eq.~(\ref{s27}). We calculate the Casimir pressure for a massive scalar field, in analogy with the massless case shown in Eq.~(\ref{s13}), and obtain,  
\begin{eqnarray}\label{s29}
&&\hspace{-9mm}P_{\mbox{\scriptsize Casimir}}(T,L) = - \frac{ m^2}{\pi^2 } \sum\limits_{n_1 = 1}^\infty\Bigg\{ \frac{1 }{8 L^2n_1^2} \left[ 3 K_{2} \left(2 n_1 m L\right) + \left(2 n_1 m L\right) K_{1} \left(2 n_1 m L\right)\right] +\nonumber \\
&&\hspace{-9mm}\sum\limits_{n_0=1}^{\infty}  \frac{\left[m T \omega_{n_0, n_1}\left(2 n_1 L\right)^2 K_{1} \left(m \beta \omega_{n_0, n_1}\right) + \left(12 n_1^2 T^2 L^2 -n_0^2\right) K_{2} \left(m \beta \omega_{n_0, n_1}\right)\right]}{\beta^2\omega_{n_0, n_1}^4} \Bigg\}  ,\nonumber\\
\end{eqnarray}
where $\omega_{n_0, n_1} = \sqrt{n_0^2+\left(2 n_1 TL\right)^2}$. The zero temperature and finite temperature correction parts, {\it i.e.}, $P_{\mbox{\scriptsize Casimir}}(0,L)$ and $\Delta P_{\mbox{\scriptsize Casimir}}(T,L)$, are associated with the two terms in Eq.~(\ref{s29}), respectively. We plot $P_{\mbox{\scriptsize Casimir}}(T,L)$ for various values of mass in Fig.~(\ref{fp9}). As can be seen, the Casimir pressure also goes to zero rapidly as the mass of scalar field increases and decreases linearly at high temperature. Moreover, as can be seen from Fig.~(\ref{fp9}), and can be shown easily from Eq.~(\ref{s29}), the massless limit of our result for the massive case coincides exactly with the massless case given by Eq.~(\ref{s13}).

The Casimir energy can be calculated using either of the following two expressions,
\begin{equation}
E_{\mbox{\scriptsize Casimir}}(T,L) = E_{\mbox{\scriptsize bounded}}(T,L) - 
E_{\mbox{\scriptsize free}}(T,L)=\frac{\partial}{\partial \beta} \left[\beta F_{\mbox{\scriptsize Casimir}}(T,L)\right].
\end{equation}
The first expression is its fundamental definition. We use the second expression to obtain,
\begin{eqnarray}\label{s2009}
 E_{\mbox{\scriptsize Casimir}}(T,L)&=& - \frac{A m^2 L}{\pi^2} \sum\limits_{n_1 = 1}^\infty \Bigg\{ \frac{K_{2} \left(2 n_1 m L\right)}{8 L^2 n_1^2} +T^2\sum\limits_{n_0=1}^{\infty} \frac{1}{\omega_{n_0, n_1}^2} \times \nonumber \\
&& \left[K_{2} \left(m \beta \omega_{n_0, n_1}\right)- \frac{m \beta n_0^2}{\omega_{n_0, n_1}} K_{3}\left( \beta m \omega_{n_0, n_1}\right)\right] \Bigg\}  . 
\end{eqnarray} 
Finally, we calculate the Casimir entropy and obtain,
\begin{eqnarray}\label{s2020}
\hspace{-2mm}S_{\mbox{\scriptsize Casimir}}(T,L) = -\frac{\partial}{\partial T} \left[F_{\mbox{\scriptsize Casimir}}(T,L)\right] =\frac{A }{\pi^2} \sum\limits_{n_0 = 1}^\infty \sum\limits_{n_1 = 1}^\infty \frac{L m^3 n_0^2 }{\omega_{n_0, n_1}^3} K_{3}\left( \beta m \omega_{n_0, n_1}\right). 
\end{eqnarray}
Equations (\ref{s27}), (\ref{s2009}), and (\ref{s2020}) lead to the following expected relation
\begin{equation}\label{FES}
	F_{\mbox{\scriptsize Casimir}}(T,L) = E_{\mbox{\scriptsize Casimir}}(T,L) - TS_{\mbox{\scriptsize Casimir}}(T,L).
\end{equation}
The high temperature limit of $E_{\mbox{\scriptsize Casimir}}(T,L)$ is zero and those of $S_{\mbox{\scriptsize Casimir}}(T,L)$ for the massive and massless cases are 
\begin{equation}\label{ShighT}
	S_{\mbox{\scriptsize Casimir}}(T,L) \xrightarrow[T \gg m]{TL\gg 1} \frac{AL}{4} \sum\limits_{n_1 = 1}^\infty \left(\frac{m}{\pi L n_1}\right)^{\frac{3}{2}} K_{\frac{3}{2}} \left(2 n_1 m L\right) \xrightarrow{mL\ll 1} \frac{A \zeta(3)}{16 \pi L^2}.
\end{equation}
That is, the Casimir entropy goes to a nonzero positive constant at high temperatures\footnote{This is in contrast to the fermionic case, where the Casimir entropy and free energy also go to zero \cite{r26Goushe.}.}. Hence, Eqs.~(\ref{FES}) and (\ref{ShighT}) show that the appearance of the classical term, {\it i.e.}, the linear temperature dependence of $F_{\mbox{\scriptsize Casimir}}(T,L)$ at high $T$, shown in Eq.~(\ref{HighTFA}), is due to the constant nonzero  limit of $S_{\mbox{\scriptsize Casimir}}(T,L)$.

\begin{center}
\begin{figure}[h!] 
\includegraphics[width=13.5cm]{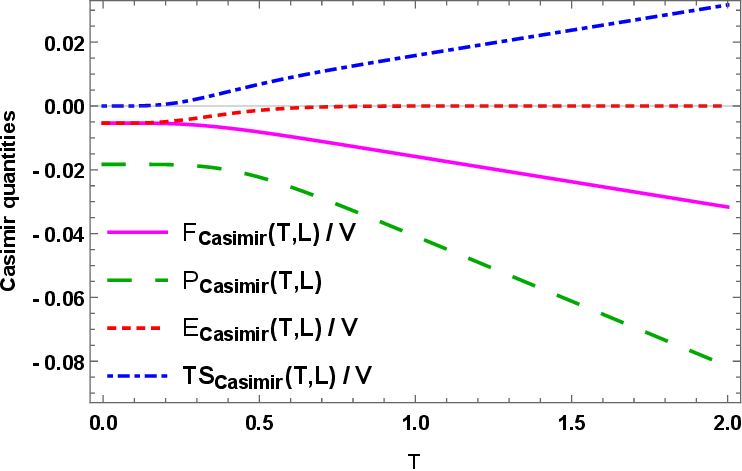}
\caption{\label{fp60} \small
The Casimir thermodynamic quantities, including the free energy, pressure, energy, and entropy, obtained using the fundamental approach, for a massive real scalar field between two parallel plates as a function of temperature, with fixed plate separation $L = 1.0$, and mass $m = 0.5$. Note that the $F_{\mbox{\scriptsize Casimir}}(T,L)$ and $P_{\mbox{\scriptsize Casimir}}(T,L)$ decrease linearly, while the $E_{\mbox{\scriptsize Casimir}}(T,L)$ goes to zero and $TS_{\mbox{\scriptsize Casimir}}(T,L)$ increases linearly, as temperature increases. So, $S_{\mbox{\scriptsize Casimir}}(T,L)$ goes to a positive constant at high temperatures.}
\end{figure}
\end{center}

In figure Fig.~(\ref{fp60}), we show all of these Casimir thermodynamic quantities. As can be seen in this figure, the Casimir free energy and pressure decrease linearly as temperature increases, while the Casimir energy goes to zero. On the other hand, in the high temperatures limit, the Casimir entropy goes to a positive constant, and hence the $TS_{\mbox{\scriptsize Casimir}}(T,L)$ increases linearly in that limit. Moreover, one can easily show that all of the Casimir thermodynamic quantities go to zero as $m$ or $L$ increases. It is worth mentioning that the Casimir entropy in our model is positive for the entire range of $T$. In the fundamental approach used here, the Casimir entropy is defined as $S_{\mbox{\scriptsize Casimir}}(T,L)= S_{\mbox{\scriptsize bounded}}(T,L)-S_{\mbox{\scriptsize free}}(T,L)$. Hence the interpretation of $S_{\mbox{\scriptsize Casimir}}(T,L)>0$ for entire range of $T$ is simply that $S_{\mbox{\scriptsize bounded}}(T,L)> S_{\mbox{\scriptsize free}}(T,L)$, which are incidentally both positive. 

\section{Massive Scalar Fields and the Generalized Zeta Function}\label{Epstein}

\indent
The zeta function approach (ZFA) has been used to calculate the Casimir free energy for the massive scalar field between two plates and some solutions have been presented (see for example in~\cite{r29Teo., r45Teo., r283Flach., r29Erdas., r29Alex.}). In this section, we compute explicitly the final results for the Casimir free energy and Casimir pressure for this problem using the ZFA, the Schl\"{o}milch formulas approach (SFA) as the second representative of the analytic continuation approach, and also using the zero temperature subtraction approach (ZTSA). We then show that, similar to the massless case, only the results of ZFA and SFA are equivalent, while neither of these results are equivalent to the one obtained in Sec.~\ref{massive} based on the fundamental approach. Most importantly, we show that, contrary to the massless case, these discrepancies cannot be fixed completely by the renormalization program mentioned before, since the extra unphysical terms are non-polynomial functions of $T$.

To use the ZFA, we start with the first form of the free energy given by Eq.~(\ref{SZeta}), and compute the sum over spatial modes using the inhomogeneous Epstein zeta function (see Appendix \ref{appendixA:The zeta function}). The expression for the free energy becomes
\begin{eqnarray}\label{s33}
&& \hspace{-4mm} F_{\mbox{\scriptsize Zeta}}(T,L) = -\frac{T A}{16 \pi}  \lim\limits_{s \to 0} \frac{\partial }{\partial s} \frac{{\mu^{2 s}}}{\Gamma (s)}   \sum\limits_{n_0 = -\infty}^\infty \Bigg\{ - \Gamma (s -1)  \left(\omega'_{n_0}\right)^{2-2s} +\nonumber\\
&&\hspace{-4mm} \frac{L \Gamma \left( s -\frac{3}{2}\right)}{ \sqrt{\pi}} \left(\omega'_{n_0}\right)^{3 - 2s}  +\frac{4L}{\sqrt{\pi}}  \sum\limits_{n_1 =1}^\infty \left(\frac{\omega'_{n_0}}{n_1 L}\right)^{\frac{3}{2} - s} K_{\frac{3}{2} - s} \left(2 n_1 L \omega'_{n_0}\right) \Bigg\},
\end{eqnarray}
where $\omega'_{n_0} = \sqrt{\left(\frac{2 n_0 \pi}{\beta}\right)^2+ m^2}$. To explore the mechanism of removal of divergences from this point forward, it is useful to compare this expression with the analogous one that we have obtained for the massless case after using the inhomogeneous zeta function on the spatial modes, {\it i.e.}, Eq.~(\ref{s21c}). There are terms in both expressions which include the divergent sum over the Matsubara frequencies and are remnants from the use of the inhomogeneous zeta function on the spatial modes. As mentioned in Sec.~\ref{zeta} for the massless case, we can obtain the analytic continuation of these divergent terms using a supplementary zeta function. So, here, we first present the sum over $n_0$ modes in terms of positive integers and a zero mode, and then use the inhomogeneous Epstein zeta function given by Eq.~(\ref{A8}). After computing this divergent sum for the first and second terms in the curly bracket in Eq.~(\ref{s33}), we obtain the following expression
\begin{eqnarray}\label{s34}
&&\hspace{-4mm}F_{\mbox{\scriptsize Zeta}}(T,L) =  -\frac{ A}{8\sqrt{\pi^3}}  \lim\limits_{s \to 0} \frac{\partial }{\partial s} \frac{{\mu^{2 s}}}{\Gamma (s)} \left\lbrace  -\frac{\Gamma \left( s - \frac{3}{2}\right) m^{3 - 2s}}{4} + \frac{L\Gamma \left( s - 2\right)m^{4 - 2s}}{4 \sqrt{\pi}}+\right. \nonumber \\
&& \left. \sum\limits_{j=1}^\infty \left[ \frac{L}{\sqrt{\pi}} \left(\frac{m2}{j \beta}\right)^{2-s} K_{2-s}\left(j m \beta\right) - \left(\frac{m2}{j \beta}\right)^{\frac{3}{2}-s} K_{\frac{3}{2}-s}\left(j m \beta\right) \right] + \right. \nonumber \\
&& \left. 2TL \sum\limits_{n_0= - \infty}^\infty \sum\limits_{n_1=1}^\infty \left(\frac{\omega'_{n_0}}{L n_1}\right)^{\frac{3}{2}-s} K_{\frac{3}{2}-s}\left(2n_1 L \omega'_{n_0}\right)  \right\rbrace.
\end{eqnarray}
As can be seen in this expression, the second term includes a divergent gamma function, {\it i.e.}, $\Gamma \left(s - 2\right)$, and is equivalent to the first term of the $F_{\mbox{\scriptsize bounded}}(T,L)$, given by Eq.~(\ref{s24}) after evaluating the integral over $p$: As shown in Eq.~(\ref{F10}), these terms are both equal to the analytic continuation of $F_{\mbox{\scriptsize free}}(0,L)$ which is equal to $[-A L m^4/(128 \pi^2)]  \left[ 3 - 4 \ln \left(\frac{m}{\mu}\right)\right]$. This is the analytic continuation embedded in the first form of free energy Eq.~(\ref{s4}), and has rendered this term finite.
Finally, evaluating $\lim\limits_{s \to 0} \frac{\partial}{\partial s}$, we can express the final result as follows
\begin{eqnarray}\label{s35d}
&&\hspace{-6mm} F_{\mbox{\scriptsize Zeta}}(T,L) = \frac{A m^3}{24 \pi}- \frac{A L m^4}{128 \pi^2}  \left[ 3 - 4 \ln \left(\frac{m}{\mu}\right) \right] +\Delta F_{\mbox{\scriptsize free}}(T,L) + \nonumber\\
&& \hspace{-6mm} A  \sum\limits_{j=1}^\infty \left(\frac{T m}{2 \pi j}\right)^{\frac{3}{2}} K_{\frac{3}{2} }\left(j m\beta \right)- \frac{A L T}{4\sqrt{\pi^3}} \sum\limits_{n_0= - \infty}^\infty \sum\limits_{n_1=1}^\infty \left(\frac{\omega'_{n_0}}{L n_1}\right)^{\frac{3}{2}} K_{\frac{3}{2}}\left(2n_1 L \omega'_{n_0}\right) .
\end{eqnarray}
The order of the terms presented above is the same as in Eq.~(\ref{s34}), {\it i.e.}, the first and second terms are the finite temperature-independent terms mentioned above, $\Delta F_{\mbox{\scriptsize free}}(T,L)$ is the thermal correction of the free case given by Eq.~(\ref{s25}), and the fourth term is an extra L-independent thermal correction term which is not in the Casimir free energy obtained in the last section. In fact, the first and fourth terms in $F_{\mbox{\scriptsize Zeta}}(T,L)$, given by Eq.~(\ref{s35d}), which are L-independent and do not contribute to the pressure, are precisely minus one half of the contribution of the zero spatial mode that is disallowed by the Dirichlet boundary conditions, and these terms make $F_{\mbox{\scriptsize Zeta}}(T,L)$ a nonextensive thermodynamic quantity (see Appendix~\ref{appendixA:The zeta function})\footnote{As discussed in Sec.~\ref{zeta}, analogous superfluous term, {\it i.e.}, $\frac{A \zeta (3)}{4 \pi} T^3$, has appeared in the expression for $F_{\mbox{\scriptsize Zeta}}(T,L)$ for the massless case given by Eq.~(\ref{s21bgd}). In fact this term is precisely the massless limit of the two terms mentioned above.}. Oftentimes, the first two terms are discarded, with reasoning that they are temperature independent. However, note that the second one contributes to the pressure. Next, we express the sum over the $n_0$ modes of the last term by a sum over positive integers and a zero mode, and evaluate it using the Abel-Plana formula, as used for Eq.~(\ref{C7}). The result is $F_{\mbox{\scriptsize Casimir}}(T,L)$ given in Eq.~(\ref{s27}). After simplifying, we can summarize the final result as follows
\begin{eqnarray}\label{s35dfd}
\hspace{-2mm}F_{\mbox{\scriptsize Zeta}}(T,L) &=& F_{\mbox{\scriptsize Casimir}}(T,L) + \Delta F_{\mbox{\scriptsize free}}(T,L)- \frac{A L m^4}{128 \pi^2}  \left[ 3 - 4 \ln \left(\frac{m}{\mu}\right) \right] +\nonumber \\ 
&&\frac{A m^3}{24 \pi} +A\sum\limits_{n_0=1}^\infty \left(\frac{m T}{2 \pi n_0}\right)^{\frac{3}{2}} K_{\frac{3}{2}} \left(n_0 \beta m\right) ,
\end{eqnarray}
where $\Delta F_{\mbox{\scriptsize free}}(T,L)$ is given by Eq.~(\ref{s25}).

As stated above, we also obtain the Casimir free energy for a massive real scalar field confined between two plates using the Schl\"{o}milch formulas approach (SFA)~\cite{r29Junji.}, as a second representative of the analytic continuation approach, similar to the massless case in Sec.~\ref{zeta}. As mentioned before, we can use this approach for obtaining only the thermal corrections of the Casimir effect, while we calculate the zero temperature part separately using the Epstein inhomogeneous zeta function. To compute the Casimir free energy in this approach, we use the second form of the free energy given by Eq.~(\ref{s5}). In the massive case, we have $\omega_{n_1, K_T} = \sqrt{\left(\frac{n_1 \pi}{L}\right)^2+K_T^2 +m^2}$ and use the generalized Schl\"{o}milch formulas to calculate sum over spatial modes of the thermal corrections part, to obtain (see Appendix~\ref{appendixB:The Schlomilch formula})
\begin{eqnarray}\label{Sch8}
F_{\mbox{\scriptsize SFA}}(T,L)&=&F_{\mbox{\scriptsize Zeta}}(T,L) ,
\end{eqnarray}
where the expression for $F_{\mbox{\scriptsize Zeta}}(T,L)$ is given by Eq.~(\ref{s35dfd}). As expected, the results obtained using the ZFA and the SFA, as two distinct methods within the analytic continuation approach, are completely equivalent.

For our last approach, which is ZTSA, we note that no new computation is needed, since  its definition holds for both massive and massless cases. That is, rewriting Eqs.~(\ref{s16}) and (\ref{s16000}), we have,
\begin{equation}\label{s31}
	F_{\mbox{\scriptsize ZTSA}}(T,L) \equiv F_{\mbox{\scriptsize bounded}} (T,L) - F_{\mbox{\scriptsize free}} (0,L)= F_{\mbox{\scriptsize Casimir}}(T,L) +\Delta F_{\mbox{\scriptsize free}}(T,L).
\end{equation}
For the massive case, we simply use Eq.~(\ref{s24}) for $F_{\mbox{\scriptsize bounded}}(T,L)$,  Eq.~(\ref{s25}) for $F_{\mbox{\scriptsize free}}(T,L)$, and Eq.~(\ref{s27}) for $F_{\mbox{\scriptsize Casimir}}(T,L)$. This shows that, by definition, the results of ZTSA differs from those of the fundamental approach by $\Delta F_{\mbox{\scriptsize free}}(T,L)$, which is zero only at zero temperature.

Now we can compare the results of the analytic continuation approach, {\it i.e.,} $F_{\mbox{\scriptsize Zeta}}$ and $F_{\mbox{\scriptsize SFA}}$ given by Eqs.~(\ref{s35dfd}) and (\ref{Sch8}), with the fundamental approach, {\it i.e.,} $F_{\mbox{\scriptsize Casimir}}$ given in Eq.~(\ref{s27}), and the zero temperature subtraction approach, {\it i.e.,} $F_{\mbox{\scriptsize ZTSA}}$ given by Eq.~(\ref{s31}). First we note that the sum of the first two terms of $F_{\mbox{\scriptsize Zeta}}$ is precisely $F_{\mbox{\scriptsize ZTSA}}$. Hence, $F_{\mbox{\scriptsize Zeta}}$ and $F_{\mbox{\scriptsize SFA}}$ contain three extra terms as compared to $F_{\mbox{\scriptsize ZTSA}}$, and four extra terms as compared to $F_{\mbox{\scriptsize Casimir}}$. The first extra term in Eq.~(\ref{s35dfd}) is $\Delta F_{\mbox{\scriptsize free}}(T,L)$. The second extra term of  is the analytic continuation of the divergent $F_{\mbox{\scriptsize free}}(0,L)$ term computed in Eq.~(\ref{F10}), which does contribute a $T$-independent term to the pressure. As mentioned above, the third and fourth extra terms in Eq.~(\ref{s35dfd}) are equivalent to the contribution of the $- F_{\mbox{\scriptsize Zeta}}^{n_1=0}(T)/2$, which is disallowed by the Dirichlet boundary condition.

An important feature of the Casimir thermodynamic quantities is their high temperature expansion. Below we display this expansion for $F_{\mbox{\scriptsize Zeta}}(T,L)$, 
\begin{eqnarray}\label{s35dfdBB}
	&&\hspace{-24mm}F_{\mbox{\scriptsize Zeta}}(T,L) \xrightarrow[T \gg m]{TL\gg 1}
	- \frac{A}{4} \left[ \sum\limits_{n_1=1}^\infty \sqrt{\frac{m^3}{L \pi^3 n_1^3}} K_{\frac{3}{2}} \left(2 n_1 m L\right) \right]T\nonumber\\
	&&\hspace{7mm}  - \left( \frac{ A L \pi^2}{90 } \right) T^4     +\left( \frac{A L m^2}{24 } \right) T^2   +\frac{A L m^4}{32 {\pi}^2}\ln \left( \frac{2 T}{m}\right)\nonumber \\
	&& \hspace{8mm} +\left( \frac{A \zeta (3)}{4 \pi} \right) T^3 -\left( \frac{A m^2}{8 \pi}\right) T \ln \left(\frac{T}{m}\right).
\end{eqnarray}
The first and second lines are the high temperature expansions of  $F_{\mbox{\scriptsize Casimir}}$ and $F_{\mbox{\scriptsize free}}$, respectively. Hence the sum of the fist two lines is the high temperature expansion of $F_{\mbox{\scriptsize ZTSA}}$ and $F_{\mbox{\scriptsize bounded}}$. That is
\begin{eqnarray}\label{ZTSAHighT}
	&&\hspace{-24mm}F_{\mbox{\scriptsize ZTSA}}(T,L) \xrightarrow[T \gg m]{TL\gg 1}
	- \frac{A}{4} \left[ \sum\limits_{n_1=1}^\infty \sqrt{\frac{m^3}{L \pi^3 n_1^3}} K_{\frac{3}{2}} \left(2 n_1 m L\right) \right]T\nonumber\\
	&&\hspace{7mm}  - \left( \frac{ A L \pi^2}{90 } \right) T^4     +\left( \frac{A L m^2}{24 } \right) T^2   +\frac{A L m^4}{32 {\pi}^2}\ln \left( \frac{2 T}{m}\right)
\end{eqnarray}
The two terms in last line of Eq.~(\ref{s35dfdBB}) are the high temperature expansion of the last extra term of $F_{\mbox{\scriptsize Zeta}}(T,L)$, given in Eq.~(\ref{s35dfd}). Note that this expansion includes $\ln(T/m)$ and  $T\ln(T/m)$, in addition to integer powers of $T$.

As we have shown here for the case of scalars, and also for the fermionic case~\cite{r26Goushe.}, it has long been recognized that the use of the ZFA yields extra unphysical terms. To remedy this, Geyer et al.~\cite{r17Geyer.} defined a renormalization program in which the polynomial terms obtained using the heat kernel coefficients with powers greater or equal to two are subtracted. In their work on the bosonic cases, they emphasized that all of the mentioned terms are of quantum character and do not include the classical term which is proportional to the temperature. We now explore the results of this renormalization program. Below, we state the renormalization program as presented in reference~\cite{r17Geyer.},
\begin{equation}\label{s35a}
F^{\mbox{\scriptsize ren}}= E_{0}^{\mbox{\scriptsize ren}} + \Delta_{\mbox{\scriptsize T}} F_{0} 
- \alpha_{0} \left( k_{\mbox{\scriptsize B}} T\right)^4 -\alpha_{1} \left( k_{\mbox{\scriptsize B}} T\right)^3  - \alpha_{2} \left(  k_{\mbox{\scriptsize B}} T\right)^2 . 
\end{equation}
We can identify the sum of the first two terms as $F_{\mbox{\scriptsize Zeta}}(T,L)$. The heat kernel coefficients $\alpha_n$ depend on geometrical characteristics of the configuration.
We calculate these coefficients in Appendix~\ref{appendixF:The Heat kernel method}, and show that they are identical to those of the high temperature expansions of $F_{\mbox{\scriptsize Zeta}}$, as given by Eq.~(\ref{s35dfdBB}).
We also show how the divergent vacuum energy at zero temperature can be obtained by the heat kernel method.
Therefore, based on this renormalization program, the physical Casimir free energy for a massive scalar field confined between two parallel plates, and its high temperature limit, obtained using zeta function are as follows 
\begin{eqnarray}\label{s36}
&&\hspace{-8mm}F_{\mbox{\scriptsize Zeta}}^{\mbox{\scriptsize ren}}(T,L)= F_{\mbox{\scriptsize Zeta}}(T,L) +
       \frac{A L \pi^2}{90 }  T^4   -   \frac{A \zeta \left(3\right)}{4 \pi } T^3  -   \frac{A L m^2}{24 } T^2 \\ 
&&\hspace{-8mm}\xrightarrow[T \gg m]{TL\gg 1}- \frac{A}{4} \left[ \sum\limits_{n_1=1}^\infty \sqrt{\frac{m^3}{L \pi^3 n_1^3}} K_{\frac{3}{2}} \left(2 n_1 m L\right) \right]T- \frac{A m^2}{8 \pi} T \ln \left(\frac{T}{m}\right)+\frac{A L m^4}{32 {\pi}^2}\ln \left( \frac{2 T}{m}\right),\nonumber
\end{eqnarray}
where $ F_{\mbox{\scriptsize Zeta}}$ is given by Eq.~(\ref{s35dfd}). Since the results of SFA and ZFA are equivalent, we can also define $F_{\mbox{\scriptsize SFA}}^{\mbox{\scriptsize ren}}(T,L) = F_{\mbox{\scriptsize Zeta}}^{\mbox{\scriptsize ren}}(T,L)$, and, henceforth, only concentrate on the latter. One can analogously define a renormalized ZTSA free energy as follows  
\begin{eqnarray}\label{s36b}
F_{\mbox{\scriptsize ZTSA}}^{\mbox{\scriptsize ren}}(T,L) &=& F_{\mbox{\scriptsize ZTSA}}(T,L) + \left( \frac{A L \pi^2}{90 } \right) T^4     -  \left( \frac{A L m^2}{24 } \right) T^2  \\
&\xrightarrow[T \gg m]{TL\gg 1}&- \frac{A}{4} \left[ \sum\limits_{n_1=1}^\infty \sqrt{\frac{m^3}{L \pi^3 n_1^3}} K_{\frac{3}{2}} \left(2 n_1 m L\right) \right]T+\frac{A L m^4}{32 {\pi}^2}\ln \left( \frac{2 T}{m}\right),\nonumber
\end{eqnarray}
where $ F_{\mbox{\scriptsize SFA}}$ and $ F_{\mbox{\scriptsize ZTSA}}$ are given by Eqs.~(\ref{Sch8}, \ref{s31}).\\

%

\begin{center}
\begin{figure}[!h] 
\includegraphics[width=13.5cm]{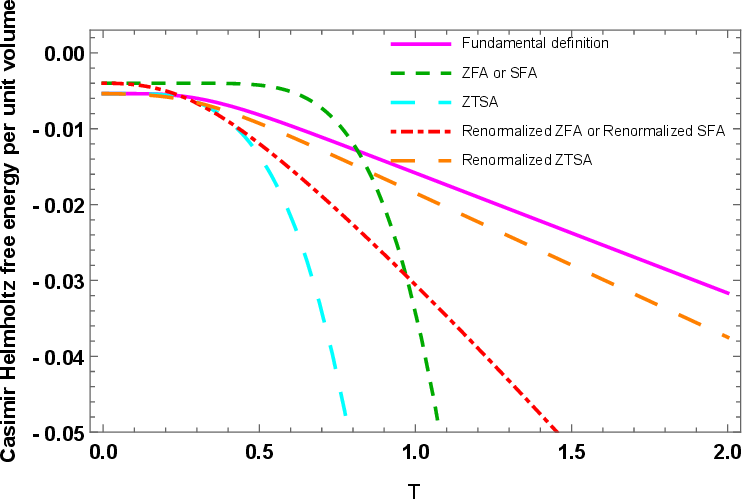}
\caption{\label{fp21} \small
The Casimir free energies per unit volume for a massive real scalar field between two parallel plates as a function of temperature with fixed plate separation $L = 1.0$, and mass $m = 0.5$, obtained using seven methods within three approaches.}
\end{figure}
\end{center}

To illustrate the differences between the five different expressions that we have obtained for the Casimir free energy and their high temperature limits, {\it i.e.}, $F_{\mbox{\scriptsize Casimir}}$ given by Eqs.~(\ref{s27}) and (\ref{HighTFA}), $F_{\mbox{\scriptsize Zeta}}$ given by  Eq.~(\ref{s35dfd}) and Eq.~(\ref{s35dfdBB}), $F_{\mbox{\scriptsize ZTSA}}$ given by  Eq.~(\ref{s31}) and (\ref{ZTSAHighT}), $F_{\mbox{\scriptsize Zeta}}^{\mbox{\scriptsize ren}}(T,L)$ given by Eq.~(\ref{s36}), and $F_{\mbox{\scriptsize ZTSA}}^{\mbox{\scriptsize ren}}(T,L)$ given by Eq.~(\ref{s36b}), we plot them in Fig.~(\ref{fp21}). As can be seen in this figure, none of these results are equivalent. Even the renormalized versions do not match $F_{\mbox{\scriptsize Casimir}}$. The free energies obtained via ZFA and the ZTSA decrease as $T^4$ at high temperatures due to the black body term, while the Casimir free energy decreases as $T$ at high temperatures due to the classical term. The renormalized versions do not match $F_{\mbox{\scriptsize Casimir}}$ even at high temperatures: $F_{\mbox{\scriptsize Zeta}}$, besides the classical term, has extra $T\ln T$ and $\ln T$ terms, while  $F_{\mbox{\scriptsize ZTSA}}$ has an extra $\ln T$ term. Only the zero temperature limit of ZTSA matches that of $F_{\mbox{\scriptsize Casimir}}$.																																																																										


So far, we have illustrated that the conventional renormalization programs for the ZFA or ZTSA, based on subtracting powers of $T^2$ and higher in the high temperature expansions, do not in general yield the correct results. However, we can still utilize the facility of the zeta function method and use the fundamental approach as a new renormalization program. To do this, we need to calculate the free energy of the free massive case by applying the zeta function. Here, we emphasize that to use the fundamental approach, the procedures for the computation of the free energy in the bounded and free cases should be equivalent. Hence, we perform the same procedure as for the bounded case, {\it i.e.}, $F_{\mbox{\scriptsize Zeta}}(T,L)$ given by Eq.~(\ref{s34}), by assuming that in the free case there are plates located at $\pm L'/2$, which we shall eventually take to infinity,
\begin{eqnarray}
&&\hspace{-4mm}F_{\mbox{\scriptsize Zeta}}^{\mbox{\scriptsize free}}(T,L') =  -\frac{ A}{8\sqrt{\pi^3}}  \lim\limits_{s \to 0} \frac{\partial }{\partial s} \frac{\mu^{2 s}}{\Gamma (s)} \left\lbrace  -\frac{\Gamma \left( s - \frac{3}{2}\right) m^{3 - 2s}}{4} + \frac{L'\Gamma \left( s - 2\right)m^{4 - 2s}}{4 \sqrt{\pi}}+\right. \nonumber \\
&& \left. \sum\limits_{j=1} \left[ \frac{L'}{\sqrt{\pi}} \left(\frac{m2}{j \beta}\right)^{2-s} K_{2-s}\left(j m \beta\right) - \left(\frac{m2}{j \beta}\right)^{\frac{3}{2}-s} K_{\frac{3}{2}-s}\left(j m \beta\right) \right] + \right. \nonumber \\
&& \left. 2TL' \sum\limits_{n_0= - \infty}^\infty \sum\limits_{n_1=1}^\infty \left(\frac{\omega'_{n_0}}{L' n_1}\right)^{\frac{3}{2}-s} K_{\frac{3}{2}-s}\left(2n_1 L' \omega'_{n_0}\right)  \right\rbrace.
 \end{eqnarray}
As can be seen, there are terms in the above expression which are proportional to the volume and, as mentioned in Sec.~\ref{FCasimir}, in the fundamental approach the difference between the free energy of the bounded and free cases is considered at the same temperature $T$ and same volume. So, we express the second and third terms of $F_{\mbox{\scriptsize Zeta}}^{\mbox{\scriptsize free}}(T,L')$ at volume $V=AL$, while the last terms goes to zero as $L'\rightarrow \infty$. In this limit, the result is
\begin{eqnarray}\label{s36eew}
 &&\hspace{-4mm}\lim\limits_{L'\gg1}F_{\mbox{\scriptsize Zeta}}^{\mbox{\scriptsize free}}(T,L) =  -\frac{ A}{8\sqrt{\pi^3}}  \lim\limits_{s \to 0} \frac{\partial }{\partial s} \frac{{\mu^{2 s}}}{\Gamma (s)} \left\lbrace  -\frac{\Gamma \left( s - \frac{3}{2}\right) m^{3 - 2s}}{4} + \frac{L\Gamma \left( s - 2\right)m^{4 - 2s}}{4 \sqrt{\pi}}+\right. \nonumber \\
&& \left. \sum\limits_{j=1} \left[ \frac{L}{\sqrt{\pi}} \left(\frac{m2}{j \beta}\right)^{2-s} K_{2-s}\left(j m \beta\right) - \left(\frac{m2}{j \beta}\right)^{\frac{3}{2}-s} K_{\frac{3}{2}-s}\left(j m \beta\right) \right]  \right\rbrace.
 \end{eqnarray}
One can see that $F_{\mbox{\scriptsize Zeta}}^{\mbox{\scriptsize free}}(T,L)$, given by Eq.~(\ref{s36eew}), is equivalent to the first four terms of $F_{\mbox{\scriptsize Zeta}}(T,L)$, given by Eq.~(\ref{s34}). This implies that
\begin{equation}
	F_{\mbox{\scriptsize Casimir}}(T,L)= F_{\mbox{\scriptsize Zeta}}^{\mbox{\scriptsize bounded}}(T,L) - F_{\mbox{\scriptsize Zeta}}^{\mbox{\scriptsize free}}(T,L),
\end{equation}
where we have denoted $F_{\mbox{\scriptsize Zeta}}(T,L)$ by $F_{\mbox{\scriptsize Zeta}}^{\mbox{\scriptsize bounded}}(T,L)$ to emphasis that this is in accord with the fundamental definition. Note that the nonextensive terms, {\it i.e.}, the first and the fourth terms, have also canceled out in this approach and the final result is extensive. This expression can be looked upon as the correct renormalization scheme, but is nothing more than an, albeit useful, expression for the fundamental definition.

One can now obtain other thermodynamic quantities based on the expressions we have obtained for the free energy using the ZFA given in Eq.~(\ref{s35dfd}), the ZTSA given in Eq.~(\ref{s31}), and their renormalized versions given in Eqs.~(\ref{s36}, \ref{s36b}). As an example, we calculate the pressure for a massive scalar field using the free energy obtained via the zeta function, in analogy with the massless case shown in Eq.~(\ref{s19}). We express the result in terms of $P_{\mbox{\scriptsize Casimir}}(T,L)$, obtained by the fundamental approach and given in Eq.~(\ref{s29}), as follows
\begin{eqnarray}\label{s35}
 P_{\mbox{\scriptsize Zeta}}(T,L) =P_{\mbox{\scriptsize SFA}}(T,L)&=&  P_{\mbox{\scriptsize Casimir}}(T,L)+\frac{ m^4}{128 \pi^2}  \left[ 3 - 4 \ln\left(\frac{m}{\mu}\right) \right]+\Delta P_{\mbox{\scriptsize free}}(T),  \nonumber \\
\hspace{1 mm} \mbox{where}  \hspace{5 mm} \Delta P_{\mbox{\scriptsize free}}(T)&=&\frac{ T^2 m^2}{2 \pi^2} \sum\limits_{j = 1}^{\infty} \frac{K_{2}\left( j \beta m\right)}{j^2}   .
\end{eqnarray}
As before, the second term is a constant term which is a remnant from the use of the zeta function, and the third term is the thermal correction to the pressure of the free case, which the zeta function fails to subtract. 
Next, we calculate the pressure using the free energy obtained via the ZTSA. We express the result in terms of $P_{\mbox{\scriptsize Casimir}}(T,L)$ as follows,
\begin{equation}\label{s61}
P_{\mbox{\scriptsize ZTSA}}(T,L) =  P_{\mbox{\scriptsize Casimir}}(T,L) +\Delta P_{\mbox{\scriptsize free}}(T) ,
\end{equation}
where $\Delta P_{\mbox{\scriptsize free}}(T)$ is given in Eq.~(\ref{s35}).
Next, we calculate the pressure obtained via the renormalized zeta function, {\it i.e.}, $F_{\mbox{\scriptsize Zeta}}^{\mbox{\scriptsize ren}}(T,L)$, and the renormalized  ZTSA, {\it i.e.}, $F_{\mbox{\scriptsize ZTSA}}^{\mbox{\scriptsize ren}}(T,L)$, given by Eqs.~(\ref{s36}, \ref{s36b}). The results are,
\begin{eqnarray}\label{s62}
 \hspace{-8mm} P_{\mbox{\scriptsize Zeta}}^{\mbox{\scriptsize ren}}(T,L) =P_{\mbox{\scriptsize SFA}}^{\mbox{\scriptsize ren}}(T,L)&=&  P_{\mbox{\scriptsize Zeta}}(T,L)  - \left( \frac{ \pi^2}{90 } \right) T^4 + \left( \frac{m^2}{24 } \right) T^2 \\
 P_{\mbox{\scriptsize ZTSA}}^{\mbox{\scriptsize ren}}(T,L) &=&  P_{\mbox{\scriptsize ZTSA}}(T,L)  - \left( \frac{ \pi^2}{90 } \right) T^4 + \left( \frac{m^2}{24 } \right) T^2   .
 \end{eqnarray}
\begin{center}
	\begin{figure}[h!] 
		\includegraphics[width=13.4cm]{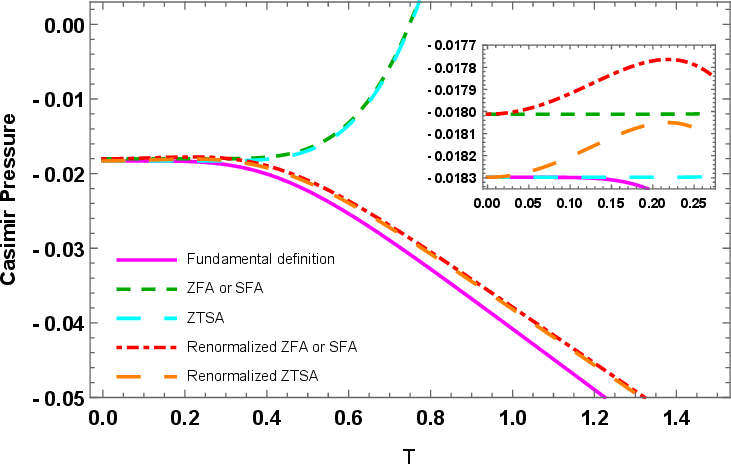}
		\caption{\label{fp50} \small
		The Casimir pressure for a massive real scalar field between two parallel plates as a function of temperature with fixed plate separation $L = 1.0$, and mass $m = 0.5$, obtained using seven methods within three approaches. The zoomed subgraph shows that at $T=0$, only the results of ZTSA coincide with that of the fundamental approach.}
	\end{figure}
\end{center}
In Fig.~(\ref{fp50}), we compare these results with the Casimir pressure obtained based on the fundamental approach, given by Eq.~(\ref{s29}). As can be seen, the pressure obtained using the ZFA and the ZTSA are negative at low temperatures and positive at high temperature, while the Casimir pressure is always negative and decreases linearly with increasing temperature. The differences between these results, besides the constant term present in $P_{\mbox{\scriptsize Zeta}}(T,L)$ and its renormalized version, are due to the thermal correction of pressure of free case $\Delta P_{\mbox{\scriptsize free}}(T)$ which is a non-polynomial function of $T$ for the massive scalar field, as presented in Eq.~(\ref{s35}). The pressure obtained using the ZFA and the ZTSA all diverge as $T^4$ at high temperatures, and their renormalized versions as $T\ln (T/m)$ and $\ln (T/m)$. At $T=0$, only the ZTSA results match the $P_{\mbox{\scriptsize Casimir}}(0,L)$.

On a side note, we can now examine the applicability and limitations of the piston method, which can be used if one is interested only in the Casimir pressure. In this approach, the pressure on the bounded and unbounded sides of the piston are calculated and subtracted (see in~\cite{r29Marac., r29Teo., r29Rand.}). The zeta function method is almost invariably used for this purpose. To trace the cancellations that occur in this subtraction, we first write Eq.~(\ref{s35}), with the labeling mentioned above, as follows 
\begin{eqnarray}\label{pressue}
		P_{\mbox{\scriptsize Zeta}}^{\mbox{\scriptsize bounded}}(T,L) &=&  P_{\mbox{\scriptsize Casimir}}(T,L) + P_{\mbox{\scriptsize Zeta}}^{\mbox{\scriptsize free}}(T), \hspace{1 cm} \mbox{where} \nonumber \\
P_{\mbox{\scriptsize Zeta}}^{\mbox{\scriptsize free}}(T)&=& \frac{ m^4}{128 \pi^2}  \left[ 3 - 4 \ln \left(\frac{m}{\mu}\right) \right]+\Delta P_{\mbox{\scriptsize free}}(T),
\end{eqnarray}
where $\Delta P_{\mbox{\scriptsize free}}(T)$ is given by Eq.~(\ref{s35}).
Now it is clear that if the zeta function is used to calculate the pressure of both the bounded and unbounded sides of the piston and the results are subtracted, the extra term cancels and one obtains the correct result, {\it i.e.} $P_{\mbox{\scriptsize Casimir}}(T,L)$. In fact, one obtains the correct expression as long as the method used for the two regions is the same, whether it is ZFA, SFA, ZTSA or the Abel-Plana formula.



\section{Summary and Discussion}\label{summary}

\indent

In this paper, we have explored the implications and results of the fundamental definition of the Casimir free energy for a scalar field, and how they compare with the results based on two general approaches in common use, {\it i.e.}, the analytic continuation approach, represented here by the zeta function approach (ZFA) and the Schl\"{o}milch formula approach (SFA), and the zero temperature subtraction approach (ZTSA). We have also included the renormalized versions of the latter two, as only the fundamental approach does not require one, since it has it built-in. Here, we have concentrated on the Casimir effects for a real scalar field between two parallel plates, separated by a distance $L$, with Dirichlet boundary condition. The fundamental definition of $F_{\mbox{\scriptsize Casimir}}$ is the difference between the free energy of the system in the presence of nonperturbative conditions or constraints, and the one with no constraints, which we have referred to as the free case, both being at the same temperature $T$ and having the same volume. That is,  $F_{\mbox{\scriptsize Casimir}}(T,L)=F_{\mbox{\scriptsize bounded}}(T,L)-F_{\mbox{\scriptsize free}}(T,L)$. Our two main tools for the computation of $F_{\mbox{\scriptsize bounded}}$ and $F_{\mbox{\scriptsize free}}$ have been the Abel-Plana formula and the Principle of the Argument Theorem. As is well known, both $F_{\mbox{\scriptsize bounded}}(T,L)$ and $F_{\mbox{\scriptsize free}}(T,L)$ have zero temperature divergent parts and finite temperature correction parts which partially cancel upon subtraction, leaving $F_{\mbox{\scriptsize Casimir}}(T,L)$ with both zero and finite temperature parts. We have found that $F_{\mbox{\scriptsize free}}(T,L) = F_{\mbox{\scriptsize free}}(0,L) + \Delta F_{\mbox{\scriptsize free}}(T,L)$ computed using the Abel-Plana formula or the Principle of the Argument Theorem, is precisely extensive, {\it i.e.}, proportional to the volume $V=AL$. In the analytic continuation approach this subtraction is to be rendered by the analytic continuation of $F_{\mbox{\scriptsize bounded}}(T,L)$. In the zero temperature subtraction approach only the zero temperature part of the free energy is subtracted, {\it i.e.}, $F_{\mbox{\scriptsize ZTSA}}(T,L)=F_{\mbox{\scriptsize bounded}}(T,L)- F_{\mbox{\scriptsize free}}(0,L)$. To ensure that the delicate cancellation of infinities has been done correctly within each general approach, we have used or outlined several different methods leading to the same results within each approach throughout the paper.

In Sec.~{\ref{FCasimir}}, we have used the fundamental approach to compute the Casimir thermodynamics quantities for the massless case, including the Casimir free energy, pressure, energy, and entropy. 
In Sec.~{\ref{massive}}, we have used the fundamental approach to compute the Casimir thermodynamics quantities for the massive case, and shown that its results in massless limit coincides with those of the massless case computed in Sec.~{\ref{FCasimir}}. We have shown that, as expected, all of the Casimir thermodynamic quantities go to zero as the mass or $L$ increases. The high temperature limit of $F_{\mbox{\scriptsize bounded}}(T,L)$ contains $T^4$, $T^2$, $T$, $\ln(T)$ terms, out of which only the linear term remains after subtracting $F_{\mbox{\scriptsize free}}(T,L)$.
In Fig.~{\ref{fp60}}, we have displayed the Casimir thermodynamic quantities as a function of $T$, which shows that they do not change sign, and only $S_{\mbox{\scriptsize Casimir}}$ is positive. In the high temperature limit, $E_{\mbox{\scriptsize Casimir}}\rightarrow 0$, $S_{\mbox{\scriptsize Casimir}}\rightarrow \mbox{constant}$, $F_{\mbox{\scriptsize Casimir}}\sim -T$, $P_{\mbox{\scriptsize Casimir}}\sim -T$. The linear temperature dependence of the latter two is attributed to the classical term: What we have shown here explicitly is that this linear $T$-dependence at high $T$ can be related to the behavior of $S_{\mbox{\scriptsize Casimir}}$ and $E_{\mbox{\scriptsize Casimir}}$, in accordance with the relation $F_{\mbox{\scriptsize Casimir}}(T,L)=E_{\mbox{\scriptsize Casimir}}(T,L)-T S_{\mbox{\scriptsize Casimir}}(T,L)$.
These results, and in particular $E_{\mbox{\scriptsize Casimir}}(\infty,L)= 0$, are obtained due to the subtraction of the free case at the same temperature, which amounts to the complete cancellation of both the zero temperature and the thermal correction parts of the bounded case which are equivalent to those of the free case.

We have also computed the Casimir thermodynamic quantities using the analytic continuation approach, represented here by ZFA and SFA, and also ZTSA. In Sec.~{\ref{zeta}}, we have concentrated on the massless case, and in Sec.~{\ref{Epstein}} on the massive case, and shown that the results of the latter in massless limit coincides with those of the massless case computed in Sec.~{\ref{zeta}}. We have shown that, as expected, the results of ZFA and SFA are always equivalent, but they differ from those of the fundamental approach. In particular $F_{\mbox{\scriptsize Zeta}}(T,L)$ contains four extra terms as compared to $F_{\mbox{\scriptsize Casimir}}(T,L)$: The analytic continuation of $F_{\mbox{\scriptsize free}}(0,L)$, $\Delta F_{\mbox{\scriptsize free}}(T,L)$, and two extra $L$-independent terms, the sum of which is equivalent to $F_{\mbox{\scriptsize Zeta}}^{n_1=0}(T)$, which is disallowed by the Dirichlet boundary conditions. It is interesting to note that the free energy of the free case computed using the zeta function, denoted by $F_{\mbox{\scriptsize Zeta}}^{\mbox{\scriptsize free}}(T,L)$, consists of the same four terms mentioned above, the last two of which make this energy nonextensive. This also shows that if we were to use the fundamental definition as a renormalization within the ZFA, we would obtain the correct results:  $F_{\mbox{\scriptsize Zeta}}(T,L)- F_{\mbox{\scriptsize Zeta}}^{\mbox{\scriptsize free}}(T,L)=F_{\mbox{\scriptsize Casimir}}(T,L)$. However, the renormalization programs thus far devised rely on the high temperature expansion, which we describe below. The high temperature expansion of  $F_{\mbox{\scriptsize Zeta}}(T,L)$ includes that of $F_{\mbox{\scriptsize Casimir}}(T,L)$ which is the classical term proportional to $T$, those of $\Delta F_{\mbox{\scriptsize free}}(T,L)$ which consists of $T^4$, $T^2$, and $\ln(T)$ terms, and the two nonextensive terms which consist of $T^3$ and $T\ln(T)$ terms. As mentioned above, it has long been recognized that the zeta function approach produces extra unphysical terms. The renormalization programs that has been devised for this purpose is to find the high temperature expansion of $F_{\mbox{\scriptsize Zeta}}(T,L)$,  using the heat kernel method, and to subtract all terms $T^n$ with $n\geq 2$. As we have shown, the heat kernel method can reproduce the coefficients of all terms in the high temperature expansion, including those of $T\ln(T)$ and $\ln(T)$, except for the linear term which we have obtained by a slight modification. As is apparent, the result of this program, denoted by $F_{\mbox{\scriptsize Zeta}}^{\mbox{\scriptsize ren}}(T,L)$, is not equal to $F_{\mbox{\scriptsize Casimir}}(T,L)$ even at high temperatures, since $T\ln(T)$ and $\ln(T)$ terms remain unsubtracted. Moreover the four extra unphysical terms of $F_{\mbox{\scriptsize Zeta}}^{\mbox{\scriptsize ren}}(T,L)$ mentioned above are nonpolynomial functions of $m$ and $T$ for the massive case, which reduce to a polynomial with $T^4$ and $T^3$ terms in the massless limit. Hence the renormalization program devised works well only in the massless case. 

Finally, the results for the ZTSA, by definition, have only one extra term as compared to the fundamental approach since $	F_{\mbox{\scriptsize ZTSA}}(T,L) =  F_{\mbox{\scriptsize Casimir}}(T,L) + \Delta F_{\mbox{\scriptsize free}}(T,L)$. Once again, $\Delta F_{\mbox{\scriptsize free}}(T,L)$ is a nonpolynomial function of $m$ and $T$, which reduces to the black-body term $T^4$ in the massless limit. Hence the renormalization program works well only in the massless case.
That is, in the massless case we have $F_{\mbox{\scriptsize Casimir} }=F_{\mbox{\scriptsize Zeta}}^{\mbox{\scriptsize ren}}=F_{\mbox{\scriptsize SFA}}^{\mbox{\scriptsize ren}}=F_{\mbox{\scriptsize ZTSA}}^{\mbox{\scriptsize ren}}$. However, as illustrated in Fig.~(\ref{fp21}) for the massive case, the five expressions for the Casimir free energy, {\it i.e.}, $F_{\mbox{\scriptsize Casimir}}$, $F_{\mbox{\scriptsize Zeta}}$ or $F_{\mbox{\scriptsize SFA}}$, $F_{\mbox{\scriptsize ZTSA}}$, $F_{\mbox{\scriptsize Zeta}}^{\mbox{\scriptsize ren}}$ or $F_{\mbox{\scriptsize SFA}}^{\mbox{\scriptsize ren}}$, and $F_{\mbox{\scriptsize ZTSA}}^{\mbox{\scriptsize ren}}$, are not equivalent at any temperature, except at $T=0$, where the ZTSA results are equal to $F_{\mbox{\scriptsize Casimir}}(0,L)$. In particular, as $T\to \infty$, $F_{\mbox{\scriptsize Casimir}}\sim  T$, $F_{\mbox{\scriptsize Zeta}}$, $F_{\mbox{\scriptsize SFA}}$ and $F_{\mbox{\scriptsize ZTSA}}$ $\sim T^4$, and $F_{\mbox{\scriptsize Zeta}}^{\mbox{\scriptsize ren}}$, $F_{\mbox{\scriptsize SFA}}^{\mbox{\scriptsize ren}}$ and $F_{\mbox{\scriptsize ZTSA}}^{\mbox{\scriptsize ren}}$ $\sim T\ln(T/m)$. These differences are also present in the Casimir pressure illustrated in Fig.~(\ref{fp50}). 





\appendix
\numberwithin{equation}{section}

\section{Calculation of the Casimir free energy of the massless and massive cases using the Abel-Plana summation formula}
\label{appendixC:FCasimir}
In the first part of this appendix, we calculate the Casimir free energy for a massless real scalar using its fundamental definition, starting with the second form of the free energy given by Eq.~(\ref{s5}), and show that the final result is equivalent to the result given in Eq.~(\ref{s12}). We first evaluate the integrals over the transverse momenta for both the bounded and free cases using the dimensional regularization, and then subtract the results according to the fundamental definition given by Eq.~(\ref{s6}), to obtain,
\begin{eqnarray}\label{C3}
&&\hspace{-8mm}F_{\mbox{\scriptsize Casimir}} (T,L) =- \frac{A \pi^2}{12 L^3}  \left[ \sum\limits_{n_{1} =  1 }^\infty \left(n_1\right)^3 - \int_{ 0 }^\infty dk' \left(k'\right)^3\right] -A \sqrt{\frac{T^3}{2L^3}} \sum\limits_{j=1}^{\infty} \frac{1}{\sqrt{j^3}} \times  \nonumber \\
&&\hspace{-8mm} \left[ \sum\limits_{n_{1} =  1 }^\infty \left(n_1 \right)^{\frac{3}{2}} K_{\frac{3}{2}} \left( \frac{j \pi n_1}{TL}\right) - \int_{  0 }^\infty  dk' \left(k'\right)^{\frac{3}{2}} K_{\frac{3}{2}} \left( \frac{j \pi k'}{TL}\right)\right],
\end{eqnarray}
where $k'=k \pi /L$. As can be seen in the above expression, the zero temperature parts of the bounded and free cases, given by the two terms in the first square bracket, are separately divergent, since the expression given in Eq.~(\ref{s5}) contains no analytic continuation. Now, using the Abel-Plana formula, given by Eq.~(\ref{abelplana}), the divergences cancel and after simplifying\footnote{Using $\left[(it)^{\frac{3}{2}} K_{\frac{3}{2}} (i t \alpha) - (-it)^{\frac{3}{2}} K_{\frac{3}{2}} (- i t \alpha)\right]= -i \pi t^{\frac{3}{2}} J_{\frac{3}{2}} (t \alpha)$.} we obtain the $F_{\mbox{\scriptsize Casimir}}$ given by Eq.~(\ref{s12}).

We can also obtain another form for the Casimir free energy. First, we expand the logarithm of thermal correction part of the free energy given by Eq.~(\ref{s5}) for large values of $\beta$, then we integrate over the transverse momenta, and finally use the Abel-Plana formula to obtain,
\begin{equation}\label{C4}
F_{\mbox{\scriptsize Casimir}}(T,L) = 
 -  \frac{\pi ^2 A}{1440  L^3}  + \frac{\pi ^2 A L T^4 }{90} - \frac{A T^3}{4\pi } \sum\limits_{j = 1}^\infty  
 \frac{ \coth\left( \frac{\pi j}{2 T L} \right)  + \frac{ \pi j}{2 T L} \csch^2 \left( \frac{\pi j}{2 T L} \right)}{j^3 }  .
\end{equation}
The first term is the zero temperature part and the rest constitute the thermal correction part. This form is equivalent to the result obtained above, {\it i.e.}, Eq.~(\ref{s12}). However, to have an accurate plot using this form, one has keep a large number of terms, otherwise the graph would show an increase relative to the classical term at high values of $T$. This is due to the high $\beta$ expansion mentioned above.

In the last part of this appendix, we use the generalized Abel-Plana summation formula to compute the Casimir free energy for a massive case based on the fundamental definition. As mentioned in Sec.~\ref{massive}, one can consider two different ways of using the Abel-Plana summation formula to calculate the Casimir free energy. In the first case, we only use this formula to calculate the sum over the spatial modes. Hence, for the bounded case, we consider $F_{\mbox{\scriptsize bounded}}(T,L)$ given by Eq.~(\ref{s241}), and for the free case, we start with the first form of the free energy, given by Eq.~(\ref{s004}), and perform the same procedure as the bounded case in Sec.~\ref{massive} and resulted in  Eq.~(\ref{s241}). Then, we express the Casimir free energy based on the fundamental definition as follows
\begin{eqnarray}\label{C5}
&&\hspace{-9mm}F_{\mbox{\scriptsize Casimir}}(T,L) = -\frac{A}{ 4\sqrt{\pi^3}} \lim\limits_{s \to 0} \frac{\partial}{\partial s} \frac{{\mu^{2 s}}}{\Gamma (s)} \Bigg\{\frac{\Gamma \left(s - \frac{3}{2}\right)}{4}\left[\sum\limits_{n_1}^\infty \omega_{n_1}^{3 - 2s} - \int_0^\infty dk' \omega_{k'}^{3 - 2s}\right]+  \nonumber \\
&&\hspace{-9mm}   \sum\limits_{{n_0} = 1}^\infty  \left[\sum\limits_{{n_1} = 1}^\infty \left(\frac{2 \omega_{n_1}}{n_0\beta}\right)^{\frac{3}{2} - s}   K_{\frac{3}{2}-s} \left(\beta {n_0} \omega_{n_1}\right) -\int_0^\infty dk' \left(\frac{2 \omega_{k'}}{n_0\beta}\right)^{\frac{3}{2}-s}   K_{\frac{3}{2}-s} \left(\beta n_0 \omega_{k'}\right)\right]\Bigg\}, \nonumber\\
\end{eqnarray}  
where $\omega_{n_1}=\sqrt{\left(\frac{n_1 \pi}{L}\right)^2+m^2}$ and $\omega_{k'}=\sqrt{\left(\frac{k' \pi}{L}\right)^2+m^2}$.
Next, we calculate the sum over $n_1$ for each bracket of Eq.~(\ref{C5}) using the generalized Abel-Plana formula, as used in Eqs.~(\ref{s10aaa}, \ref{C3}) for the massless case, and obtain
\begin{eqnarray}\label{C6}
&&\hspace{-8mm}F_{\mbox{\scriptsize Casimir}} (T,L) =-\frac{ALm}{ 4\sqrt{\pi^3}} \lim\limits_{s \to 0} \frac{\partial}{\partial s} \frac{{\mu^{2 s}}}{\Gamma(s)} \left[\frac{m^{3-2s}}{ 2\Gamma \left(\frac{5}{2} - s\right)} \int_1^\infty \frac{dt \left(t^2 - 1\right)^{\frac{3}{2} - s}}{ \left(e^{2 m L t} - 1\right)} +\right. \nonumber \\
&&\hspace{-7mm} \left. \sum\limits_{{n_0} = 1}^\infty \left(\frac{2 m}{n_0\beta}\right)^{\frac{3}{2}-s} \int_0^\infty dt  \frac{ t^{\frac{5}{2}-s} J_{\frac{3}{2} -s} \left(  n_0 \beta m t\right) }{\sqrt{t^2+1} \left(e^{2 m L \sqrt{t^2 +1}} - 1\right)} \right].
\end{eqnarray}
After evaluating the integral over $t$ and simplifying, we obtain the Casimir free energy given by Eq.~(\ref{s27}).\\
In the second case, we first calculate the sum over Matsubara modes and then the sum over the remaining spatial modes using the Abel-Plana summation formula for both of them. To do this, we start with the first form of the free energy given by Eq.~(\ref{SZeta}), express the sum over the Matsubara frequencies as the positive integer and zero modes, and then evaluate the sum over $n_0$ for the bounded and free cases using the Abel-Plana summation formula\footnote{Here, we have used the following simple form of the Abel-Plana summation formula to evaluate the sum over $n_0$ modes:

\hspace{1cm}$\sum\limits_{n=0}^\infty f(n) = \int_0^\infty dx f(x) +\frac{1}{2}f(0) + i \int_0^\infty dt\frac{f(it) - f(-it)}{e^{2 \pi t} - 1} $
}. The resulting expression is as follows
\begin{eqnarray}\label{C7}
&&\hspace{-9mm}F_{\mbox{\scriptsize Casimir}} (T,L) =-\frac{TA}{ 4\pi} \lim\limits_{s \to 0} \frac{\partial}{\partial s} \frac{{\mu^{2 s}}}{\Gamma(s)} \Bigg\{ \Gamma \left(s-1\right)  \int_0^\infty dk_0 \left[ \sum\limits_{n_1}^\infty \omega_{n_1, k_0}^{1 - s} - \int_0^\infty dk' \omega_{{k'}, k_0}^{1 - s}\right]+\nonumber \\
&&\hspace{-9mm}  \frac{\beta}{\sqrt{\pi}} \sum\limits_{j = 1}^\infty  \left[\sum\limits_{{n_1} = 1}^\infty \left(\frac{2\omega_{n_1}}{j\beta} \right)^{\frac{3}{2} - s}   K_{\frac{3}{2}-s} \left(\beta j \omega_{n_1}\right) -\int_0^\infty dk' \left(\frac{2\omega_{k', k_0}}{j\beta}\right)^{\frac{3}{2}-s}   K_{\frac{3}{2}-s} \left(\beta j \omega_{k', k_0}\right)\right]\Bigg\}, \nonumber\\
\end{eqnarray}
where $\omega_{n_1, k_0}=\sqrt{\omega_{n_1}^2+\left(k_0 2 \pi T\right)^2}$ and $\omega_{k', k_0}=\sqrt{\omega_{k'}^2+\left(k_0 2 \pi T\right)^2}$. Then, we use the generalized Abel-Plana formula for the expressions in the bracket of Eq.~(\ref{C6}) which contains the sum over the spatial modes, as used in Eq.~(\ref{C5}), and after simplifying obtain the same expression given by Eq.~(\ref{s27}).

%
%

\section{Calculation of the Casimir free energy of the massless and massive cases using the Principle of the Argument theorem}
\label{appendixB:The Argument Principle}

In this appendix we compute the Casimir free energy of a real scalar field, for both the massless and massive cases, using its fundamental definition and utilizing the Principle of the Argument theorem. In particular, we use this theorem to sum over the spatial modes of the free energy. As mentioned in Sec.~\ref{Helmholtz free energy}, this summation includes the regular modes which are the roots of $f(k_{n_1})$ in Eq.~(\ref{s1000}). To show the consistency of our method, we use this theorem for various expressions that we have obtained for the free energy, all yielding equivalent results.

The Principle of the Argument theorem relates the difference between the number of zeros and poles of a meromorphic function $f(z)$, to a contour integral of the logarithmic derivative of the function~\cite{r49Ahlf.}. In this paper, we use the generalized form of the Principle of the Argument theorem which is as follows~\cite{r49Ahlf.}
\begin{equation}\label{BB1}
\sum_{n} g(a_n) - \sum_{m} g(b_m) = \frac{1}{2 \pi i} \oint_C g(z)  d\left[ \ln(f(z)) \right],
\end{equation}
where ${a_n}$ and ${b_m}$ are the zeroes and poles of $f(z)$ inside the closed contour $C$, respectively, and $g(z)$ is assumed to be an analytic function in the region enclosed by the contour $C$. In applying this theorem to our problem, we find it convenient to use the following generalization of Eq.\ (\ref{BB1})~\cite{r26Goushe.}
\begin{equation}\label{BB2}
\sum_{n} g(a_n) - \sum_{m} g(b_m) = \frac{1}{2 \pi i} \oint_C g(z)  d\left[ \ln(f(z)  h(z))\right],
\end{equation}
with the condition that the function $h(z)$ should be analytic and have no zeros in the region enclosed by the contour $C$. 

\begin{figure}[h!] 
\centering
\includegraphics[scale=.5]{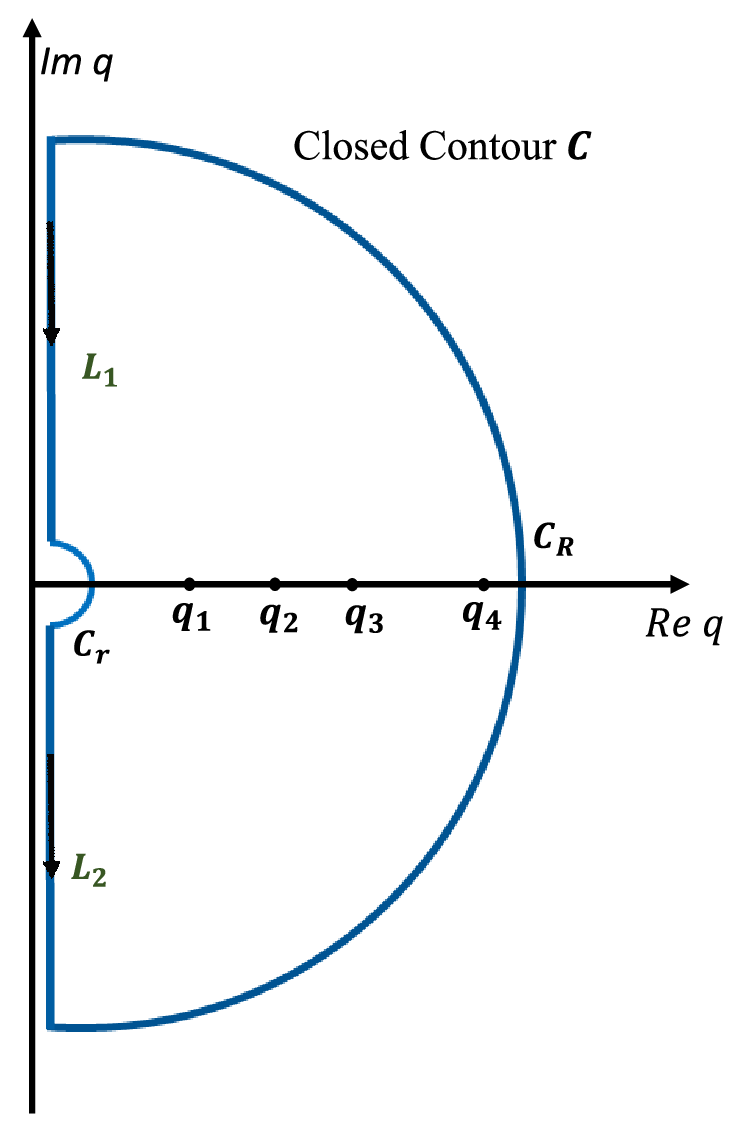}
\caption{\label{fp46a} \small
The closed integration contour $C$ for evaluating the free energy of the massless and massive scalar between two parallel plates, referred to in Eq.~(\ref{BBAA3}) and Eqs.~(\ref{BB3}), respectively. The modes $q_i$ in the massless case are given by $q_i^2 = k_i^2 + 4 \pi^2 n_0^2 T^2$, and in the massive case by $q_i^2 = k_i^2 + m^2$. As $R \to \infty $ and $r \to 0$, only segments $L_1$ and $L_2$ give nonzero contributions.}
\end{figure}

In the first part of this appendix, we compute the Casimir free energy for a massless real scalar field. We use the Principle of the Argument theorem to evaluate the sum over the spatial modes for various forms of the $F_{\mbox{\scriptsize bounded}}(T,L)$ in four different ways. The first three methods are based on the first forms of the free energy, given by Eqs.~(\ref{SZeta}, \ref{s004}), and the fourth is based on the second form given by Eq.~(\ref{s5}). Then, we show that they all yield equivalent results. In the first and second methods within the fundamental approach, we start with the form of the free energy given by Eq.~(\ref{SZeta}) and, after using this theorem for the sum over $n_1$, obtain  
\begin{eqnarray}\label{BBAA3}
F_{\mbox{\scriptsize bounded}}(T,L) &=& -\frac{TA}{8 \pi} \sum\limits_{n_0= - \infty}^{\infty} \frac{1}{2 \pi i}  \oint_{C}   \lim\limits_{s \to 0} \frac{\partial}{\partial s} \frac{\Gamma (s - 1)}{{\mu^{- 2 s}}\Gamma (s)} q^{2 - 2s} \times \nonumber \\
&&d\left\{\ln\left[ \frac{2   }{i  } \sin\left(\sqrt{q^2 - \left(2 \pi n_0 T\right)^2} L \right)\right]\right\},  
\end{eqnarray}
where $g(q_{n_1}^2) = g({k_{n_1}^2} + 4 \pi^2 n_0^2  T^2)$ is the summand in Eq.~(\ref{SZeta}), while $g(q)$ is the integrand defined in Eq.~(\ref{BB2}). We have chosen $h(q) = 2  / i$\footnote{In the fermionic case, $h(q)$ turns out to be nontrivial~\cite{r26Goushe.}.}. The closed contour $C$ in the complex $q$-plane should enclose all of the roots of $f(k_{n_1})$. As can be seen in Fig.~(\ref{fp46a}), the closed contour $C$ is composed of two arcs, $C_{R}$ and  $C_{r}$, and also two straight line segments $L_{1}$, and $L_{2}$. To compute this contour integral over $q$, we replace the $q^{2 - 2s}$ term by the following integral representation
\begin{eqnarray}\label{BBAA4}
q^{2 - 2s} = \int_{0}^{\infty} \frac{e^{- t q^2}dt}{t^{2-s} \Gamma \left(s - 1\right)}.
\end{eqnarray}
Next, we integrate by parts. In the limit $R \to \infty $ and $r \to 0$, only $L_1$ and $L_2$ give nonzero contributions, which can be written as follows 
%
%
\begin{eqnarray}\label{BBAA5}
\hspace{-4mm} F_{\mbox{\scriptsize bounded}}(T,L)   &=& - \frac{i TA}{8\pi^2 } \sum\limits_{n_0 = - \infty}^{\infty} \lim\limits_{s \to 0} \frac{\partial}{\partial s}  \frac{{\mu^{2 s}}}{\Gamma(s)} \int_{- i \infty}^{i \infty}  dq  \int_{0}^{\infty} \frac{dt  e^{- t q^2} q}{t^{1 - s} }   \times \nonumber \\
&&\ln\left[ \frac{2}{i} \sin\left(\sqrt{q^2 - \left(2 \pi n_0 T\right)^2} L \right)\right].  \nonumber \\
\end{eqnarray}
After changing variable $q = ip$, and evaluating the integral over $t$, we obtain
\begin{eqnarray}\label{BBAA6}
F_{\mbox{\scriptsize bounded}}(T,L)   &=&   \frac{TA}{8 \pi^2 } \sum\limits_{n_0 = - \infty}^{\infty} \lim\limits_{s \to 0} \frac{\partial}{\partial s} \frac{{\mu^{2 s}}}{\Gamma(s)}  \int_{0}^{ \infty}  dp \Gamma(s) \left[ \left(i p\right)^{1 - 2s} +  \left(- i p\right)^{1 - 2s} \right] \times    \nonumber\\
&&  \ln\left[ e^{L\sqrt{p^2 + \left(2 \pi n_0 T\right)^2}  } \left(1 - e^{- 2 L \sqrt{p^2 + \left(2 \pi n_0 T\right)^2}}\right)\right] .  
\end{eqnarray}
Then we simplify\footnote{ We use the following identity:    \hspace{.3cm} $$\Gamma(-z) \sin (\pi z) = -\frac{\pi}{\Gamma (z+1)}   \hspace{.5cm} \mbox{for}\hspace{.2cm} z \notin \mathbb{Z}$$ } and obtain,
\begin{eqnarray}\label{BBAA6v}
F_{\mbox{\scriptsize bounded}}(T,L) &=&   \frac{TAL}{4 \pi} \sum\limits_{n_0 =- \infty}^{\infty} \lim\limits_{s \to 0} \frac{\partial }{\partial s}  \frac{{\mu^{2 s}}}{\Gamma(s)\Gamma\left(1 - s\right) } \int_{0}^\infty dp \Bigg\lbrace p^{1 - 2s} \omega_{n_0} (p) - \nonumber \\
&& \frac{p^{3 - 2s}}{(1-s) \omega_{n_0} (p)}\frac{1}{e^{ 2 L \omega_{n_0}(p)} - 1} \Bigg\rbrace ,
\end{eqnarray}
where $\omega_{n_0} (p) = \sqrt{p^2 +(2 \pi n_0 T)^2}$. Next, we use the Poisson summation formula to evaluate the sum over temperature modes, and after simplifying we evaluate $\lim \limits_ {s \to 0} \partial / \partial s$ for all terms except for the divergent integral term which appear in the zero temperature part and obtain
\begin{eqnarray}\label{BBAA6v2}
\hspace{-8mm}F_{\mbox{\scriptsize bounded}}(0,L) &=& -\frac{ A L}{16\sqrt{ \pi^5}} \lim\limits_{s \to 0}   \frac{\partial }{\partial s}\frac{\Gamma \left(s- \frac{3}{2}\right)}{{\mu^{- 2 s}} \Gamma (s)} \int_0^\infty dp p^{3-2s} - \frac{A \pi^2}{1440 L^3} \nonumber\\
\hspace{-8mm}\Delta F_{\mbox{\scriptsize bounded}}(T,L) &=&-\frac{A L \pi^2}{90} T^4 - \frac{2 A L T^4 }{\pi^2} \sum\limits_{n_0=1}^{\infty} \sum\limits_{j=1}^\infty \frac{1}{\left[n_0^2 + \left(2 j T L\right)^2\right]^2}.
\end{eqnarray}
To use the fundamental definition, we calculate the contribution of the free case by starting with the same form of the free energy, given by Eq.~(\ref{SZeta}), expressed as follows
\begin{eqnarray}\label{BBAA6v3}
F_{\mbox{\scriptsize free}}(T,L)  = - \frac{TAL}{8 \pi^2 } \sum\limits_{n_0 = - \infty}^{\infty} \lim\limits_{s \to 0} \frac{\partial}{\partial s} \frac{\Gamma (s-1)}{{\mu^{-2 s}}\Gamma(s)}  \int_{0}^{ \infty}  dk  \left[ k^2 + \left(2 \pi n_0 T\right)^2\right]^{1 - s} .
\end{eqnarray}
we evaluate the sum over $n_0$, similarly to the bounded case, using the Poisson summation formula and obtain             
\begin{eqnarray}\label{BBAA6v4}
\hspace{-8mm}F_{\mbox{\scriptsize free}}(0,L) &=& -\frac{ A L}{16\sqrt{ \pi^5}} \lim\limits_{s \to 0}   \frac{\partial }{\partial s}\frac{\Gamma \left(s- \frac{3}{2}\right)}{{\mu^{-2 s}}\Gamma (s)} \int_0^\infty dk k^{3-2s}  \nonumber\\
\hspace{-8mm}\Delta F_{\mbox{\scriptsize free}}(T,L) &=&-\frac{A L \pi^2}{90} T^4 .
\end{eqnarray}
As can be seen from Eq.~(\ref{BBAA6v4}), the zero and finite temperature correction terms of the free case is identical to the first terms of the bounded case given in Eq.~(\ref{BBAA6v2}) which after subtracting, these terms completely cancel and we obtain the same expression for the Casimir free energy as given by Eq.~(\ref{s11}).

In the second method, we rewrite the sum over the Matsubara frequencies in Eqs.~(\ref{BBAA6v}, \ref{BBAA6v3}) for the bounded and free cases in terms of the positive integers and zero, and then evaluate this sum using the same form of the Abel-Plana formula which is used for Eq.~(\ref{C7}). After subtracting the free energy of the bounded from the free case, we obtain the same Casimir free energy as given by Eq~(\ref{s12}).
%
%

For the third method, we start with the expression for $F_{\mbox{\scriptsize bounded}}(T,L)$ given in Eq.~(\ref{s8}), evaluate the sum over the Matsubara frequencies using the Poisson summation formula, and use the Principle of the Argument theorem to sum over the spatial modes to obtain
\begin{eqnarray}\label{BBAA7}
&& \hspace{-4mm}F_{\mbox{\scriptsize bounded}}(T,L) = -\frac{A}{2 \sqrt{\pi^3}} \frac{1}{2 \pi i} \oint_{C} d\left\{\ln\left[ \frac{2   }{i  } \sin\left( k_{n_1} L \right)\right]\right\} \times\nonumber \\
&&\Bigg\{ \lim\limits_{s \to 0} \frac{\partial}{\partial s} \frac{\Gamma \left(s - \frac{3}{2}\right)}{{\mu^{-2 s}}8\Gamma (s)}   k_{n_1}^{3 - 2s}+\sqrt{2T^3} \sum\limits_{n_0=1}^{\infty}  \sqrt{\frac{k_{n_1}^3}{n_0^3}} K_{\frac{3}{2}}\left( n_0 \beta k_{n_1}\right)\Bigg\} .
\end{eqnarray}
We can now follow the same steps as above, use the same contour shown in Fig.~(\ref{fp46a}) for the variable $q_{n_1}^2=k_{n_1}^2$, and simplify\footnote{Using $\sqrt{(i p)^3} K_{\frac{1}{2}} \left(i p a \right) +\sqrt{(- i p)^3} K_{\frac{1}{2}} \left(- i p a \right) = \pi \sqrt{p^3} J_{\frac{1}{2}} \left(p a \right) $.} the results to obtain
\begin{eqnarray}\label{BBAA8}
&& \hspace{-4mm}F_{\mbox{\scriptsize bounded}}(T,L) = -\frac{A}{2 \sqrt{\pi^3}} \int_{0}^\infty dp \left[Lp + \ln\left( 1 - e^{- 2 p L}\right)\right] \times\nonumber \\
&&\Bigg\{ \lim\limits_{s \to 0} \frac{\partial}{\partial s} \frac{\Gamma \left(s - \frac{1}{2}\right)}{{\mu^{-2 s}}4\pi\Gamma (s)}   p^{2 - 2s}- \sum\limits_{n_0=1}^{\infty} \sqrt{\frac{T p^3}{2 n_0}} J_{\frac{1}{2}}\left( n_0 \beta p\right)\Bigg\} .
\end{eqnarray}
Only one of the four terms in the integrand, after carrying out the multiplication, is divergent. For the other three terms, evaluating the integral over $p$ and $\lim\limits_{s \to 0} \frac{\partial}{\partial s}$, where applicable, we obtain
\begin{eqnarray}\label{BBAA9}
\hspace{-4mm}F_{\mbox{\scriptsize bounded}}(T,L) &=& -\frac{AL}{8 \sqrt{\pi^5}} \lim\limits_{s \to 0} \frac{\partial}{\partial s} \frac{\Gamma \left(s - \frac{1}{2}\right)}{{\mu^{- 2 s}}\Gamma (s)} \int_0^\infty dp  p^{3 - 2s} +\nonumber \\
&& F_{\mbox{\scriptsize Casimir}}(T,L) +\Delta F_{\mbox{\scriptsize free}}(T,L),
\end{eqnarray}
where $F_{\mbox{\scriptsize Casimir}}(T,L)$ and $\Delta F_{\mbox{\scriptsize free}}(T,L)$ are given by Eqs.~(\ref{s11}, \ref{s10aa}). Next, we calculate $F_{\mbox{\scriptsize free}}(T,L))$ analogously to the bounded case, by starting with Eq.~(\ref{s4}) for the massless case and computing the sum over $n_0$ using the Poisson summation formula. After simplifying, the free energy of the free case includes the zero and finite temperature correction parts which are exactly equal to the first and last term of the free energy of the bounded case, given by Eq.~(\ref{BBAA9}). This shows that Eq.~(\ref{BBAA9}) can be rewritten as  $F_{\mbox{\scriptsize bounded}}(T,L)=F_{\mbox{\scriptsize Casimir}}(T,L)+ F_{\mbox{\scriptsize free}}(T,L)$, as expected.

In the fourth method within the fundamental approach, we start with the second form of the free energy, given by Eq.~(\ref{s5}), compute the sum over the spatial modes using the Principle of the Argument theorem, and obtain
\begin{eqnarray}\label{BBAA11}
 \hspace{-4mm}F_{\mbox{\scriptsize bounded}}(T,L) &=& \frac{A}{4\pi} \int_0^\infty dK_T K_T \frac{1}{2 \pi i} \oint_{C} d\left\{\ln\left[ \frac{2   }{i  } \sin\left( \sqrt{q^2 - K_T^2} L \right)\right]\right\}\nonumber \\
 && \Bigg\{q + 2 T \ln \left(1 - e^{- \beta q}\right) \Bigg\},
\end{eqnarray} 
where $q_{n_1}^2 = K_T^2 + k_{n_1}^2$. After following the same steps as above, we obtain
\begin{eqnarray}\label{BBAA12}
 \hspace{-4mm}F_{\mbox{\scriptsize bounded}}(T,L) &=& \frac{A}{4\pi^2} \int_0^\infty dK_T K_T \int_{0}^\infty dp\left[L \omega_{K_T} (p)+ \ln \left(1 - e^{-2L \omega_{K_T} (p) }\right)\right] \nonumber \\
 &&\left[1+2\sum\limits_{j =1}^\infty \cos \left(\beta j p\right)\right],
\end{eqnarray}
where $\omega_{K_T} (p)=\sqrt{p^2+ K_T^2} $. Then, we evaluate the integrals over $K_T$ and $p$ for all terms, except for the first one resulting from the multiplication of the brackets of Eq.~(\ref{BBAA12}), which contains a divergent integral. Next, we calculate $F_{\mbox{\scriptsize free}}(T,L)$ by starting with the same second form of the free energy as has been used for the bounded case, and obtain
\begin{eqnarray}\label{BBAA11v}
F_{\mbox{\scriptsize free}}(T,L) = \frac{A}{4\pi} \int_0^\infty dK_T K_T \int_0^\infty \frac{dk L}{\pi} \left[\omega_{K_T}(k) +2 T \ln \left(1 - e^{- \beta \omega_{K_T}(k)}\right)\right],
\end{eqnarray}
where $\omega_{K_T}(k)=\sqrt{k_T^2 + k^2}$. As can be seen the first terms of Eqs.~(\ref{BBAA12}) and (\ref{BBAA11v}) are divergent and identical. After subtracting these two expression, according to  Eq.~(\ref{s6}), we obtain the same expression for the $F_{\mbox{\scriptsize Casimir}}(T,L)$ as in Eq.~(\ref{s11}).  
%
%

In the last part of this appendix, we use this theorem for the computations related to the massive case, as stated in Sec.~\ref{massive}. We start with Eq.~(\ref{s241}) which contains a sum over the regular spatial modes which are the roots of $f(k_{n_1})$ in Eq.~(\ref{s1000}). We use the Principle of the Argument theorem, as expressed in Eq.~(\ref{BB2}), to compute this sum and obtain  
\begin{eqnarray}\label{BB3}
&&\hspace{-8mm}F_{\mbox{\scriptsize bounded}}(T,L)   = -\frac{A}{4\sqrt{\pi^3}} \frac{1}{2 \pi i}  \oint_{C}   \lim\limits_{s \to 0} \frac{\partial}{\partial s} \frac{{\mu^{2 s}}}{\Gamma (s)} \left[\frac{\Gamma\left( s - \frac{3}{2}\right)}{4} q^{3 - 2s}
+  \right.\nonumber \\
&&\hspace{-6mm} \left. \sum\limits_{{n_0} = 1}^\infty   \left(\frac{2 q}{ n_0 \beta}\right)^{\frac{3}{2} - s} 
 K_{\frac{3}{2} - s}\left(\beta {n_0} q\right)  \right]   d\left\{\ln\left[ \frac{2}{i}   \sin\left(\sqrt{q^2 - m^2} L \right)\right]\right\},  
\end{eqnarray}
where $g(q_{n_1}^2) = g({k_{n_1}^2} + m^2)$ is the summand in Eq.~(\ref{SZeta}), while $g(q)$ is the integrand defined in Eq.~(\ref{BB2}). We have also chosen $h(q) = 2  / i$ for the massive case, and consider the same closed contour C in the complex $q$-plane as shown in Fig.~(\ref{fp46a}). After integrating by parts and taking the limit $R \to \infty$ and $r \to 0$, Eq.~(\ref{BB3}) becomes
\begin{eqnarray}\label{BB4}
\hspace{-4mm} F_{\mbox{\scriptsize bounded}}(T,L)   &=& - \frac{i A}{4\sqrt{\pi^5} }\lim\limits_{s \to 0} \frac{\partial}{\partial s}   \frac{{\mu^{2 s}}}{\Gamma(s)}\int_{- i \infty}^{i \infty}  dq  \left[ \int_{0}^{\infty} \frac{dt  e^{- t q^2} q}{4 \sqrt{t^{3-2s}}  }   
+  \sum\limits_{{n_0} = 1}^\infty   \left(\frac{2 }{n_0}\right)^{\frac{1}{2} - s}  \times   \right.  \nonumber\\
&&\hspace{-2mm}  \left. T^{\frac{1}{2} - s}q^{\frac{3}{2} - s} K_{\frac{1}{2} - s}\left(\beta {n_0} q\right) \right] \left\{\ln\left[ \frac{2}{i} \sin\left(\sqrt{q^2 - m^2} L \right)\right]\right\}.  
\end{eqnarray}
After changing variable $q = ip$, and evaluating the integral over $t$, we express the results as follows
\begin{eqnarray}\label{BB5}
&&\hspace{-8mm}F_{\mbox{\scriptsize bounded}}(T,L)   =   \frac{A}{4\sqrt{ \pi^5} } \lim\limits_{s \to 0} \frac{\partial}{\partial s} \frac{{\mu^{2 s}}}{\Gamma (s)}  \int_{0}^{ \infty}  dp  \Bigg\{  \frac{\Gamma \left( s - \frac{1}{2} \right)}{4}\left[ \left(i p\right)^{2 - 2s} +  \left(- i p\right)^{2 - 2s} \right]+    \nonumber\\
&&\hspace{-6mm} \sum\limits_{{n_0} = 1}^\infty  \left(\frac{2T}{n_0}\right)^{\frac{1}{2} - s} \left[ \left(i p\right)^{ \frac{3}{2} - s} K_{ \frac{1}{2} - s} \left( i p \beta n_0\right) + \left(- i p\right)^{ \frac{3}{2} - s} K_{ \frac{1}{2} - s} \left(- i p \beta n_0\right) \right] \Bigg\}  \times \nonumber\\
&&\hspace{-6mm} \left\{ \ln\left[ e^{L \sqrt{p^2 + m^2} } \left(1 - e^{- 2 L \sqrt{p^2 + m^2}}\right)\right] \right\}.  
\end{eqnarray} 
We finally can simplify this expression, as done in Eq.~(\ref{BBAA7}), to obtain the free energy for the bounded case given by Eq.~(\ref{s1003}).


\section{Calculation of the free energy using the generalized zeta function}
\label{appendixA:The zeta function}

The most commonly-used approach for calculating the Casimir effects is the zeta function approach (ZFA). The generalized zeta function~\cite{r47Elizald.} is given dy the following expression, 
\begin{equation}\label{A1}
Z_p^{M^2}(s ; a_1 ,..., a_p ; c_1 ,..., c_p) = \sum_{n_{1} =  - \infty }^\infty  ...\sum_{n_{p} =  - \infty }^\infty  	{\left[ a_{1} {\left(n_{1} - c_{1}\right)}^2 + ... + a_{p} {\left(n_{p} - c_{p}\right)}^2  + M^2\right]^{ - s}}.
\end{equation}
The above expression yields finite results for $\mathrm{Re} (s) > \frac{p}{2}$, and admits an analytic continuation for $\mathrm{Re} (s) < \frac{p}{2}$,~\cite{r47Elizald., r10Wolf.}. This form is also referred to as the inhomogeneous generalized zeta function. If we set the parameters ${c_1} ,..., {c_p}$ to zero, we obtain a special form of the inhomogeneous generalized zeta function. An important special form called the homogeneous zeta function is obtained when the parameters ${c_1} ,..., {c_p}$, and $M$ are set to zero.  For this case, there is a constraint that the sums should not include the $({n_1}=0, ... ,{n_p}=0)$ mode. Obviously, for the massive case we have to use the inhomogeneous form, while, as shown in the text, both forms can be used for the massless case. 

%
%
%
In the first part of this appendix, we show explicitly four different ways of using the zeta function for calculating the free energy of the massless case, as outlined in Sec.~\ref{zeta}, obtaining two equivalent expressions summarized in the form given by Eq.~(\ref{s23}) with two different forms for $F_{\mbox{\scriptsize Casimir}} (T,L)$ given by Eqs.~(\ref{s12}, \ref{C4}). In the first method, we do the double sum simultaneously, so as to obtain the final result shown in Eq.~(\ref{s23}). The sums in the expression that we have obtained for $F(T,L)$, given by Eq.~(\ref{s21}), are over only positive definite integers, so we use the homogeneous form of the generalized inhomogeneous Epstein zeta function~\cite{r48Kris2.}, given by 
\begin{equation}\label{A4}
{E_p^{M^2}}(s ; {a_1} ,..., {a_p}) = \sum_{n_{1} =  1 }^\infty  {...\sum_{n_{p} = 1}^\infty  {\left[ a_1 {n_1}^2 + a_2 {n_2}^2 + ... + a_p {n_p}^2  + M^2\right]}^{ - s}}     .
\end{equation}
That is, we use $E_p^{0}$, which is usually denoted by $E_p$, and express $F_{\mbox{\scriptsize Zeta}} (T,L)$ as follows
\begin{eqnarray}\label{A3a}
F_{\mbox{\scriptsize Zeta}} (T,L) =- \frac{T A}{8 \pi }  \lim\limits_{s \to 0} \frac{\partial }{\partial s} \frac{\Gamma (s - 1)}{{\mu^{-2 s}}\Gamma (s)} 
\left\lbrace  E_{1} \left( s - 1 ;  \frac{\pi^2}{L^2}  \right)  + 2   E_{2} \left( s - 1 ; \frac{4\pi^2}{\beta^2} , \frac{\pi^2}{ L^2} \right)      \right\rbrace . \nonumber \\
\end{eqnarray}
To compute the second part of Eq.~(\ref{A3a}), we use the following relation for the Epstein zeta function, $E_2$,  
\begin{eqnarray}\label{A6}
E_2(s;{a_1},{a_2}) =   - \frac{\zeta (2s)}{2 {a_1}^{s}} + \sqrt {\frac{\pi }{{a_2}}}  
\frac{\Gamma (s - \frac{1}{2}) \zeta (2s - 1)}{2  \Gamma (s)  {a_1}^{\left(s - \frac{1}{2} \right)}} + \nonumber \\ 
\frac{2 \pi^s}{\sqrt{{a_2}^{\left( s + \frac{1}{2} \right)}} \Gamma (s) \sqrt{{a_1}^{\left( s - \frac{1}{2} \right)}}}   \sum\limits_{{m_1} = 1}^\infty  {\sum\limits_{{m_2} = 1}^\infty  {\left( \frac{{m_2}}{{m_1}} \right)}^{\left(s - \frac{1}{2}\right)} K_{ \frac{1}{2} - s} \left( 2 \pi {m_1}{m_2} \sqrt{\frac{{a_1}}{{a_2}}} \right)}.
\end{eqnarray}
Now we set $a_1=\frac{4\pi^2}{\beta^2}$ and $a_2=\frac{\pi^2}{ L^2}$ to obtain
\begin{eqnarray}\label{A6b} 
&&\hspace{-9mm}F_{\mbox{\scriptsize Zeta}} (T,L) =- \frac{T A}{8 \pi }  \lim\limits_{s \to 0} \frac{\partial }{\partial s} \frac{{\mu^{2 s}}}{\Gamma (s)}  \left\lbrace \left[\left(\frac{\pi}{L}\right)^{2 - 2s} - \left(\frac{2\pi}{\beta}\right)^{2 - 2s}\right] \Gamma (s-1) \zeta (2s -2) +  \frac{L}{\sqrt{\pi}}\times\right.\nonumber\\
&&\hspace{-9mm} \left. \frac{\left(2\pi\right)^{3 - 2s} \zeta (2s -3)}{\beta^{3 - 2s} } \Gamma \left(s - \frac{3}{2} \right)  +\frac{8\left(2\right)^{\frac{1 - 2s}{2}}}{\left( L \beta^3\right)^{\frac{1 - 2s}{2}}\pi^{s -1}}  \sum\limits_{{n_0} = 1}^{\infty} \sum\limits_{{n_1} = 1}^{\infty}  \left(\frac{n_0}{n_1}\right)^{\frac{3}{2} - s} K_{\frac{3}{2} - s} \left(4 \pi n_0 {n_1} L T\right) \right\rbrace .\nonumber\\
\end{eqnarray} 
Now, an analytic continuation may be implemented by the application of the following zeta function reflection formula~\cite{r12Kris., r48Elizald.,  r482Elizald.},
\begin{equation}\label{A323}
\pi ^{- \frac{s'}{2}}   \Gamma \left(\frac{s'}{2} \right) \zeta \left(s'\right) =  
\pi ^{\frac{s' -1}{2}}   \Gamma \left(\frac{1 - s' }{2}\right) \zeta \left(1 - s'\right).
\end{equation}
Using the reflection formula for the first two terms of Eq.~(\ref{A6b}), evaluating $\lim\limits_{s \to 0} \frac{\partial}{\partial s}$, the expression for the free energy becomes
\begin{eqnarray}\label{A3c} 
F_{\mbox{\scriptsize Zeta}} (T,L)&=& -\frac{A T}{16 \pi L^2} \zeta (3) + \frac{A T^3}{4 \pi} \zeta (3) - \frac{A L T^4}{ \pi^2 } \zeta (4) - \nonumber\\ 
&&A \sqrt{\frac{2 T^5}{L}}\sum\limits_{{n_0} =  1 }^\infty \sum\limits_{{n_1} =  1 }^\infty \left(\frac{n_0 }{n_1}\right)^{\frac{3}{2}} K_{\frac{3}{2}} \left( 4 \pi n_0 n_1 T L \right) .
\end{eqnarray}
After computing the sum over $n_0$\footnote{We have used the following identities,\\ $\sum\limits_{m = 1}^\infty  \sqrt {m^3}  K_{\frac{3}{2}}(ma) = 	\sqrt {\frac{\pi }{ 2 a^3}} \frac{\left(a + 1\right)  e^a - 1}{{\left( e^a - 1 \right)}^2} = \sum\limits_{m = 1}^\infty  \sqrt {m^3}  K_{ - \frac{3}{2}}(ma)$.} and simplifying the expression, we obtain our final result given by Eq.~(\ref{s23}), where the expression for $F_{\mbox{\scriptsize Casimir}}$ is given by Eq.~(\ref{s12}). We have checked that, as expected, reversing the labels, i.e., $a_2=\frac{4\pi^2}{\beta^2}$, $a_1=\frac{\pi^2}{ L^2}$ and $m_1 \leftrightarrow m_2$, does not alter the results.

As mentioned in Sec.~\ref{zeta}, this final result, given by Eq.~(\ref{s23}), includes an extra L-independent thermal correction term which is related to the finite value of the free energy of the zero spatial mode. This is in spite of the fact that the expressions for free energy in our problem, starting with Eq.~(\ref{s04}), do not contain the zero spatial mode. To illustrate this point, we first start with the first form of the free energy given in Eq.~(\ref{SZeta}) for the case of $n_1=0$, and after simplifying we obtain
\begin{eqnarray}\label{A3ccc} 
F_{\mbox{\scriptsize bounded}}^{n_1=0} (T)= -\frac{A T}{4 \pi} \sum\limits_{{n_0} =  1 }^\infty \lim\limits_{s \to 0} \frac{\partial}{\partial s} \frac{\Gamma (s - 1)}{{\mu^{-2 s}}\Gamma (s)} \left(\frac{2 n_0 \pi}{\beta}\right)^{2 - 2s}.
\end{eqnarray}
Next, we use the reflection formula of the zeta function, given by Eq.~(\ref{A323}), and after computing the sum over $n_0$, we obtain
\begin{eqnarray}\label{A3cccd} 
F_{\mbox{\scriptsize Zeta}}^{n_1=0} (T)= -\frac{A \zeta(3)}{2 \pi} T^3 .
\end{eqnarray}
Comparing the above result with the last term of $F_{\mbox{\scriptsize Zeta}}(T,L)$ given by Eq.~(\ref{s23}), we observe that the extra term of $F_{\mbox{\scriptsize Zeta}}$ is equal to minus one half of the contribution of the zero spatial mode. 
      
Next we show that, as mentioned in Sec.~\ref{zeta}, the final results are independent of the order of calculation of the summations. To do this, we again start with the expression for $F_{\mbox{\scriptsize Casimir}} (T,L)$, given by Eq.~(\ref{s21}), and set $a_1=\frac{\pi^2}{L^2}$ and $a_2=\frac{4\pi^2}{\beta^2}$ to get an alternative expression for  Eq.~(\ref{A3a}). After simplifying the result, using the reflection formula given by Eq.~(\ref{A323}), evaluating $\lim\limits_{s \to 0} \frac{\partial}{\partial s}$, the expression for the free energy becomes
\begin{eqnarray}\label{A44c} 
F_{\mbox{\scriptsize Zeta}} (T,L)= -\frac{A }{16 \pi^2 L^3} \zeta (4) - A \sqrt{\frac{T^3}{2 L^3}} \sum\limits_{{n_0} =  1 }^\infty \sum\limits_{{n_1} =  1 }^\infty  \left(\frac{n_1 }{n_0}\right)^{\frac{3}{2}} K_{\frac{3}{2}} \left( \frac{ \pi n_0 n_1}{ T L} \right) . 
\end{eqnarray}
By calculating the sum over $n_1$ modes first, we obtain
\begin{eqnarray}\label{A45c} 
F_{\mbox{\scriptsize Zeta}} (T,L)&=& -\frac{A \pi^2}{1440 L^3} +  \frac{A T^3}{4\pi} \zeta (3) - \nonumber\\
&&\frac{A T^3}{4 \pi} \sum\limits_{{n_0} =  1 }^\infty  \left[ \coth \left(\frac{n_0 \pi}{2 T L}\right) + \frac{n_0 \pi}{2 T L}  \csch^2 \left( \frac{ \pi n_0}{2 T L} \right)\right] . 
\end{eqnarray} 
The above result is equivalent to the form given by Eq.~(\ref{s23}) for $F_{\mbox{\scriptsize Zeta}} (T,L)$ with $F_{\mbox{\scriptsize Casimir}} (T,L)$ given by Eq.~(\ref{C4}).

Next, we compute the free energy of the massless case using the inhomogeneous Epstein zeta function to do the sums separately. To do this, we use the free energy given by Eq.~(\ref{s21}) which includes the sums over positive integers.

For our first case, which constitutes our second method, we first calculate the sum over Matsubara modes, and then the sum over the remaining spatial modes. To do this, we consider the spatial modes, {\it i.e.}, $n_1^2 \pi^2 /L^2$, as the constant term of Eq.~(\ref{A4}). Then, we use the following expression for $E_1^{M^2}(s ; a)$ ~\cite{r48Kris2.} 
\begin{eqnarray}\label{A8}
E_1^{M^2}(s ; a) &=& -  \frac{1}{2 M^{2s}}  +
\sqrt {\frac{\pi }{a}}   \frac{1}{2 \Gamma{(s)}  M^{2s - 1}}
\Bigg[ \Gamma{(s - \frac{1}{2})} +  \nonumber \\
&&\left. 4 \sum_{j = 1}^\infty  {\left( \frac{\sqrt{ a}}{\pi j M} \right)}^{\left( \frac{1}{2} - s\right)}  K_{\frac{1}{2} - s}{\left(\frac{2 \pi j M}{\sqrt{a}}\right)}\right],
\end{eqnarray}
to compute the free energy. Using this expression in  Eq.~(\ref{s21}), we obtain,
\begin{eqnarray}\label{A9}
F_{\mbox{\scriptsize Zeta}} (T,L) &=&-\frac{A }{8 \pi }  \lim\limits_{s \to 0} \frac{\partial }{\partial s} \frac{{\mu^{2 s}}}{\Gamma (s)}  \sum\limits_{{n_1} = 1 }^\infty \Bigg\{\frac{ \Gamma \left(s -\frac{3}{2}\right)}{2\sqrt{\pi}}\left( \frac{{n_1} \pi}{L} \right)^{3 - 2 s} + \nonumber \\
&&  2 \pi^{1-s} \sum\limits_{n_0 = 1 }^\infty  \left( \frac{2{n_1} }{n_0 \beta L} \right)^{\frac{3}{2} - s} K_{\frac{3}{2} - s} \left( \frac{n_0 n_1 \pi}{TL} \right)  \Bigg\} .
\end{eqnarray}
Evaluating $\lim\limits_{s \to 0} \frac{\partial}{\partial s}$ for the second term, the expression for the free energy becomes,
\begin{eqnarray}\label{A9ddd}
F_{\mbox{\scriptsize Zeta}} (T,L) &=&-\frac{A }{16 \sqrt{\pi^3} }  \lim\limits_{s \to 0} \frac{\partial }{\partial s} \frac{\Gamma \left(s -\frac{3}{2}\right)}{{\mu^{-2 s}}\Gamma (s)}  \sum\limits_{{n_1} = 1 }^\infty \left( \frac{{n_1} \pi}{L} \right)^{3 - 2 s} + \nonumber \\
&&  \frac{A}{\sqrt{2}} \sum\limits_{{n_1} = 1 }^\infty \sum\limits_{n_0 = 1 }^\infty  \left( \frac{n_1 T }{n_0 L} \right)^{\frac{3}{2}} K_{\frac{3}{2} } \left( \frac{n_0 n_1 \pi}{TL} \right)   .
\end{eqnarray}
This expression is equivalent to Eq.~(\ref{s21b}) and, as mentioned in Sec.~\ref{zeta}, we calculate the divergent sum over the spatial modes using the analytic continuation embedded in $\zeta(-3)$. The final result is displayed in Eq.~(\ref{s21bgd}).

For our second case, which constitutes our third method, we first calculate the sum over spatial modes and then the sum over the remaining Matsubara frequencies. To do this, we can start with the expression for free energy given by Eq.~(\ref{s21}) and consider the Matsubara modes, {\it i.e.}, $4 n_0^2 \pi^2 /\beta^2$, as a constant term of Eq.~(\ref{A4}). However, we prefer to backtrack and start with Eq.~(\ref{SZeta}), in which the sum over Matsubara frequencies is not broken to two pieces. We have checked that the final results are the same either way. Then, we use Eq.~(\ref{A8}) to compute the free energy, obtaining the following expression,
\begin{eqnarray}\label{A11}
&&\hspace{-2mm}F_{\mbox{\scriptsize Zeta}} (T,L)= -\frac{ AT}{16 \pi}  \lim\limits_{s \to 0} \frac{\partial }{\partial s} \frac{\Gamma(s-1)}{{\mu^{-2 s}} \Gamma(s)} \sum\limits_{n_{0} = -\infty}^{\infty } \left[- \left(  \frac{ 4 n_0^2 \pi^2 }{\beta^2 }\right)^{1 - s} + \frac{\Gamma\left(s - \frac{3}{2} \right)}{\Gamma (s-1)}\times\right.    \nonumber\\ 
&&\left. \frac{L}{\sqrt{\pi}} \left(  \frac{ 4 n_0^2 \pi^2 }{\beta^2 }\right)^{\frac{3}{2} - s} + \sum\limits_{j = 1}^{\infty }\frac{4L \left(  \frac{ 4 n_0^2 \pi^2 }{\beta^2 j^2 L^2}\right)^{\frac{3 - 2s}{4} }}{\sqrt{\pi} \Gamma (s-1)}  K_{\frac{3 - 2s}{2}} \left( 2 L j \sqrt{\frac{4\pi^2n_0^2}{\beta^2}}\right)\right] .
\end{eqnarray}
The two first terms in the above expression are divergent and can be written as the homogeneous zeta function. For the last term, we evaluate $\lim\limits_{s \to 0} \frac{\partial}{\partial s}$, and then express the sum over temperature modes in terms of positive integes and a zero mode, which gives a nonzero contribution\footnote{We have used the following expansion: \hspace{1cm} $\lim\limits_{n \to 0} \left(\sqrt{n^3} K_{\frac{3}{2}} (n a)\right) = \sqrt{\frac{\pi}{2 a^3}} + O(n^2)$}. We can now express the result as follows
\begin{eqnarray}\label{A11dd}
&&\hspace{-8mm}F_{\mbox{\scriptsize Zeta}} (T,L)= \frac{ AT}{16 \pi}  \lim\limits_{s \to 0} \frac{\partial }{\partial s} \frac{{\mu^{2 s}}}{\Gamma(s)} \left[\Gamma(s-1) Z_{1} \left( s - 1 ;  \frac{4 \pi^2}{\beta^2}  \right) - \frac{L\Gamma\left(s - \frac{3}{2} \right)}{\sqrt{\pi}} \times\right.    \nonumber\\ 
&&\hspace{-8mm}\left. Z_{1} \left( s - \frac{3}{2} ;  \frac{4 \pi^2}{\beta^2}  \right)\right] - A \sum\limits_{n_{0} = 1}^{\infty } \sum\limits_{j = 1}^{\infty } \sqrt{\frac{2 T^5 n_0^3 }{ j^3 L}}  K_{\frac{3 }{2}} \left(4 \pi j n_0 T L\right)+\frac{A T }{16 \pi L^2} \sum\limits_{j = 1}^{\infty }\frac{1}{j^3},\nonumber\\
\end{eqnarray}
where the last term in Eq.~(\ref{A11dd}) is the contribution of the zero mode of the third term of Eq.~(\ref{A11}). For the first two terms of Eq.~(\ref{A11dd}), an analytic continuation may be implemented by the application of the following zeta function reflection formula~\cite{r12Kris., r48Elizald.,  r482Elizald.},
\begin{equation}\label{A3}
\pi ^{- s'}   \Gamma (s') Z_p \left(s'  ;  {a_1},...,{a_p}\right) = 
\frac{ \pi ^{- \frac{p}{2} + s'} }{\sqrt{{a_1}{a_2}...{a_p}}}  \Gamma (\frac{p}{2} - s') 
Z_p \left( \left(\frac{p}{2} - s'\right) ; \frac{1}{{a_1}},...,\frac{1}{{a_p}}\right).
\end{equation}
The application of the reflection reduces  Eq.~(\ref{A11dd}) to Eq.~(\ref{s21c}) and, as mentioned in Sec.~\ref{zeta}, we calculate the divergent sum over the Matsubara modes using the analytic continuation of zeta function rendered by its reflection formula given by Eq.~(\ref{A323}), which yields our final result given by Eq.~(\ref{s23}).
%
%

For our fourth case, we do the double sums simultaneously using the homogeneous generalized zeta function. As mentioned in Sec.~\ref{zeta}, we first with start the free energy given by Eq.~(\ref{SZeta}), leading to the expression for the free energy of the massless case given by Eq.~(\ref{s21dd}), which can be expressed in terms of homogeneous generalized zeta functions as follows, 
\begin{eqnarray}\label{A12aab}
F_{\mbox{\scriptsize Zeta}} (T,L) = \frac{T A}{16 \pi }  \lim\limits_{s \to 0} \frac{\partial }{\partial s} \frac{\Gamma (s - 1)}{{\mu^{-2 s}} \Gamma (s)} 
\left\lbrace  Z_{1} \left( s - 1 ;  \frac{4 \pi^2}{\beta^2}  \right)   -     Z_{2} \left( s - 1 ; \frac{4\pi^2}{\beta^2} , \frac{\pi^2}{ L^2} \right)      \right\rbrace . \nonumber \\
\end{eqnarray}
Here, we use the zeta function reflection formula given by Eq.~(\ref{A3}) for the first and second terms of Eq.~(\ref{A12aab}), as follows,
\begin{eqnarray}\label{A3b} 
Z_{1} \left( s - 1 ;  \frac{4 \pi^2}{\beta^2}  \right)  &=&  \frac{\beta \Gamma \left(\frac{3}{2} - s\right)}{2 \Gamma (s - 1) {\pi}^{\frac{7}{2} - 2s}}  Z_{1} \left( \frac{3}{2} - s; \frac{\beta^2}{4\pi^2} , \frac{ L^2}{\pi^2} \right)\nonumber \\
Z_{2} \left( s - 1 ; \frac{4\pi^2}{\beta^2} , \frac{\pi^2}{L^2} \right) &=& \frac{\beta L \Gamma (2 - s)}{2 \Gamma (s - 1) {\pi}^{5 - 2s}}  Z_{2} \left( 2 - s; \frac{\beta^2}{4\pi^2} , \frac{ L^2}{\pi^2} \right) \nonumber \\
&=&\frac{ \beta L \Gamma (2 - s)}{2\Gamma (s - 1) {\pi}^{5 - 2s}} \sum\limits_{n_{0} =  - \infty }^\infty  {\sum\limits_{n_{1} =  - \infty }^{\infty'}  \left[ {\left(\frac{{n_0} \beta }{2\pi }\right)}^2 + {\left(\frac{{n_1} L}{\pi }\right)}^2\right]^{(s - 2)}}.
\end{eqnarray}
Using these analytic continuations for the two terms of Eq.~(\ref{A12aab}) and evaluating $\lim\limits_{s \to 0} \frac{\partial}{\partial s}$, the expression for the free energy becomes
\begin{eqnarray}\label{A3cc} 
F_{\mbox{\scriptsize Zeta}} (T,L) &=& \frac{A}{4 \pi} \sum\limits_{{n_0} =  1 }^\infty \frac{1}{(n_0 \beta)^3}  - \frac{AL}{\pi^2} \sum\limits_{{n_0} =  1 }^\infty \frac{1}{(n_0 \beta)^4} - \frac{AL}{16 \pi^2} \sum\limits_{{n_1} =  1 }^\infty \frac{1}{(n_1 L)^4} - \nonumber\\
&&  \frac{2 A L}{ \pi^2 } \sum\limits_{{n_0} =  1 }^\infty  \sum\limits_{{n_1} = 1}^{\infty}  \left( n_0^2 \beta^2 + 4 n_1^2 L^2 \right)^{- 2} .
\end{eqnarray}
After computing the sums over $n_0$ modes in Eq.~(\ref{A3cc}) and simplifying, we obtain our result given by Eq.~(\ref{s23}), where the expression for $F_{\mbox{\scriptsize Casimir}}$ is given by Eq.~(\ref{s12}).

In the last part of this appendix, we use the inhomogeneous generalized zeta functions to compute the free energy of the massive case. As mentioned in Sec.~\ref{Epstein}, we start with the first form of the free energy given by Eq.~(\ref{SZeta}), and compute the sum over $n_1$ modes using the inhomogeneous Epstein zeta function. To do this, we consider the mass term and the regular Matsubara frequencies, {\it i.e.}, $\frac{4{n_0}^2 \pi^2}{\beta^2} + m^2$, as the constant term of Eq.~(\ref{A4}), {\it i.e.}, $M^2$, and use Eq.~(\ref{A8}) to obtain the free energy given by Eq.~(\ref{s33}).
Moreover, as mentioned in Sec.~\ref{Epstein}, the final result given by Eq.~(\ref{s35d}) includes extra terms, which the first and fourth ones are related to a finite value of the free energy for a zero spatial mode. To clarify this point, we first start the first form of the free energy given by Eq.~(\ref{SZeta}) for the case of $n_1=0$, similar to what is done for the massless case, given by Eq.~(\ref{A3ccc}), and after simplifying we obtain
\begin{eqnarray}\label{A10bc} 
F_{\mbox{\scriptsize bounded}}^{n_1=0} (T,L) &=& -\frac{TA}{8 \pi} \lim\limits_{s \to 0} \frac{\partial}{\partial s} \frac{\Gamma (s - 1)}{{\mu^{-2 s}} \Gamma (s)} \Bigg\{ m^{2 - 2s} + \nonumber\\
&&2\sum\limits_{{n_0} =  1 }^\infty \left[\left(\frac{2 \pi n_0}{\beta}\right)^2 +m^2 \right]^{1-s}\Bigg\}.
\end{eqnarray}
Next, we use the inhomogeneous Epstein zeta function given by Eq.~(\ref{A8}) to compute the sum over Matsubara frequencies, and after taking the limit $s \to 0$, we obtain
\begin{eqnarray}\label{A10bcc} 
F_{\mbox{\scriptsize Zeta}}^{n_1=0} (T,L) = -\frac{A m^3}{12 \pi} -A \sum\limits_{j =  1 }^\infty \sqrt{\frac{m^3 T^3}{2 j^3 \pi^3}} K_\frac{3}{2} \left(\beta j m\right).
\end{eqnarray}
Comparing the above result with the first and last terms of $F_{\mbox{\scriptsize Zeta}}(T,L)$ given by Eq.~(\ref{s35dfd}), we observe that these extra terms of $F_{\mbox{\scriptsize Zeta}}$ is equal to minus one half of the contribution of the zero spatial mode.


\section{Calculation of the free energy using the generalized Schl\"{o}milch formulas}
\label{appendixB:The Schlomilch formula}

The original Schl\"{o}milch formula~\cite{r50Sch1., r50Sch2.}, which can be used for evaluating sums is the following,
\begin{eqnarray}\label{B1}
&&\hspace{-4mm}\alpha \sum\limits_{k=1}^{\infty} \frac{k}{e^{2 \alpha k} - 1} + \beta \sum\limits_{k=1}^{\infty} \frac{k}{e^{2 \beta k} - 1} = \frac{\alpha + \beta}{24} - \frac{1}{4},
\end{eqnarray}
where $\alpha, \beta > 0$, and $\alpha \beta= \pi^2$. In this paper, we use the following expression for the generalized form of the Schl\"{o}milch formula~\cite{r50Sch3., r29Junji.} to calculate the Casimir free energy for both massless and massive cases,
\begin{eqnarray}\label{B2}
&&\hspace{-8mm}\sum\limits_{n=1}^{\infty} \ln \left(1 - e^{- \alpha \sqrt{(\theta n)^2 + m^2}}\right)= \sum\limits_{n=1}^{\infty} \ln \left(1 - e^{- \frac{2 \pi}{\theta} \sqrt{\left(\frac{2 \pi n}{\alpha}\right)^2 + m^2}}\right) - \frac{1}{2} \ln \left(1 - e^{- \alpha m}\right) +\nonumber \\
&&\hspace{-8mm} \frac{1}{\theta}  \left[\int_{0}^{\infty} dx \ln \left(1 - e^{- \alpha \sqrt{x^2 + m^2}}\right) + \alpha \int_{m}^{\infty} \frac{dy \sqrt{y^2 - m^2}}{\left(e^{\frac{2 \pi y }{\theta} } - 1\right)}\right]+ \frac{1}{2} \ln \left(1 - e^{-\frac{2 \pi m}{ \theta}}\right).
\end{eqnarray}

In the first part of this appendix, we consider the massless case. As mentioned in Sec.~\ref{zeta}, we start with the second form of the free energy given by Eq.~({\ref{s5}}) and use this method only for the thermal corrections part. To do this, we consider the transverse momenta, {\it i.e.}, $K_T$, of this part as the constant term of Eq.~(\ref{B2}), {\it i.e.}, $m$, and obtain
\begin{eqnarray}\label{B3}
&&\hspace{-4mm}\Delta F_{\mbox{\scriptsize SFA}}(T,L) = \frac{TA}{2 \pi} \int_0^{\infty} dK_T K_T \Bigg\{ \sum\limits_{n = 1}^{\infty}  \ln \left(1 - e^{- 2 L \sqrt{K_T^2 + \left(2 \pi n T\right)^2}}\right) - \frac{1}{2}  \times \nonumber \\
&& \hspace{-4mm}\ln \left(1 - e^{- \beta K_T}\right) + \frac{L}{\pi} \left[\int_0^\infty dx  \ln \left(1 - e^{- \beta \sqrt{K_T^2 + x^2}}\right) + \beta \int_{K_T}^\infty \frac{dy \sqrt{y^2 - K_T^2}}{e^{2 L y} -1}\right] + \nonumber\\
&& \hspace{-4mm}\frac{\ln \left(1 - e^{- 2 L K_T}\right)}{2} \Bigg\}.
\end{eqnarray}
We have denoted the thermal correction part of the Casimir free energy obtained by the Schl\"{o}milch formula approach as $\Delta F_{\mbox{\scriptsize SFA}}$, to distinguish it from the one obtained using the fundamental definition, and the ZFA. Next, we evaluate the integrals over $K_T, x, y$ for all terms and simplify them to obtain
\begin{eqnarray}\label{B4}
 \Delta F_{\mbox{\scriptsize SFA}}(T,L)  &=& -A \sqrt{\frac{2 T^5}{L}} \sum\limits_{n = 1}^\infty 
\sum\limits_{j = 1}^\infty \left(\frac{n}{j}\right)^{\frac{3}{2}} K_{\frac{3}{2}} \left( 4 \pi {n}{j} TL\right) + \frac{A \zeta (3)}{4 \pi} T^3+\nonumber \\
&&\Delta F_{\mbox{\scriptsize free}}(T,L) - F_{\mbox{\scriptsize Casimir}}(0,L) - \frac{A \zeta (3)}{16 \pi L^2} T,  
\end{eqnarray}
where $\Delta F_{\mbox{\scriptsize free}}(T,L)$ is the thermal correction term of the massless free case, given by Eq.~(\ref{s10aa}), and $F_{\mbox{\scriptsize Casimir}}(0,L)= - A \pi^2/\left(1440 L^3\right)$ is the Casimir free energy at zero temperature. After evaluating the sum over ${n}$\footnote{We have used the following identity: \hspace{.3cm} $$\sum\limits_{n = 1}^\infty  \sqrt {n^3}  K_{\frac{3}{2}}(n\alpha) = 	\sqrt {\frac{\pi }{ 8 \alpha^3}} \left[\coth \left(\frac{\alpha}{2}\right) -1 + \left(\frac{\alpha}{2}\right) \csch^2 \left(\frac{\alpha}{2}\right)\right]$$}, and simplifying the above result, we obtain 
\begin{eqnarray}\label{B5}
 \Delta F_{\mbox{\scriptsize SFA}}(T,L)  = F_{\mbox{\scriptsize Casimir}}(T,L) +\frac{A \zeta (3)}{4 \pi} T^3+\Delta F_{\mbox{\scriptsize free}}(T,L) - F_{\mbox{\scriptsize Casimir}}(0,L) ,  
\end{eqnarray}
where $F_{\mbox{\scriptsize Casimir}}$ is given by Eq.~(\ref{s12}). As mentioned in Sec.~\ref{zeta}, the zero temperature part of the free energy should be calculated separately. To do this, we start with the zero temperature part of Eq.~({\ref{s5}}) and compute the integral over $K_T$ using the dimensional regularization, and then evaluate the sum over ${n_1}$ using the analytic continuation of the zeta function, {\it i.e.}, $\zeta(-3)$, obtaining $F_{\mbox{\scriptsize SFA}}(0,L)=F_{\mbox{\scriptsize ZFA}}(0,L) = - A \pi^2/(1440 L^3)$. The final result is, as expected, identical to the free energy obtained using the ZFA given by Eq.(\ref{s23}). That is,
\begin{eqnarray}\label{B51}
	F_{\mbox{\scriptsize SFA}}(T,L)&= F_{\mbox{\scriptsize SFA}}(0,L)+\Delta F_{\mbox{\scriptsize SFA}}(T,L)  =F_{\mbox{\scriptsize ZFA}}(T,L) \\ \nonumber
	&= F_{\mbox{\scriptsize Casimir}}(T,L) +\Delta F_{\mbox{\scriptsize free}}(T,L)+\frac{A \zeta (3)}{4 \pi} T^3,  
\end{eqnarray}
%
%

In the last part of this appendix, we consider the massive case. As mentioned in Sec.~\ref{Epstein}, we start with the second form of the free energy given by Eq.~(\ref{s5}) and use Eq.~(\ref{B2}) to calculate the sum over the spatial modes in the thermal corrections part. To do this, we consider the transverse momenta and mass term, {\it i.e.}, $M^2=m^2+K_T^2$, of this part as the constant term of Eq.~(\ref{B2}), {\it i.e.}, $m$, and obtain  
\begin{eqnarray}\label{B6}
&&\hspace{-8mm} \Delta F_{\mbox{\scriptsize SFA}}(T,L)  =  \frac{A T}{2 \pi} \int_0^\infty dK_T K_T \Bigg\{
\sum\limits_{n = 1}^\infty \ln \left(1 - e^{ - 2 L\omega_{n, K_T} }\right) - \frac{ \ln \left(1 - e^{ - \beta \omega_{K_T}}\right)}{2}  \nonumber \\
&&\hspace{-8mm} - \sum\limits_{j=1}^\infty \frac{L\omega_{K_T} }{\pi j} \left[ K_1 \left(\beta j \omega_{K_T}\right) - \frac{\beta}{2L} K_1 \left(2 L j \omega_{K_T}\right)\right] + \frac{\ln \left(1 - e^{- 2 L \omega_{K_T}}\right)}{2}\Bigg\} ,  
\end{eqnarray}
where $\omega_{n, K_T}=\sqrt{\frac{4 \pi^2 n^2}{\beta^2}+ K_T^2 +m^2}$ and $\omega_{K_T}=\sqrt{K_T^2 +m^2}$. Next, we evaluate all integrals and simplify them to obtain the following expression for the thermal corrections part
\begin{eqnarray}\label{B7}
&&\hspace{-8mm} \Delta F_{\mbox{\scriptsize SFA}}(T,L)  = - \frac{AT}{2}	\Bigg\{L\sum\limits_{j=1}^\infty  \sum\limits_{n=1}^\infty	 \left(\frac{\omega_{n}}{\pi j L}\right)^{\frac{3}{2}} K_{\frac{3}{2}} \left(2 j L \omega_{n}\right) -	\nonumber \\
&&\hspace{-8mm} \sqrt{\frac{T m^3}{2 \pi^3}} \sum\limits_{j=1}^\infty \frac{K_{\frac{3}{2}} \left(j m \beta\right)}{\sqrt{j^3}} + \frac{m^2T L}{ \pi^2} \sum\limits_{j=1}^\infty \frac{K_2 \left( j m \beta\right)}{j^2}\Bigg\} - F_{\mbox{\scriptsize Casimir}}(0,L) -\nonumber \\
&&\hspace{-8mm} \frac{AT}{4} \sqrt{\frac{m^3}{\pi^3 L} } \sum\limits_{j=1}^\infty \frac{K_{\frac{3}{2}} \left(2 j m L\right)}{\sqrt{j^3}},  
\end{eqnarray}
where $\omega_{n} = \sqrt{\left(2 \pi n T\right)^2+ m^2}$, and $F_{\mbox{\scriptsize Casimir}}(0,L)$ is given by Eq.~(\ref{s28}). To simplifying this result, we can write the last term of this expression as a half of the contribution of $n=0$ of the first term. 

As mentioned in Sec.~\ref{Epstein}, we calculate the zero temperature part of the free energy of the massive case separately. Therefore, we start with the zero temperature part of Eq.~(\ref{s5}) and after evaluating the integral over $K_T$ using the dimensional regularization, we compute the sum over the spatial modes using the inhomogeneous Epstein zeta function to obtain
\begin{eqnarray}\label{B8}
\hspace{-8mm}  F_{\mbox{\scriptsize Zeta}}(0,L)  =\frac{ Am^3}{24 \pi} -\frac{A L m^4}{128 \pi^2} \left[3 - 4 \ln \left(\frac{m}{\mu}\right) \right]
+ F_{\mbox{\scriptsize Casimir}}(0,L).
\end{eqnarray}
After considering the zero and finite temperature contributions of the free energy given by Eqs.~(\ref{B7}, \ref{B8}), $F_{\mbox{\scriptsize SFA}}(T,L)$ becomes
\begin{eqnarray}\label{B9}
&&\hspace{-8mm} F_{\mbox{\scriptsize SFA}}(T,L)  =  \frac{ Am^3}{24 \pi} -\frac{A L m^4}{128 \pi^2} \left[3 - 4 \ln \left(\frac{m}{\mu}\right) \right] - 	\sum\limits_{j=1}^\infty  \sum\limits_{n=0}^\infty	 \frac{A\sqrt{\omega_{n}^3} K_{\frac{3}{2}} \left(2 j L \omega_{n}\right)}{2\sqrt{\pi^3 j^3 L}\beta} +	\nonumber \\
&&\hspace{-8mm} A \sqrt{\frac{T^3 m^3}{8 \pi^3}} \sum\limits_{j=1}^\infty \frac{K_{\frac{3}{2}} \left(j m \beta\right)}{\sqrt{j^3}} + \Delta F_{\mbox{\scriptsize free}}(T,L)- \frac{AT}{4} \sqrt{\frac{m^3}{\pi^3 L} } \sum\limits_{j=1}^\infty \frac{K_{\frac{3}{2}} \left(2 j m L\right)}{\sqrt{j^3}}.  
\end{eqnarray}
We can finally use the Abel-Plana summation formula to evaluate the sum over $n$ of the third term of this result, and after simplifying, as expected, the final result is identical to $F_{\mbox{\scriptsize Zeta}}(T,L)$ given by Eq.~(\ref{s35dfd}).


\section{Calculation of the Casimir free energy using the Dimensional Regularization}
\label{appendixD:The Dimensional Regularization}

In this appendix, we calculate the Casimir free energy for a massive real scalar field, based on its fundamental definition, using the second form of the free energy given by Eq.~(\ref{s5}). For the bounded region, we evaluate the integral over the transverse momenta using the dimensional regularization, and obtain 
\begin{eqnarray}\label{D1}
F_{\mbox{\scriptsize bounded}}(T,L) &=&- \frac{A}{2} \sum\limits_{n_1=1}^\infty\lim_{D\to2}\Bigg\{ \frac{\Gamma \left(-\frac{D+1}{2}\right)}{ \sqrt{\left(4 \pi\right)^{D+1}}} \omega_{n_1}^{D+1} +\nonumber \\
&& 4\sum\limits_{j=1}^{\infty}  \left(\frac{\omega_{n_1}T}{2 \pi j}\right)^{\frac{D+1}{2}} K_{\frac{D+1}{2}} \left( \beta j \omega_{n_1}\right) \Bigg\}, 
\end{eqnarray}
where $\omega_{n_1}=\sqrt{\frac{n_1^2 \pi^2}{L^2}+m^2}$. Then, we evaluate the sum over the regular spatial modes using the Principle of the Argument theorem, which is the same procedure as done for the bounded case in Sec.~\ref{massive} given by Eq.~(\ref{s241}) (see Appendix~\ref{appendixB:The Argument Principle}), and obtain
\begin{eqnarray}\label{D2}
F_{\mbox{\scriptsize bounded}}(T,L) &=&A  \lim_{D\to2} \int_0^\infty dp\Bigg\{ 
 \left[\frac{p^D}{\Gamma \left( \frac{D+1}{2}\right) \sqrt{\left(4 \pi\right)^{D+1}}} + \left(\frac{p}{2 \pi}\right)^{\frac{D+1}{2}}\times \right. \nonumber \\
&& \left.  \sum\limits_{j=1}^{\infty}  \frac{J_{\frac{D-1}{2}} \left( \beta j p\right)}{\sqrt{\left(\beta j\right)^{D-1}}}\right] \left[L \omega (p) +\ln \left(1 - e^{- 2 L \omega (p)}\right)\right]  \Bigg\}, 
\end{eqnarray}
where $\omega (p) = \sqrt{p^2 + m^2}$. After evaluating the integral over $p$ for all terms of the above expression, and simplifying, the free energy for the bounded case becomes
\begin{eqnarray}\label{D3}
\hspace{-2mm}F_{\mbox{\scriptsize bounded}}(T,L) &=&- \frac{A L}{\left(2 \pi\right)^{\frac{D+2}{2}}}  \lim_{D\to2} \Bigg\{\frac{m^{D+2} \Gamma \left(- \frac{D+2}{2}\right)}{2^{\frac{D+4}{2}}} +2 \left(\frac{m}{\beta}\right)^{\frac{D+2}{2}}   \times\nonumber \\
&&  \sum\limits_{n_0=1}^{\infty}  \frac{K_{\frac{D+2}{2}} \left( \beta n_0 m\right)}{ n_0^{\frac{D+2}{2}}} +2 \left( \frac{m}{2L}\right)^{\frac{D+2}{2}}  \sum\limits_{n_1=1}^{\infty}  \frac{K_{\frac{D+2}{2}} \left( 2 n_1 mL\right)}{ n_1^{\frac{D+2}{2}}} + \nonumber \\
&& 4 \left( \frac{m}{\beta}\right)^{\frac{D+2}{2} } \sum\limits_{n_0=1}^{\infty}  \sum\limits_{n_1=1}^{\infty} \frac{K_{\frac{D+2}{2}} \left( \beta m \omega_{n_0, n_1}\right)}{ \left(\omega_{n_0, n_1}\right)^{\frac{D+2}{2}}}\Bigg\} ,
\end{eqnarray}
where $\omega_{n_0, n_1} = \sqrt{n_0^2 + \left(2 n_1 T L\right)^2}$. Next, we calculate the free energy of the free case at finite temperature, by starting with the second form of the free energy, and using dimensional regularization to calculate the integrals over momenta and obtain
\begin{equation}\label{D4}
\hspace{-4mm}F_{\mbox{\scriptsize free}}(T,L) =- AL  \lim_{D\to2} \Bigg\{\frac{m^{D+2} \Gamma \left(- \frac{D+2}{2}\right)}{2 \left(4 \pi\right)^{\frac{D+2}{2}}} +2 m^{\frac{D+2}{2}} \sum\limits_{n_0=1}^{\infty}  \frac{K_{\frac{D+2}{2}} \left( \beta n_0 m\right)}{ \left(2 \pi \beta n_0\right)^{\frac{D+2}{2}}} \Bigg\} .
\end{equation}
As can be seen, the first two terms of $F_{\mbox{\scriptsize bounded}}(T,L)$, given by Eq.~(\ref{D3}), are identical to the two terms of $F_{\mbox{\scriptsize free}}(T,L)$, given by Eq.~(\ref{D4}). Notice that the first terms actually diverge as $\Gamma \left( -2\right)$, in contrast to the analogous terms in Eqs.~(\ref{s24}, \ref{s25}) which are obtained using the first form of the free energy Eq.~(\ref{s4}), which has an embedded analytic continuation. After subtracting these terms, and taking the limit $D \to 2$, we obtain the same expression for the Casimir free energy as in Eq.~(\ref{s27}).

In this appendix, we also show that if we use the generalized Abel-Plana summation formula to calculate the sum over the regular spatial modes in Eq.~(\ref{D1}) for the bounded case, the Casimir free energy is identical to the expression given by Eq.~(\ref{s27}). For do this, we also consider $F_{\mbox{\scriptsize free}}(T,L)$ in the second form of the free energy and evaluate the integral over the transverse momenta using the dimensional regularization, and present the Casimir free energy as follows
\begin{eqnarray}\label{D5}
&&\hspace{-9mm}F_{\mbox{\scriptsize Casimir}}(T,L) =\lim_{D\to2}\Bigg\{ -\frac{A \Gamma \left(-\frac{D+1}{2}\right)}{2 \sqrt{\left(4 \pi\right)^{D+1}}} \left[\sum\limits_{n_1=1}^\infty \omega_{n_1}^{D+1} -\int_0^\infty dk'\omega_{k'}^{D+1} \right]- \frac{2 A}{\left(2 \pi \right)^{\frac{D+1}{2}}}  \times\nonumber \\
&&\hspace{-8mm} \sum\limits_{j=1}^{\infty}  \sqrt{\frac{T^{D+1}}{ j^{D+1}}} \left[\sum\limits_{n_1=1}^\infty \sqrt{\omega_{n_1}^{D+1}} K_{\frac{D+1}{2}} \left( \beta j \omega_{n_1}\right) -\int_0^\infty dk'\sqrt{\omega_{k'}^{D+1}} K_{\frac{D+1}{2}} \left( \beta j \omega_{k'}\right)\right]\Bigg\}, 
\end{eqnarray}
where $\omega_{n_1}=\sqrt{\frac{n_1^2 \pi^2}{L^2}+m^2}$ and $\omega_{k'}=\sqrt{\frac{{k'}^2 \pi^2}{L^2}+m^2}$.
Notice that each term in the first bracket of Eq.~(\ref{D5}) is divergent which one can easily calculate and show that the divergent part of each term is exactly identical to the first term of Eqs.~(\ref{D3}, \ref{D4}).\\
By evaluating the sum over $n_1$ modes of Eq.~(\ref{D5}) using the generalized Abel-Plana summation formula, as the same procedure as done in Appendix \ref{appendixC:FCasimir} to calculate the sum over the spatial modes for the massive case, we obtain
\begin{eqnarray}\label{D6}
\hspace{-3mm}F_{\mbox{\scriptsize Casimir}}(T,L) &=& -ALm\lim_{D\to2}\Bigg\{-\frac{m^{D+1}}{ \sqrt{\left(4 \pi\right)^{D+1}}  \Gamma \left(\frac{D+3}{2}\right)} \int_1^\infty dt \frac{\sqrt{\left(t^2 -1\right)^{D+1}}}{e^{2mLt}-1} + \nonumber \\
&&\hspace{-6mm}  \sum\limits_{j=1}^{\infty} \sqrt{\left( \frac{Tm}{ 2 \pi j}\right)^{D+1}} \int_0^\infty dt \frac{2\sqrt{t^{D+3}} J_{\frac{D+1}{2}} \left(\beta j m t\right)}{\sqrt{t^2+1} \left(e^{2mL \sqrt{t^2+1}} - 1\right)} \Bigg\}.  
\end{eqnarray}
Then, by computing the integral over $t$, simplifying, and taking $D\to 2$, we obtain the same expression for the Casimir free energy as in Eq.~(\ref{s27}).


 \section{Calculation of the Casimir free energy for massive scalars using the Boyer method}
\label{appendixE:The Boyer method}

\begin{figure}[h!] 
	\centering
	\includegraphics[scale=.4]{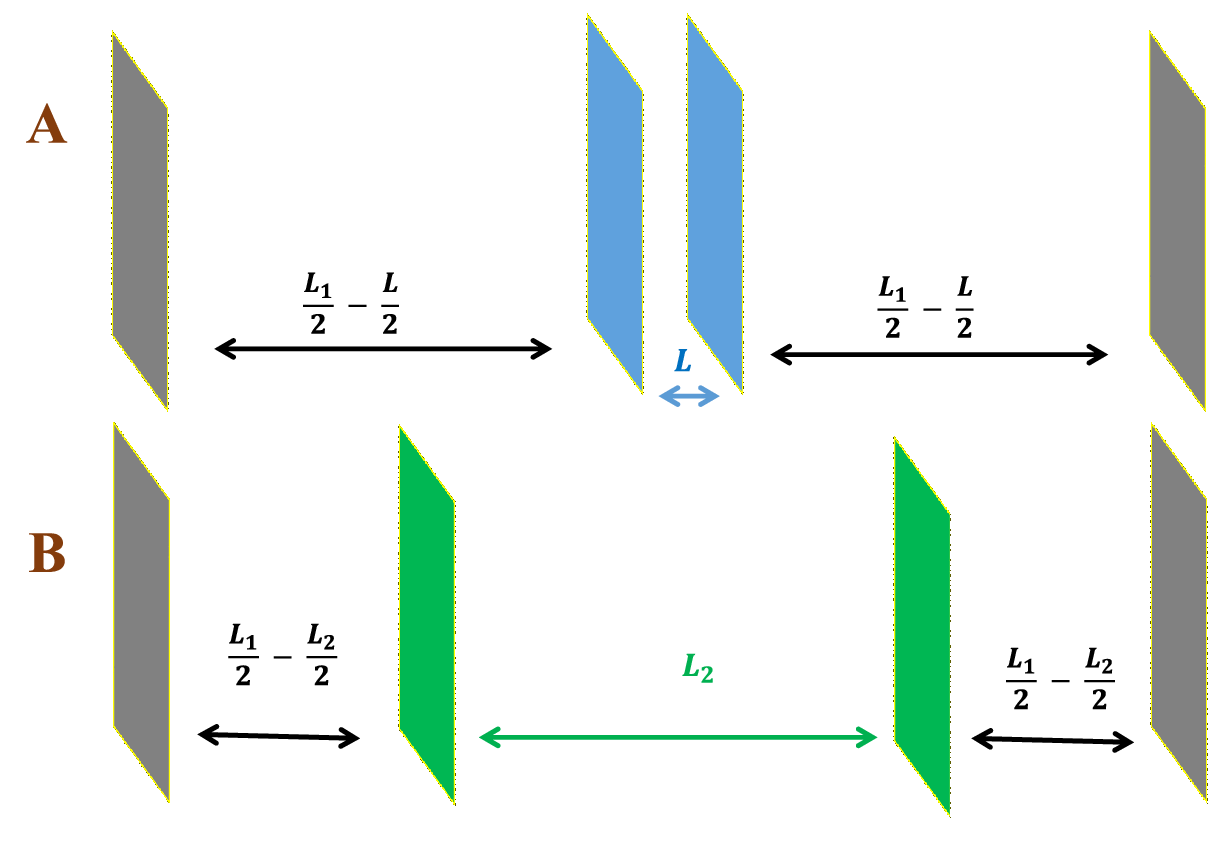}
	\caption{\label{fp100} \small
		The geometry of the two different configurations whose free energies for a massive real scalar field are to be subtracted. In the upper (lower) configuration, labeled $A$ ($B$), two inner plates are located at $z=\pm L/2$ ($z=\pm L_2/2$) surrounded by two outer plates located at $z=\pm L_1 /2$. We impose the Dirichlet boundary conditions on all plates.}
\end{figure}

In this appendix, we calculate the Casimir free energy for a massive real scalar field using the Boyer method~\cite{r50Boyer.} and show that the final result is equivalent to the result given in Eq.~(\ref{s27}), obtained using the fundamental approach. In this method, we subtract the free energies of two configurations, at the same temperature, and obtain the Casimir free energy of our original system by taking appropriate limits. Configuration $A$ consists of two inner plates located at $z=\pm L/2$ surrounded by two outer plates located at $z=\pm L_1 /2$. Configuration $B$ is similar to $A$ except the two inner plates are located $z=\pm L_2 /2$, with $ L< L_2 <L_1$, as depicted in Fig.~(\ref{fp100}).  

The Casimir free energy can be defined in terms of the difference between the free energies of configurations $A$ and $B$ as follows
\begin{eqnarray}\label{E1}
F_{\mbox{\scriptsize Casimir}}(T,L) =\lim\limits_{L_2 \to \infty} \Big\lbrace \lim\limits_{L_1 \to \infty} \left[ F_A(T,L,L_1) - F_B(T,L_2,L_1)\right] \Big\rbrace ,
\end{eqnarray}
where $F_A(T,L,L_1) = F_{\mbox{\scriptsize bounded}}^{I}(T,L)+2 F_{\mbox{\scriptsize bounded}}^{II}(T,L,L_1)$ and $F_B(T,L_2,L_1) = F_{\mbox{\scriptsize bounded}}^{I}(T,L_2)+2 F_{\mbox{\scriptsize bounded}}^{II}(T,L_2,L_1)$. Moreover, $F_{\mbox{\scriptsize bounded}}^{I}$ denotes the free energy between two inner plates and the $F_{\mbox{\scriptsize bounded}}^{II}$ denotes the free energy of bounded regions adjacent to the inner plates.
In fact, the Boyer method can be thought of as a rigorous implementation of the fundamental definition, provided the two configurations are taken to be at the same temperature, in which any possible contributions from the regions outside of the bounded region is also taken into account.


To calculate the free energy for each of six regions shown in Fig.~(\ref{fp100}), we use the result obtained in the Sec.~\ref{massive} for the $F_{\mbox{\scriptsize bounded}}$, given by Eq.~(\ref{s24}). For example, the free energy for the outer bounded regions of configuration $B$ becomes
\begin{eqnarray}\label{E2}
&&\hspace{-8mm}F_{\mbox{\scriptsize bounded}}^{II}(T,L_2,L_1) =  -\frac{A (L_1 - L_2)}{16	\sqrt{\pi^5}} \lim\limits_{s \to 0} \frac{\partial}{\partial s}\frac{ \Gamma \left(s - \frac{1}{2}\right)}{{\mu^{-2 s}}\Gamma (s)} \int_{ 0}^{ \infty}  p^{2 - 2s}  \omega(p) dp -\frac{A m^2}{ \pi^2} \times\nonumber \\ 
&&\hspace{-8mm} \Bigg\{\frac{ \left(L_1 - L_2\right)}{4 \beta^2} \sum\limits_{{n_0} = 1}^\infty \frac{K_2 \left(n_0 \beta m\right)}{  n_0^2 } + \frac{1}{4 \left(L_1 - L_2\right)}\sum\limits_{{n_1} = 1}^\infty \frac{K_2 \left(n_1 m \left(L_1 - L_2\right)\right)}{ n_1^2} +
\frac{T^2 }{2}\times\nonumber \\ 
&&\hspace{-8mm} \left(L_1 - L_2\right) \sum\limits_{{n_0} = 1}^\infty \sum\limits_{{n_1} = 1}^\infty \frac{K_2 \left(\beta m {\omega}'_{n_0, n_1}\right)}{\left({\omega}'_{n_0, n_1}\right)^2} \Bigg\},
\end{eqnarray} 
where ${\omega}'_{n_0, n_1} = \sqrt{n_0^2 + n_1^2 T^2 \left( L_1 - L_2\right)^2}$.
As mentioned before, only the first term of the above expression contains a divergent part. Moreover, $F_{\mbox{\scriptsize bounded}}^{II}(T,L_2,L_1)=F_{\mbox{\scriptsize bounded}}^{II}(T,L_1-L_2)$ and the first two terms are linear in $L_1-L_2$. Adding the contributions of the three regions of configuration $B$ we obtain
\begin{eqnarray}\label{E3}
&&\hspace{-8mm }F_B(T,L_2,L_1) =  -\frac{A L_1}{8\sqrt{\pi^5}} \lim\limits_{s \to 0} \frac{\partial}{\partial s}\frac{ \Gamma \left(s - \frac{1}{2}\right)}{{\mu^{-2 s}}\Gamma (s)} \int_{ 0}^{ \infty}  p^{2 - 2s}  \omega(p) dp	-\frac{A m^2}{ \pi^2} \times\nonumber \\ 
&&\hspace{-8mm} \Bigg\{\frac{ L_1}{2 \beta^2} \sum\limits_{{n_0} = 1}^\infty \frac{K_2 \left(n_0 \beta m\right)}{  n_0^2 } +\sum\limits_{{n_1} = 1}^\infty \left[\frac{K_2 \left(2n_1 m L_2\right)}{ 8 L_2 n_1^2}+\frac{K_2 \left(n_1 m \left(L_1 - L_2\right)\right)}{ 2 \left(L_1 - L_2\right) n_1^2}\right] +\nonumber \\ 
&&\hspace{-8mm} T^2\sum\limits_{{n_0} = 1}^\infty \sum\limits_{{n_1} = 1}^\infty \left[  L_2  \frac{K_2 \left(\beta m {\omega}''_{n_0, n_1}\right)}{\left({\omega}''_{n_0, n_1}\right)^2}+\left(L_1- L_2\right)  \frac{K_2 \left(\beta m {\omega}'_{n_0, n_1}\right)}{\left({\omega}'_{n_0, n_1}\right)^2} \right]\Bigg\}   ,
\end{eqnarray} 
where ${\omega}''_{n_0, n_1} = \sqrt{n_0^2 +  \left( 2 n_1 T L_2\right)^2}$.
We can compute $F_A(T,L,L_1)$ similarly. Upon using Eq.~(\ref{E1}) to calculate $F_{\mbox{\scriptsize Casimir}}(T,L)$, the first two terms of $F_A(T,L,L_1)$ and $F_B(T,L_2,L_1)$, which include divergent integrals, cancel even before we take the limits and we obtain
\begin{eqnarray}\label{E4}
&&F_{\mbox{\scriptsize Casimir}}(T,L) = - \frac{Am^2}{\pi^2} \sum\limits_{n_1=1}^{\infty} \lim\limits_{L_2 \to \infty} \Bigg\{ \lim\limits_{L_1 \to \infty} \left[ \frac{K_2 \left(2n_1 m L\right)}{ 8 L n_1^2}-\frac{K_2 \left(2n_1 m L_2\right)}{ 8 L_2 n_1^2} +\right. \nonumber \\
&& \hspace{-9mm}\left. \frac{K_2 \left(n_1 m \left(L_1 - L\right)\right)}{ 2 \left(L_1 - L\right) n_1^2}-\frac{K_2 \left(n_1 m \left(L_1 - L_2\right)\right)}{ 2 \left(L_1 - L_2\right) n_1^2} + T^2\sum\limits_{n_0 = 1}^\infty \left[  \frac{L K_2 \left(\beta m {\omega}_{n_0, n_1}\right)}{\left({\omega}_{n_0, n_1}\right)^2} -L_2 \times  \right. \right.   \nonumber \\
&&  \hspace{-9mm} \left.  \left. \frac{ K_2 \left(\beta m {\omega}''_{n_0, n_1}\right)}{\left({\omega}''_{n_0, n_1}\right)^2} +  \frac{\left(L_1- L\right)K_2 \left(\beta m {\omega}'''_{n_0, n_1}\right)}{\left({\omega}'''_{n_0, n_1}\right)^2} -  \frac{\left(L_1- L_2\right) K_2 \left(\beta m {\omega}'_{n_0, n_1}\right)}{\left({\omega}'_{n_0, n_1}\right)^2}
 \right] \right]\Bigg\},
\end{eqnarray}
where ${\omega}'''_{n_0, n_1} = \sqrt{n_0^2 + n_1^2 T^2 \left( L_1 - L\right)^2}$.
Finally, upon taking the limits $L_1 \to \infty$, and $L_2 \to \infty$, sequentially, we obtain the same expression for the Casimir free energy given by Eq.~(\ref{s27}). This proves that, when using the fundamental approach, the contributions from the outer regions in the bounded case are precisely canceled by the corresponding contributions of the free case.


\section{The heat kernel coefficients}
\label{appendixF:The Heat kernel method}

The heat kernel expansion is an important tool in the computations of the Casimir effects, which can be used to obtain the high and low temperature limits of the Casimir thermodynamics quantities, including the divergences in the vacuum energy~\cite{r31Bord2., r12Kris., r17Geyer.}. In the first part of this appendix, we obtain the divergent term of the energy at zero temperature for our model by calculating the heat kernel coefficients. To obtain the energy, we use the partition function at zero temperature for a massive free real scalar in path integral representation:
\begin{eqnarray}\label{F1}
Z[0] &=& \int  D \phi \exp\left\{ i \int {d^4 x} \frac{1}{2}\left[ \left( \partial_{\mu}\phi\right)^2 -m^2 \phi^2 \right] \right\}  \nonumber \\ 
&=& \left[\det \left( \frac{\partial^2 + m^2}{\mu^2} \right)\right]^{-\frac{1}{2}},
\end{eqnarray}
where $\mu$ is an arbitrary mass scale introduced for dimensional reasons, as explained in Sec.~\ref{Helmholtz free energy}.
Using the effective action, the vacuum energy for time-independent boundaries is obtained as \cite{r31Bord2.}
\begin{eqnarray}\label{F2}
E =  \frac{i}{T } \ln(Z[0]) &=&   -\frac{i}{2T } \ln\left[\det \left(\frac{\partial^2 + m^2 }{\mu^2}\right)\right]= - 
\frac{i}{2T}  \Tr \left[ \ln\left(\frac{ -P^{2} + m^2 }{\mu^2}\right) \right],\nonumber \\
\end{eqnarray}
where $T$ is the total time and the trace indicates the summation over eigenvalues of Klein-Gordon operator in the momentum space representation. The explicit form of the energy at zero temperature is
\begin{eqnarray}\label{F3}
&&\hspace{-9mm}E _{\mbox{\scriptsize bounded}}(0,L)= \frac{i}{2 T} \int_{-\infty}^{\infty} \frac{Td\omega}{2 \pi} \int \frac{Ad^2 K_T}{\left(2 \pi \right)^2}  \sum\limits_{n_1} \lim\limits_{s \to 0}   \frac{\partial }{\partial s}  \frac{\left[ - \omega^2 + \omega _{{n_1} , {K_T}}^2 \right]^{-s}}{{\mu^{-2s}}}\nonumber \\
&&\hspace*{+9mm} = \frac{i}{2} \int_{-\infty}^{\infty} \frac{d\omega}{2 \pi} \int \frac{Ad^2 K_T}{\left(2 \pi \right)^2}  \sum\limits_{n_1} \lim\limits_{s \to 0}   \frac{\partial }{\partial s} \int_{0}^\infty \frac{dt e^{-t\left( - \omega^2 + \omega _{{n_1} , {K_T}}^2\right)}}{{\mu^{-2s}} t^{1-s}\Gamma (s)},
\end{eqnarray}
where $\omega_{{n_1} , {K_T}}=\sqrt{K_T^2+ k_{n_1}^2+m^2}$. Due to the Dirichlet boundary conditions at the plates for the massive case, the longitudinal momentum $k_{n_1}$ takes on discrete regular values which are solutions to Eq.~(\ref{s1000}) and given by Eq.~(\ref{s1b}).
Note that the second form of energy given by Eq.~(\ref{F3}) has an embedded analytic continuation, as mentioned in Sec.~\ref{Helmholtz free energy}. After performing a wick rotation on $\omega$, we evaluate its integral and, to obtain the nonzero heat kernel coefficients for our model at zero temperature, present the result in terms of the spatial heat kernel as 
\begin{equation}\label{F4}
E _{\mbox{\scriptsize bounded}}(0,L)= -\lim\limits_{s \to 0} \frac{\partial }{\partial s} \frac{{\mu^{2s}}}{\Gamma (s)}\int_{0}^\infty \frac{dt e^{ -tm^2}}{4\sqrt{\pi}t^{\frac{3}{2} - s}} \mathbf{K}(t) \equiv -\lim\limits_{s \to 0} \frac{\partial }{\partial s} 
\tilde{E} _{\mbox{\scriptsize bounded}}(s),
\end{equation}
%
%
where the spatial heat kernel for our model is the following
	\begin{equation}\label{F5}
		\mathbf{K}(t)=\int \frac{Ad^2 K_T}{\left(2 \pi \right)^2} \sum\limits_{n_1} e^{ - t \left(K_T^2+k_{n_1}^2\right) } = \frac{A}{4\pi t} \sum\limits_{n_1} e^{-t k_{n_1}^2}.
\end{equation}
In the second part of  Eq.~(\ref{F4}) we have defined $\tilde{E} _{\mbox{\scriptsize bounded}}(s)$. The spatial heat kernel, which we shall henceforth refer to it simply as the heat kernel, obeys the heat conduction equation with the initial condition $\mathbf{K}(\bold{r}, \bold{r'}| t=0) = \delta^3 (\bold{r}-\bold{r'})$. Hence, its behavior for small $t$ describes the divergences in the vacuum energy. It has th following expansion for small $t$~\cite{r31Bord2.,r3Vassil},
\begin{equation}\label{F6}
		\hspace{-5mm}	\mathbf{K}(t) \xrightarrow{t< 1}\frac{1}{(4\pi t)^{\frac{3}{2}}} \sum\limits_{n=0}^\infty a_{\frac{n}{2}} t^{\frac{n}{2}}.
\end{equation}
where $ a_{n/2}$ are the heat kernel coefficients for the massless case. The overall coefficient of the sum is actually $t^{-d/2}$, where $d=3$ is the dimension of space. The integrand in Eq.~(\ref{F4}) has a pole at $t=0$. So, to obtain the divergent part of this integral, we divide the interval of the integration into $t \in [0, 1]$ and $t \in (1,\infty)$. Then we need to only evaluate the integral over $t$ in the first interval, for which we use the expansion of $\mathbf{K}(t)$, given by Eq.~(\ref{F6}), and the exponential mass term as follows:
\begin{eqnarray}\label{F7}
\mathbf{K}(t) e^{-t m^2} \xrightarrow{t< 1} \frac{1}{(4\pi t)^{\frac{3}{2}}} \sum\limits_{n=0}^\infty a_{\frac{n}{2}} t^{\frac{n}{2}} \sum\limits_{k=0}^\infty \frac{(-1)^{k} m^{2k} t^k}{k!}= \nonumber\\
\frac{1}{(4\pi t)^{\frac{3}{2}}} \sum\limits_{j=0}^\infty \left(\sum\limits_{k=0}^{[\frac{j}{2}]} a_{\frac{j}{2} - k} \frac{(-1)^k  m^{2k}}{k!}\right) t^{\frac{j}{2}}\equiv
\frac{1}{(4\pi t)^{\frac{3}{2}}}\sum\limits_{j=0}^\infty \alpha_{\frac{j}{2}} t^{\frac{j}{2}},
\end{eqnarray}
where the second expression has been obtained by combining powers of $t$. In the last expression we have defined the heat kernel coefficients for the massive case as $\alpha_{\frac{j}{2}}$. Then we insert this expression into Eq.~(\ref{F4}) and integrate to obtain 
 \begin{equation}\label{F8}
\tilde{E} _{\mbox{\scriptsize bounded}}(s)= -\frac{1}{32 \pi^2} \sum\limits_{j=0}^\infty \left[\frac{\alpha_{\frac{j}{2}}}{s-2+\frac{j}{2}}\right].
\end{equation}
The expression in the bracket in Eq.~(\ref{F8}) has a simple pole at $j=4$, with coefficient $\alpha_2$. Therefore its divergent part is
\begin{equation}\label{F9}
\tilde{E} _{\mbox{\scriptsize bounded}}^{\mbox{\scriptsize div. part}}(s)= -\frac{1}{64 \pi^2}   \left[\frac{2 a_2-2 a_1 m^2+ a_0 m^4}{s}\right].
\end{equation}
Now, we obtain the heat kernel coefficients, {\it i.e.}, $a_{\frac{n}{2}}$, which include a sum of two local integrals, one over the volume and the other over the surface~\cite{r31Bord2.}.
According to our model, which includes two parallel plates with Dirichlet boundary conditions, the surface part of the plates and the volume part give a nonzero contribution for each part. So, the only nonzero coefficients for the volume and surface parts are $a_0=AL$ and $a_{\frac{1}{2}} = - A \sqrt{\pi}$, respectively. Hence, the divergence of the integral part of the expression for the energy at zero temperature, given by Eq.~(\ref{F4}), can be inferred from Eq.~(\ref{F9}), which yields $-\frac{ALm^4}{64 \pi^2s}$.
Note that this is precisely the divergent term that appears in $F _{\mbox{\scriptsize bounded}}(0,L)$ and $F _{\mbox{\scriptsize free}}(0,L)$ shown Eq.~(\ref{F10}), and also in the second term in Eq.~(\ref{s34}) for $F_{\mbox{\scriptsize Zeta}}(T,L)$.
%
%

In the last part of this appendix, we obtain the nonzero heat kernel coefficients for our model at high temperatures. We first, present Eq.~(\ref{s4}) in terms of the spatial heat kerne, and express the sum over Matsubara frequencies as the sum over the positive integers and a zero mode to obtain
\begin{equation}\label{F11}
F_{\mbox{\scriptsize bounded}}(T,L) = -\frac{T}{2} \lim\limits_{s \to 0} \frac{\partial}{\partial s}  \int_0^\infty \frac{dt e^{-t m^2} }{\Gamma (s) t^{1-s}} \mathbf{K}(t) \left[1 + 2 \sum\limits_{n_0 = 1}^\infty e^{-t \left(\frac{2 n_0 \pi}{\beta}\right)^2}\right],
\end{equation}
where the $\mathbf{K}(t)$ is given by Eq.~(\ref{F5}) for our model. As stated in~\cite{r31Bord2.}, the behavior of the above expression as $T \to \infty$, is determined by the behavior of the heat kernel $\mathbf{K}(t)$ as $t \to 0$. We evaluate the integral over $t$ using the expansion of $e^{-t m^2}\mathbf{K}(t)$ given by Eq.~(\ref{F7}), and obtain the following expression in the high temperature limit
\begin{eqnarray}\label{F12}
\hspace{-9mm} F_{\mbox{\scriptsize bounded}}(T,L) &\xrightarrow[T\gg m]{TL\gg 1}& -\frac{T}{8\sqrt{\pi^3}} \sum\limits_{j=0}^{\infty}  \lim\limits_{s \to 0} \frac{\partial}{\partial s} \frac{1}{\Gamma (s) }  \mathbf{\alpha}_{\frac{j}{2}} \left[ \frac{1}{j + 2s -3} + \right. \nonumber \\ 
&& \left. \left(\frac{2\pi}{\beta}\right)^{3-2s-j} \Gamma \left(\frac{2s-3+j}{2}\right)  \zeta (2s+j-3) \right].
\end{eqnarray}
As can be seen, the first term in the bracket in Eq.~(\ref{F12}), which is proportional to $T$, has only one simple pole at $j=3$, whereas the second term in the bracket includes one pole in the gamma function and one in the zeta function. Evaluating $\lim\limits_{s \to 0}\partial/\partial s$, we obtain the following expression for the high temperature expansion of the free energy
\begin{eqnarray}\label{F13}
\hspace{-9mm}F_{\mbox{\scriptsize bounded}}(T,L) &\xrightarrow[T\gg m]{TL\gg 1}& - \frac{ \pi^2 \mathbf{\alpha}_0}{90}  T^4   - \frac{ \zeta (3) \mathbf{\alpha}_{\frac{1}{2}}}{4 \sqrt{\pi^3}} T^3 - \frac{\mathbf{\alpha}_1}{24} T^2 -\frac{\mathbf{\alpha}_{\frac{3}{2}}}{8 \sqrt{\pi^3}}T \left[\gamma +\ln \left(T\right)\right] -\nonumber\\
&&\frac{\mathbf{\alpha}_2}{16 \pi^2}\left(\gamma - \ln (4 \pi T)\right) +... .
\end{eqnarray} 
As mentioned above, the only nonzero heat kernel coefficients for our model are $a_0$ and $a_{\frac{1}{2}}$. So, we can easily obtain the heat kernel coefficients $\mathbf{\alpha}_{\frac{j}{2}}$ that appear in Eq.~(\ref{F13}), using their definition given in Eq.~(\ref{F7}), in the high temperature limit and express them as follows
\begin{eqnarray}\label{F14}
\mathbf{\alpha}_0=AL,   \quad  &&\mathbf{\alpha}_{\frac{1}{2}}= - \sqrt{\pi} A,    \quad\quad   \mathbf{\alpha}_1=- m^2 AL,    \quad\quad  \mathbf{\alpha}_{\frac{3}{2}} = \sqrt{\pi}  m^2 A,  \nonumber \\
&&  \mathbf{\alpha}_{2}= \frac{m^4}{2} AL . 
\end{eqnarray}
Now we can compare the high temperature expansion obtained here for $F_{\mbox{\scriptsize bounded}}(T,L)$, using the heat kernel method, with that of $F_{\mbox{\scriptsize Zeta}}(T,L)$ given in Eq.~(\ref{s35dfdBB}). We note that the coefficients of all of the temperature-dependent terms of the former are correct except for the linear or the classical term. This term is the result of the first term in the square bracket in Eq.~(\ref{F11}), and has to be computed separately and without any expansions. After evaluating the integral over $t$ for this term, we obtain
\begin{eqnarray}\label{F18}
 F_{\mbox{\scriptsize bounded}}^{\mbox{\scriptsize class. term}}(T,L) = -\frac{TA}{8 \pi} \lim\limits_{s \to 0} \frac{\partial}{\partial s} \frac{ \Gamma (s -1 )}{{\mu^{-2s}}\Gamma (s)} E_1^{m^2} \left(s-1 ; \frac{\pi^2}{L^2}\right).
\end{eqnarray}
Next, we use Eq.~(\ref{A8}) to compute this term in high temperature limit, and obtain 
	\begin{eqnarray}\label{F19}
		F_{\mbox{\scriptsize bounded}}^{\mbox{\scriptsize class. term}}(T,L)\xrightarrow[T \gg m]{TL\gg 1}  -\frac{TAL}{4 \sqrt{\pi^3}} \lim\limits_{s \to 0} \frac{\partial}{\partial s} \frac{{\mu^{2s}}}{\Gamma (s)} \sum\limits_{j=1}^{\infty} \left(\frac{m}{jL}\right)^{ \frac{3}{2}- s} K_{\frac{3}{2}- s} \left(2jmL\right) 
	\end{eqnarray}
After taking the $\lim\limits_{s \to 0} \partial/\partial s$, the $\mu$ factor disappears, and the result is identical to the classical term in $F_{\mbox{\scriptsize Zeta}}$ in the high temperature limit, given by Eq.~(\ref{s35dfdBB}). On a side note, the temperature-independent term in Eq.~(\ref{F13}) is, not surprisingly, incorrect.


\end{document}